\pdfoutput=1
\documentclass[11pt,twoside,a4paper,cmspaper,final,collab]{cms-tdr}
\def\svnVersion{c5de115}\def\svnDate{2021/12/20}\def\cmsCernNoTag{CERN-EP-2021-171}\def\cmsCernDate{\today}\def\cmsMessage{Published in the Journal of High Energy Physics as \href{http://dx.doi.org/10.1007/JHEP01(2022)177}{\doi{10.1007/JHEP01(2022)177}.}}
\begin{document}\cmsNoteHeader{SMP-20-007}

\providecommand{\cmsTable}[1]{\resizebox{\textwidth}{!}{#1}}

\providecommand{\cmsLeft}{left\xspace} \providecommand{\cmsRight}{right\xspace}

\newlength\cmsTabSkip\setlength{\cmsTabSkip}{1ex}

\newcommand{\ptst}{\ensuremath{p_{\text{T,1st}}}\xspace}
\newcommand{\ptnd}{\ensuremath{p_{\text{T,2nd}}}\xspace}
\newcommand{\ptrd}{\ensuremath{p_{\text{T,3rd}}}\xspace}
\newcommand{\ptth}{\ensuremath{p_{\text{T,4th}}}\xspace}

\newcommand{\DS}{\ensuremath{\Delta S}\xspace}
\newcommand{\DSDPS}{\ensuremath{\Delta S(\text{DPS template})}\xspace}
\newcommand{\DY}{\ensuremath{\Delta Y}\xspace}
\newcommand{\DptS}{\ensuremath{\Delta p_{\text{T,Soft}}}\xspace}
\newcommand{\Dphimin}{\ensuremath{\Delta \phi_\mathrm{3j}^\text{min}}\xspace}
\newcommand{\phiij}{\ensuremath{\phi_{ij}}\xspace}
\newcommand{\DphiS}{\ensuremath{\Delta \phi_\text{Soft}}\xspace}

\newcommand{\DScuts}{\ensuremath{\mathbf{p}_T^{\DS}}\xspace}
\newcommand{\cuts}{\ensuremath{\mathbf{p}_T}\xspace}

\newcommand{\joinR}{\hspace{-.1em}}
\newcommand{\RomanI}{I}
\newcommand{\RNum}[1]{\uppercase\expandafter{\romannumeral #1\relax}}
\newcommand{\regioni}{\ensuremath{Region}\,{\RomanI}\xspace}
\newcommand{\regionii}{\ensuremath{Region\,{\mbox{\RomanI\joinR\RomanI}}}\xspace}

\newcommand{\sigmaeff}{\ensuremath{\sigma_\text{eff}}\xspace}
\newcommand{\sigmaDPS}{\ensuremath{\sigma_\mathrm{A,B}^\mathrm{DPS}}\xspace}
\newcommand{\sigmaA}{\ensuremath{\sigma_\mathrm{A}}\xspace}
\newcommand{\sigmaB}{\ensuremath{\sigma_\mathrm{B}}\xspace}

\newcommand{\ptvecst}{\ensuremath{\vec{p}_\mathrm{T,1}}\xspace}
\newcommand{\ptvecnd}{\ensuremath{\vec{p}_\mathrm{T,2}}\xspace}
\newcommand{\ptvecrd}{\ensuremath{\vec{p}_\mathrm{T,3}}\xspace}
\newcommand{\ptvecth}{\ensuremath{\vec{p}_\mathrm{T,4}}\xspace}

\newcommand{\MEtwo}{\ensuremath{2 \to 2}\xspace}
\newcommand{\MEthree}{\ensuremath{2 \to 3}\xspace}
\newcommand{\MEfour}{\ensuremath{2 \to 4}\xspace}
\newcommand{\MEmixed}{\ensuremath{2 \to 2,3,4}\xspace}

\newcommand{\ej}{\ensuremath{\epsilon_\mathrm{4j}}\xspace}

\newcommand{\fDPS}{\ensuremath{f_\mathrm{DPS}}\xspace}

\newcommand{\sigmaI}{\ensuremath{\sigma_\mathrm{\RomanI}}\xspace}
\newcommand{\sigmaII}{\ensuremath{\sigma_\mathrm{\RomanI\joinR\RomanI}}\xspace}
\newcommand{\CASCADE} {{\textsc{Cascade}}\xspace}
\newcommand{\POWHEGBOX} {{\textsc{PowhegBox}}\xspace}
\newcommand{\KATIE} {{\textsc{KaTie}}\xspace}
\newcommand{\VINCIA} {\textsc{Vincia}\xspace}

\cmsNoteHeader{SMP-20-007} 

\title{Measurement of double-parton scattering in inclusive production of four jets with low transverse momentum in proton-proton collisions at \texorpdfstring{$\sqrt{s} = 13\TeV$}{sqrt(s) = 13 TeV}}

\author*[inst1]{CMS experiment}

\date{\today}

\abstract{
A measurement of inclusive four-jet production in proton-proton collisions at a center-of-mass energy of 13\TeV is presented. The transverse momenta of jets within $\abs{\eta} < 4.7$ are required to exceed 35,~ 30,~ 25,~ and 20\GeV for the first-, second-, third-, and fourth-leading jet, respectively. Differential cross sections are measured as functions of the jet transverse momentum, jet pseudorapidity, and several other observables that describe the angular correlations between the jets. 
The measured distributions show sensitivity to different aspects of the underlying event, parton shower modeling, and matrix element calculations. In particular, the interplay between angular correlations caused by parton shower and double-parton scattering contributions is shown to be important. The double-parton scattering contribution is extracted by means of a template fit to the data, using distributions for single-parton scattering obtained from Monte Carlo event generators and a double-parton scattering distribution constructed from inclusive single-jet events in data. The effective double-parton scattering cross section is calculated and discussed in view of previous measurements and of its dependence on the models used to describe the single-parton scattering background.
}

\hypersetup{
pdfauthor={CMS Collaboration},
pdftitle={Measurement of double-parton scattering in inclusive production of four jets with low transverse momentum in proton-proton collisions at \texorpdfstring{$\sqrt{s} = 13\TeV$}{sqrt(s) = 13 TeV}},
pdfsubject={CMS},
pdfkeywords={CMS, physics, QCD, double-parton scattering, DPS, jets}}

\maketitle 

\clearpage 

\section{Introduction}

Quantum chromodynamics (QCD), the theory of strong interactions, provides a good description of the production of hadron jets with large transverse momentum (\PT) in high-energy proton-proton ($\Pp\Pp$) collisions. This is achieved by factorizing the cross section into a perturbatively calculable matrix element describing the scattering between partons, and parton distribution functions (PDFs) that provide the probability to find a parton with given properties within the proton. The PDFs cannot be perturbatively calculated and are obtained by fitting available data. This fitting process includes nonperturbative effects such as the underlying event, hadronization, and parton showering. Measurements of the cross section for the production of inclusive high-\PT jets have been performed by the CMS collaboration at various center-of-mass energies and show agreement with perturbative QCD predictions at next-to-leading-order (NLO) accuracy~\cite{Chatrchyan:2014gia,Khachatryan:2016mlc,Khachatryan:2016wdh}. However, final states with multiple jets are not as well understood~\cite{Chatrchyan:2013qza}, suggesting a need for additional theoretical treatments of the strong interaction.
	
\begin{figure}[!htb] 
\centering
\subfloat{\includegraphics[width=0.44\textwidth]{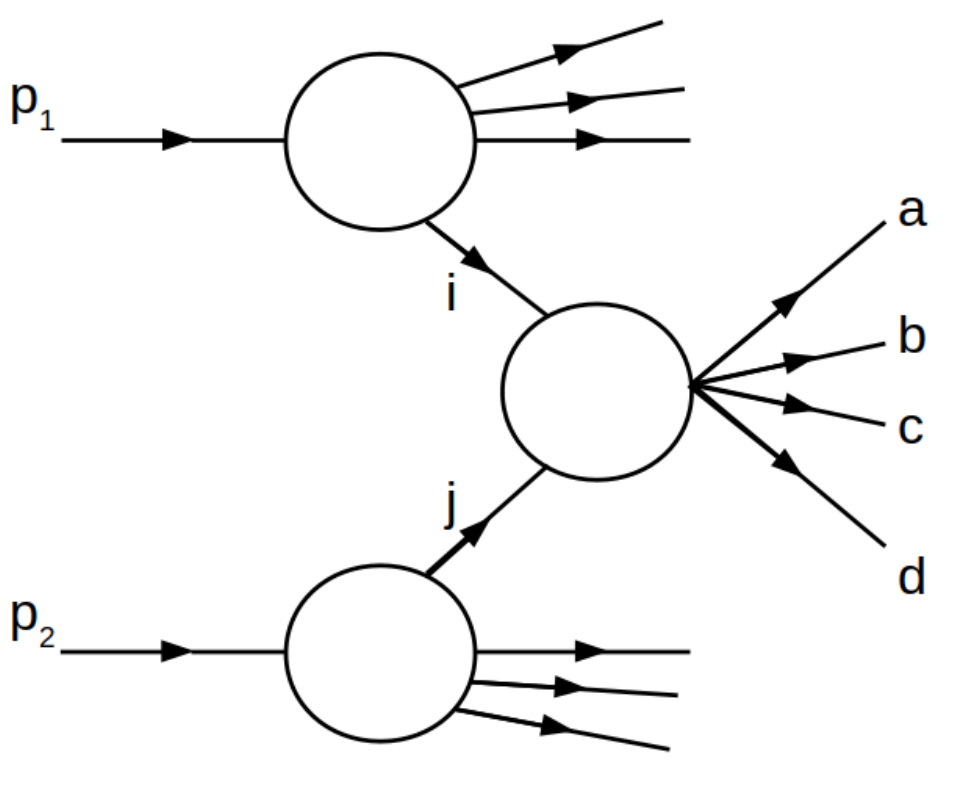}} \hspace*{\fill}
\subfloat{\includegraphics[width=0.44\textwidth]{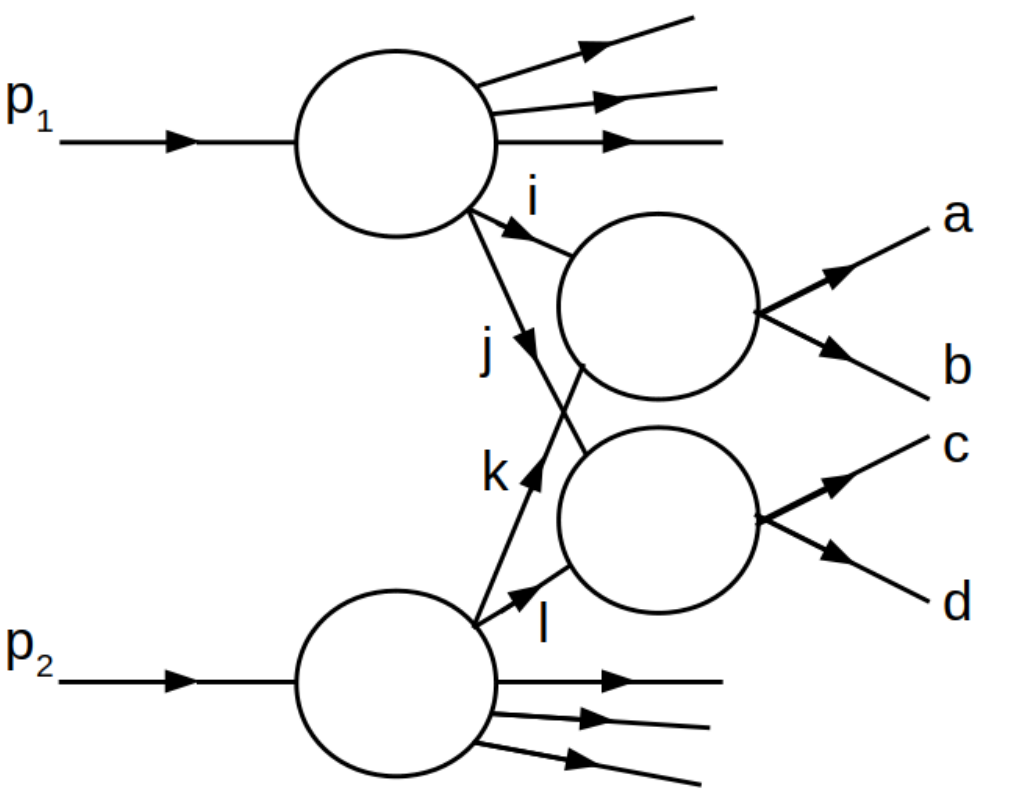}} 

\caption{A schematic depiction of inclusive four-jet production through SPS (\cmsLeft) and DPS (\cmsRight). In the case of SPS, one hard scattering produces the jets $a$ through $d$, whereas two independent hard scatterings create two jets each in the case of DPS. Since the two jet pairs are created independently in a DPS event, they are expected to show different kinematic correlations compared with the four jets originating from an SPS event.}
\label{fig:SPSvsDPS}
\end{figure}

Multijet final states can be produced in a single-parton scattering (SPS).  Depending on the order of the matrix element in the strong coupling, two or more jets can be produced in SPS.  Radiation before and/or after the interaction between the partons, as described by parton shower models, can  contribute additional jets to the final state. Thus, predictions for multijet processes in SPS provide an important test of  the matching between fixed-order matrix element calculations and the parton-shower formalism. A different approach introduces an additional hard scattering in the $\Pp\Pp$ collision, which also contributes a number of jets to the final state. Such processes are in general referred to as double-parton scattering (DPS), and they represent the simplest case of multiple-parton interactions (MPI). A schematic depiction of inclusive four-jet production through SPS and DPS is shown in Fig.~\ref{fig:SPSvsDPS}.

The cross section of a DPS process, $\sigmaDPS$, where A and B denote two processes with their own respective cross sections $\sigmaA$ and $\sigmaB$, can be expressed as:

\begin{align}
	\sigmaDPS = \frac{m}{2} \frac{\sigmaA \sigmaB}{\sigmaeff}. \label{eq:DPSpocketformula}
\end{align}

The factor $m$ is a combinatorial factor, which is equal to 1 for identical processes and 2 for nonidentical processes.  The effective cross section (\sigmaeff) reflects how strongly the occurrence of A and B is correlated~\cite{Diehl:2011yj}. For fully uncorrelated production of A and B, \sigmaeff tends to the total inelastic $\Pp\Pp$ cross section, whereas a small \sigmaeff indicates an enhanced simultaneous occurrence of processes A and B.  For multijet production, SPS processes often exhibit strong kinematic correlations between all jets, whereas DPS processes will manifest a distinctly different behavior. Indeed, the jets resulting from DPS are more often produced in two independent pairs, each in a back-to-back configuration in the transverse plane.  
The relevance of DPS rises with increasing center-of-mass energy; at higher energy and for fixed \PT of the jets, smaller values of the momentum fraction of the protons carried by the partons are probed, resulting in a strong increase of the gluon density and a larger probability for DPS.  A study of the extent to which DPS processes can supplement various SPS models is therefore beneficial for a complete description of hadronic interactions.  

Various DPS measurements at different center-of-mass energies and for various final states have been performed. Studies including one or two photons in the final state have been published in Refs.~\cite{PhysRevLett.79.584,PhysRevD.56.3811,PhysRevD.81.052012,Abazov:2014fha,Abazov:2015nnn}. Signatures involving one or two vector bosons have been measured by the ATLAS and  CMS Collaborations~\cite{Aad:2013bjm,Chatrchyan:2013xxa,Sirunyan:2017hlu,Sirunyan:2019zox,CMS:2021wfx}. Other studies have opted to include the production of heavy flavors~\cite{Aaij:2012dz,Aaij:2014hea,PhysRevD.90.111101,Aaij:2015wpa,Aaij:2016bqq}. Earlier measurements in the four-jet final state have been performed by the UA2 and CDF experiments~\cite{ALITTI1991145,Abe:1993rv}. The ATLAS and CMS Collaborations have more recently also performed DPS measurements with four jets~\cite{Aad:2015nda,Sadeh:2013wka,Aaboud:2016dea,Chatrchyan:2013qza} at a center-of-mass energy of 7\TeV. The CMS Collaboration additionally performed at 7\TeV~\cite{Khachatryan:2016rjt} a measurement of the final state with two b-tagged jets in combination with two light-flavored jets.

This paper presents an analysis of the inclusive production of four-jet events in $\Pp\Pp$ collisions at a center-of-mass energy of 13\TeV.  The data correspond to an integrated luminosity of 42 \nbinv and were collected with the CMS detector at the CERN LHC in 2016 during a special data-taking period with a low probability for several $\Pp\Pp$ interactions occurring within the same or neighbouring bunch crossings (pileup).  This avoids the challenges posed by pileup and enables us to include jets with low \PT. As a result of the low-\PT jets, a custom calibration of the jet energy scale is required. Data are corrected for detector efficiency and resolution effects by means of an unfolding procedure.

Several aspects of multijet production are studied by comparing the distributions of DPS-sensitive observables predicted by various Monte Carlo (MC) event generators with the distributions measured in data.
These observables all exploit differences in the kinematic correlations between the jets expected for SPS and DPS.
The DPS cross section is extracted with a template method. A pure DPS signal template is reconstructed from data by randomly mixing two inclusive single-jet events into one inclusive DPS four-jet event. This is then fitted together with several SPS-only background MC models to the distributions obtained from inclusive four-jet data. Finally, the effective cross section is computed using Eq.\@~\eqref{eq:DPSpocketformula}, with $\sigma_{\rm A}$ and $\sigma_{\rm B}$ measured from data.

Tabulated results are provided in the HEPData record for this analysis~\cite{hepdata}.

This paper is organized as follows. In Section~\ref{sec:observables}, the observables of interest are defined. Section~\ref{sec:cmsdetector} gives a brief overview of the CMS detector. The MC models used in the comparison with data are detailed in Section~\ref{sec:models}, whereas the data samples, event selection, and correction procedure are discussed in Section~\ref{sec:dataanalysis}. The strategy for the extraction of the DPS cross section and \sigmaeff is detailed in Section~\ref{sec:extraction}.  Systematic uncertainties for each of the unfolded observables are discussed in Section~\ref{sec:systematics}. Section~\ref{sec:results} contains a discussion of the results, which are summarized in Section~\ref{sec:summary}.

\section{Observables}
\label{sec:observables}

Six observables are defined to study DPS in four-jet production processes.   Many of these variables have been used in earlier measurements~\cite{PhysRevLett.79.584,PhysRevD.56.3811,PhysRevD.81.052012,Abazov:2014fha,Abazov:2015nnn,Aad:2013bjm,Chatrchyan:2013xxa,Sirunyan:2017hlu,Sirunyan:2019zox,Aaij:2012dz,Aaij:2014hea,PhysRevD.90.111101,Aaij:2015wpa,Aaij:2016bqq,
Aad:2015nda,Sadeh:2013wka,Aaboud:2016dea,Chatrchyan:2013qza,Khachatryan:2016rjt} and in phenomenological studies~\cite{Domdey:2009bg,Berger:2009cm,Maciula:2015vza,Kutak:2016mik,Kutak:2017zlp,Kutak:2016ukc}.  
The four leading jets are ordered with decreasing \PT. Based on the azimuthal angle ($\phi$), pseudorapidity ($\eta$), and transverse momentum vector ($\vec{p}_\mathrm{T}$), the variables studied in this paper can be described as follows.

\begin{itemize}

\item The azimuthal angular difference between the two softest jets:
\begin{align*}
\DphiS = \vert \phi_{3} - \phi_{4} \vert. 
\end{align*}
The two softest jets are more likely to be in a back-to-back configuration when produced by DPS since there is an increased probability that the two softest jets are produced in an independent scattering fron the two hardest jets and since the momentum should be conserved in the transverse plane. The increased probability leads to an enhanced DPS contribution around $\DphiS = \pi$.

\item The minimal combined azimuthal angular range of three jets:
\begin{align*}
\Dphimin = \mathrm{min}\left\lbrace \vert \phi_i - \phi_j \vert + \vert \phi_j - \phi_k \vert \middle\vert i,j,k \in [1,2,3,4], i\neq j\neq k \right\rbrace.
\end{align*}
In DPS, at least two out of three jets are more likely to be in a back-to-back configuration, while SPS processes have a more random distribution in their azimuthal angular difference.  Therefore, a DPS process is prone to yield larger values of \Dphimin~\cite{Kutak:2016ukc}.

\item The maximum $\eta$ difference between two jets:
\begin{align*}
\DY = \mathrm{max} \left\lbrace \vert \eta_i - \eta_j \vert \middle\vert i,j \in [1,2,3,4], i\neq j \right\rbrace. 
\end{align*}
As the maximum separation in $\eta$ between two jets becomes larger, the probability for the two jets to originate from two different parton interactions increases.

\item The azimuthal angular difference between the jets with the largest $\eta$ separation:
\begin{align*}
\phiij = \vert \phi_i - \phi_j \vert \quad \mathrm{for} \quad \DY = \vert \eta_i - \eta_j \vert 
\end{align*}
Since the jets with the largest $\eta$ separation are more likely to be produced in separate DPS subprocesses, a decorrelation in the distribution of the azimuthal angular difference of these jets is expected, whereas the jets will show stronger correlations in a SPS event.  

\item The transverse momentum balance of the two softest jets:
\begin{align*}
\DptS = \frac{\vert \ptvecrd + \ptvecth \vert}{\vert \ptvecrd \vert + \vert \ptvecth \vert}. 
\end{align*}
When the two softest jets originate from a DPS process,  they are more likely to be in a back-to-back configuration rendering the value for \DptS small. In SPS processes, the two softest jets do not necessarily balance.

\item The azimuthal angular difference between the hard and the soft jet pairs:
\begin{align*}
\DS &= \arccos{\left( \frac{(\ptvecst+\ptvecnd) \cdot (\ptvecrd+\ptvecth)}{\vert \ptvecst+\ptvecnd \vert \  \vert \ptvecrd+\ptvecth\vert} \right)}. 
\end{align*}
In a SPS process, the four jets must balance so that the \DS distribution peaks around $\pi$, while in DPS the two jet pairs are more likely to be independently produced, yielding a less correlated $\DS$ distribution. Thus we anticipate that DPS events tend toward lower values of \DS.	

\end{itemize}

\section{Measuring jets with the CMS detector}
\label{sec:cmsdetector}

The central feature of the CMS apparatus is a superconducting solenoid of 6\unit{m} internal diameter, providing a magnetic field of 3.8\unit{T}. Within the solenoid volume are a silicon pixel and strip tracker, a lead tungstate crystal electromagnetic calorimeter (ECAL), and a brass and scintillator hadron calorimeter (HCAL), each composed of a barrel and two endcap sections.  Forward calorimeters extend the $\eta$ coverage provided by the barrel and endcap detectors. Muons are detected in gas-ionization chambers embedded in the steel flux-return yoke outside the solenoid.  

The silicon tracker measures charged particles within the range $\abs{\eta} < 2.5$. During the LHC running period when the data used in this paper were recorded, the silicon tracker consisted of 1440 silicon pixel and 15\,148 silicon strip detector modules. 

The ECAL consists of 75\,848 lead tungstate crystals, which provide coverage in $\abs{\eta} < 1.48 $ in a barrel region and $1.48 < \abs{\eta} < 3.0$ in the two endcap regions. Preshower detectors consisting of two planes of silicon sensors interleaved with a total of $3 X_0$ of lead are located in front of each endcap detector.

In the region $\abs{\eta} < 1.74$, the HCAL cells have widths of 0.087 in $\eta$ and 0.087 in azimuth ($\phi$). In the $\eta$-$\phi$ plane, and for $\abs{\eta} < 1.48$, the HCAL cells map on to $5 \times 5$ arrays of ECAL crystals to form calorimeter towers projecting radially outwards from close to the nominal interaction point. For $\abs{ \eta } > 1.74$, the coverage of the towers increases progressively to a maximum of 0.174 in $\Delta \eta$ and $\Delta \phi$. Within each tower, the energy deposits in ECAL and HCAL cells are summed to define the calorimeter tower energies.

The forward hadron (HF) calorimeter uses steel as an absorber and quartz fibers as the sensitive material. The two halves of the HF are located 11.2\unit{m} from the interaction region, one on each end, and together they provide coverage in the range $3.0 < \abs{\eta} < 5.2$. They also serve as luminosity monitors.

Events of interest are selected using a two-tiered trigger system. The first level (L1), composed of custom hardware processors, uses information from the calorimeters and muon detectors to select events at a rate of around 100\unit{kHz} within a fixed latency of about 4\mus~\cite{Sirunyan:2020zal}. The second level, known as the high-level trigger (HLT), consists of a farm of processors running a version of the full event reconstruction software optimized for fast processing, and reduces the event rate to around 1\unit{kHz} before data storage~\cite{Khachatryan:2016bia}.

A global event reconstruction (particle flow) algorithm~\cite{CMS-PRF-14-001} reconstructs and identifies each individual particle in an event, with an optimized combination of all subdetector information. Jets are clustered from these reconstructed particles using the infrared and collinear safe anti-\kt algorithm~\cite{Cacciari:2008gp, Cacciari:2011ma} with a distance parameter of 0.4. The jet momentum is determined as the vectorial sum of the momenta of all particles in the jet, and is typically within 5 to 10\% of the true momentum over the whole \PT spectrum and detector acceptance, based on simulation.  Jet-energy corrections are derived from simulation studies so that the average measured energy of jets becomes identical to that of particle-level jets. In situ measurements of the momentum balance in dijet, photon+jet, Z+jet, and multijet events are used to determine any residual differences between the jet-energy scale in data and in simulation, and appropriate corrections are made~\cite{Khachatryan:2016kdb}. Additional selection criteria are applied to each jet to remove those potentially dominated by instrumental effects or reconstruction failures. The jet-energy resolution varies with rapidity and transverse momentum and typically amounts in the central region to 20--25\% at 20\GeV, 10\% at 100\GeV, and 5\% at 1\TeV~\cite{Khachatryan:2016kdb}.

A more detailed description of the CMS detector, together with a definition of the coordinate system  and the kinematic variables, can be found in Ref.~\cite{Chatrchyan:2008zzk}.

\section{Monte Carlo event generators}
\label{sec:models}

\subsection{The \texorpdfstring{\PYTHIA{}8}{PYTHIA8}, \texorpdfstring{\HERWIGpp}{HERWIG++}, and \texorpdfstring{\HERWIG{}7}{HERWIG7} models}
\label{subsec:P8H}

The \PYTHIA{}8~\cite{Sjostrand:2014zea},  \HERWIGpp~\cite{Bahr:2008pv}, and  \HERWIG{}7~\cite{Bellm:2110873} MC event generators use \MEtwo leading order (LO) matrix elements, matched to a DGLAP evolution~\cite{Gribov:1972ri,Dokshitzer:1977sg,ALTARELLI1977298} at leading logarithmic level for the simulation of the parton shower. By default, a \PT-ordered parton shower is implemented in \PYTHIA{}8, whereas an angular-ordered parton shower is used in \HERWIGpp and \HERWIG{}7 (jointly referred to as \HERWIG). For hadronization, \PYTHIA{}8 uses the Lund string model~\cite{ANDERSSON198331}, whereas \HERWIG relies on the cluster model~\cite{AMATI197987}.  The generators are interfaced with different sets of predetermined parameters (or ``tunes'') for the description of the underlying event, including MPI. These  tunes are obtained from fitting predictions to data. Off all the generated samples, two have been passed through the detector simulation program \GEANTfour~\cite{AGOSTINELLI2003250}. These two samples will be used to correct the data for detector effects by means of an unfolding procedure. The configurations used  in this paper to generate events with \PYTHIA{}8 and \HERWIG, are listed below.

\begin{itemize}

\item The \PYTHIA{}8.240 generator is interfaced with three different tunes, \ie, the {CUETP8M1} tune~\cite{Skands:2014pea,Khachatryan:2015pea}, the {CP5} tune~\cite{Sirunyan:2019dfx} and the {CDPSTP8S1-4j} tune~\cite{Khachatryan:2015pea}.

 The \PYTHIA{}8 sample, interfaced with the {CUETP8M1} tune, uses the {NNPDF2.3\_LO} PDFs~\cite{Ball:2017nwa}. The simulation of the detector effects has been applied to these generated events since they will be used to correct the date through an unfolding procedure.

 The \PYTHIA{}8 sample interfaced with the {CP5} tune uses the {NNPDF3.1 next-to-next-to-leading order (NNLO)} PDFs~\cite{Ball:2017nwa}. 

 The {CDPSTP8S1-4j} tune~\cite{Khachatryan:2015pea} is a CMS DPS tune, based on the standard {Tune 4C}~\cite{Corke:2010yf}, where parameters related to MPI and DPS have been altered to fit  predictions to the \DptS and \DS distributions obtained from an inclusive four-jet measurement at a center-of-mass energy of 7\TeV~\cite{Khachatryan:2015pea}. This tune uses the {CTEQ6L1} PDFs~\cite{Pumplin:2002vw}.

\item An additional \PYTHIA{}8.301 sample is generated with \VINCIA~\cite{Fischer:2016vfv} activated, which replaces the \PT-ordered parton shower from \PYTHIA{}8 with a dipole-antenna shower. The default parameter values of \PYTHIA{}8.301 and the {NNPDF2.3\_LO} PDFs are used. 

\item The \HERWIGpp~2.7.1 generator is used in combination with the {CUETHS1} tune~\cite{Khachatryan:2015pea} and the {CTEQ6L1} PDFs.
This sample is also passed through the detector simulation program \GEANTfour and is used in the unfolding procedure.

\item Two tunes are used for the \HERWIG~7.1.5 MC event generator.

 The {SoftTune} is the default tune provided by the \HERWIG{}7 authors and uses the {MMHT2014lo68cl} PDFs~\cite{Harland-Lang:2014zoa}.

 The {CH3} tune has been obtained by the CMS Collaboration from a study of underlying event measurements~\cite{CMS-PAS-GEN-19-001}.  It is used in combination with the {NNPDF2.3\_NNLO} PDF.

\end{itemize}

\subsection{Multijet models}

A second group of models, referred to as the multijet models, uses higher-order matrix elements to produce more than two jets in the hard parton scattering. These MC event generators are interfaced with \PYTHIA, \HERWIG, or \CASCADE~\cite{Jung:2010si} to include a description of the underlying event. Details of the generated event samples are given below.

\begin{itemize}

\item \MGvATNLO (version 2.6.5)~\cite{Alwall:2014hca} is a generator with the ability to compute tree-level and NLO matrix elements for arbitrary processes. Two LO samples and one NLO sample are generated, as listed below. 

The LO samples combine a \MEtwo, a \MEthree, and a \MEfour matrix element, referred to as \MEmixed. An $\HT > 50\GeV$ generation condition is used, where \HT is defined as the scalar sum of the transverse momenta of the produced partons, and all partons must have $\abs{\eta} < 5$.  For one  sample the description of the underlying event, parton shower, and  hadronization is  handled by \PYTHIA{}8.240, using the {CP5} tune, whereas the other sample is interfaced to \PYTHIA{}8.301 with \VINCIA. The former uses the {NNPDF2.3\_NNLO} PDFs, whereas the latter uses the {NNPDF2.3\_LO} PDFs. The MLM scheme~\cite{Alwall:2007fs} is used to match jets produced via matrix-element calculations with those from parton showers, using the matching \PT scale of 18\GeV, which was optimized by analyzing the differential jet-rate distributions. 

The \MEtwo NLO sample is interfaced with \PYTHIA{}8.240, using the {CP5} tune with the {NNPDF2.3\_NNLO} PDFs and an MLM matching scheme. The two leading partons are required to lie within $\abs{\eta} < 5$ and have a \PT above 25 and 20\GeV, respectively.

\item \POWHEGBOX version 3633 (2019.02.25)~\cite{Nason:2004rx,Frixione:2007vw,Alioli:2010xd} is a framework for implementing NLO corrections in MC event generators. Each event is constructed by producing the Born configuration, on which the real phase space is built afterwards. Two different samples are generated, both are interfaced with \PYTHIA{}8.240, using the {CP5} tune along with the {NNPDF2.3\_NNLO} PDFs.

A first sample is generated with a \MEtwo NLO matrix element~\cite{Alioli:2010xa}. The factorization and renormalization scales are set to the \PT of the underlying Born configuration.

A second sample was generated with a \MEthree NLO matrix element~\cite{Kardos:2014dua}. The generator-level minimal \PT of the underlying Born configuration is set to 10\GeV, and the factorization and renormalization scales are set to \HT/2.

\item \KATIE version 23April2019~\cite{vanHameren:2016kkz} is a LO parton-level event generator, based on \kt-factorization~\cite{Catani:1994sq,Deak:2009xt,Sapeta:2015gee}, allowing for on-shell and off-shell production. In the case of the latter, the initial partons are generated with a nonzero intrinsic \kt, which can alter the momentum balance of the jets, yielding various topologies and correlations between the jets compared with on-shell production. A \MEfour matrix element is used for all samples generated with \KATIE. The generator-level requirements for the \pt of the four partons produced by the matrix element are 35, 30, 25, and 20\GeV, and their rapidities are limited to $\vert \eta \vert < 5.0$. Since the \pt requirements are introduced at parton level, the effective \pt thresholds for the resulting hadron-level jets are typically 5 to 10\GeV lower.  The factorization and renormalization scales are set to \HT/2. Two on-shell and two off-shell samples are produced.

The two on-shell samples are interfaced with \PYTHIA{}8.240 and \HERWIG~7.1.5, along with the {CP5} and {CH3} tunes, respectively. Both samples use the {NNPDF2.3\_NNLO} PDFs. 

For the two off-shell samples, the showering and hadronization is handled by \CASCADE~3.0.01-beta1. Two different transverse-momentum-dependent (TMD) PDFs are used: the {MRW-CT10nlo} TMD PDFs (MRW)~\cite{Bury:2017jxo} and the {PB-NLO-HERAI+II-2018-set2} TMD PDFs (PBTMD)~\cite{Martinez:2018jxt}.

\end{itemize}

\subsection{SPS+DPS samples}
\label{sec:total_samples}

The \PYTHIA{}8.240 and \KATIE MC event generators both produce two \MEtwo matrix elements per event, resulting in a pure DPS sample. MPIs are also present as part of the underlying event. In \KATIE, \sigmaeff is a parameter that directly determines the size of the DPS contribution relative to the SPS cross section. A value of 21.3\unit{mb} for \sigmaeff is adopted, as in~\cite{Khachatryan:2015pea}.  For \PYTHIA{}8, it is not possible to set \sigmaeff, because it is determined by the underlying event parameters, and therefore the second \MEtwo process is simply added with the same kinematic requirements, without any additional scaling of the cross section.

Four samples with an explicit DPS contribution are used in this paper.
	
\begin{itemize}

\item A \PYTHIA{}8.240 sample is generated with the {CP5} tune and the {NNPDF2.3\_NNLO} PDFs. It is the aforementioned \PYTHIA{}8 sample to which a pure DPS sample, obtained by overlaying two \MEtwo matrix elements, is added.

\item The second \PYTHIA{}8 sample, which includes an explicit DPS contribution, is the one already mentioned in Section~\ref{subsec:P8H}, since the {CDPSTP8S1-4j} tune has been fitted to DPS-sensitive observables.

\item An on-shell \KATIE sample is generated with an explicit DPS contribution that is obtained by overlaying two \MEtwo matrix elements, with the exact same generation parameters as the on-shell \KATIE LO sample from the multijet models. Showering and hadronization are handled by \PYTHIA{}8.240 with the {CP5} tune and the {NNPDF2.3\_NNLO} PDFs. 

\item Two off-shell \KATIE samples with an explicit DPS contribution are generated using the same TMD PDFs as for the multijet samples. Since \CASCADE cannot handle two \MEtwo matrix elements, nonperturbative corrections have been derived from the on-shell SPS and DPS \KATIE samples, and are applied to the off-shell DPS \KATIE parton level sample. The nonperturbative corrections range from 1--4\% for all observables, except for the \DS observable for which corrections up to 11\% were found.	

\end{itemize}

\section{Event selection and unfolding}
\label{sec:dataanalysis}

This analysis uses data from $\Pp\Pp$ collisions at a center-of-mass energy of 13\TeV, collected during a data taking period at low luminosity, with an average pileup of 1.3 and an integrated luminosity of $42\,\nbinv$. The online selection of multi-jet events was based on four single-jet triggers each requiring the presence of at least one jet with a \PT above 30, 50, 80, or 100\GeV, and within $\abs{\eta} < 4.7$. Because the triggers have been prescaled, they are used in disjoint \PT ranges.  Offline requirements are imposed to ensure that  the triggers are fully efficient, except for the trigger with the lowest \PT threshold. In this last case, a correction as a function of the jet \PT is applied to the selected event, effectively altering its weight. The trigger efficiency was determined by comparing the performance of the jet trigger with a minimum-bias trigger serving as an unbiased reference.

Events are selected offline by requiring exactly one primary vertex, so effects of pileup can be neglected. The correction of the event yield is based on the measurement of the average pileup and has negligible uncertainty. A systematic uncertainty due to a possible contamination of events containing two or more $\Pp\Pp$ collisions is nevertheless included in the results. Two phase space regions defined by selections on jet \PT are used.  In \regioni, the four leading jets within $\abs{\eta} < 4.7$ are required to exceed  \PT thresholds of 35, 30, 25, and 20\GeV.  Asymmetric thresholds have been chosen over symmetric ones because the latter tend to dampen the DPS contribution with respect to the SPS fraction, according to higher-order calculations or calculations performed in the \kt-factorization framework~\cite{Kutak:2016mik,Kutak:2016ukc}.  
The \DS distribution is obtained in \regionii, with \PT thresholds of 50, 30, 30, and 30\GeV. On the one hand, the resolution of the \DS observable is improved by imposing higher \PT cuts. On the other hand, the \DS observable can now be used to perform the extraction of \sigmaeff, using the lowest jet-\PT trigger threshold of 30\GeV. The second set of selections is needed to obtain the cross sections \sigmaA and \sigmaB, as detailed in Section~\ref{sec:extraction}.

The measured distributions are corrected for detector effects with the  {TUnfold} program~\cite{Schmitt:2012kp,Schmitt:2016orm}, which is based on a least squares fit and Tikhonov regularization~\cite{Tikhonov:1963}. The regularization is necessary to avoid possible instabilities in the inversion of the matrix describing the migrations within the phase space.  Bin-to-bin migrations are kept to a minimum by choosing a bin width that is two times larger than the resolution of the considered variable, as obtained from a simulation study with \PYTHIA{}8. The migration matrix, as well as the probabilities for migration into and out of the phase space, are obtained from the \PYTHIA{}8 and \HERWIGpp MC models.

\section{Extraction of the effective cross section}
\label{sec:extraction}

The DPS formula~\eqref{eq:DPSpocketformula} allows the calculation of \sigmaeff if the DPS cross section, as well as the cross sections of the two Processes A and B, are known. In the simplest case, the Processes A and B would both be dijet production, resulting in a four-jet final state with uncorrelated jet pairs, as depicted in Fig.~\ref{fig:SPSvsDPS}. However, initial- and final-state radiation, and higher-order interactions can produce final states with more than two jets, yielding additional possibilities to form a four-jet topology. 

To avoid additional model dependencies, a DPS signal template is constructed from data, following an approach similar to the one laid out in Ref.~\cite{Aaboud:2016dea}. The Processes A and B are both defined as inclusive single-jet production. Combining two events of Type A and B will result in a multijet final state. Whenever at least four jets are found in the combined final state originating from the same process, the combined event is labeled as SPS process, otherwise the event is labeled as a DPS process. 

\regionii is used for the extraction of \sigmaeff. The choice is motivated by Ref.~\cite{Kutak:2016mik} where it is suggested that such asymmetric cuts could boost the DPS signature. The cross sections of the Processes A and B are defined as inclusive single jet production with
\begin{align*}
\sigmaA &= \sigma_{\rm jet} \left( p_{\rm T} \geq 50 \right), \\
\sigmaB &= \sigma_{\rm jet} \left( p_{\rm T} \geq 30 \right).
\end{align*}

Combining two inclusive single-jet processes results in final states with at least four jets in only a fraction of the cases.  A ``four-jet efficiency'' (\ej) has been obtained from the combined sample as detailed below.

From the event sample with at least one jet with $\PT > 30\GeV$, two events are drawn at random with the second event containing at least one jet with $\PT > 50\GeV$. The two selected events are combined to form one single event.  A combined event is discarded whenever two or more jet axes spatially coincide.  This veto condition is formulated as $R_{ij} = ((\phi_i - \phi_j)^2 + (\eta_i - \eta_j)^2)^{1/2} \geq 0.4$,  where the indices $i$ and $j$ indicate jets belonging to an event from the first and second data sample, respectively. The newly constructed combined event sample is then subjected to the four-jet selection criteria of \regionii. The four-jet efficiency was estimated to be 

\begin{align}
\ej = 0.324^{+0.037}_{-0.065}\,\mathrm{(syst)} \label{eq:four-jet-eff},
\end{align}

where the statistical uncertainty is negligible and the systematic error is detailed in the next section.  

The \DS observable is chosen for the extraction of the DPS cross section and \sigmaeff because it is the least affected by parton shower effects.  The signal template is taken from the combined data sample, which is used to extract the \DS distribution in exactly the same manner as before, including the correction for detector effects by means of unfolding.

The SPS MC models are taken as background templates.  To avoid contamination by MPI,  additional samples are provided where an event is omitted if it contains a generator level parton with a $\PT > 20\GeV$ that originates from a MPI. This selection ensures that no hard jets originating from MPI enter the four-jet analysis, and will be referred to as ``hard MPI removed''. 

The fraction of DPS events, \fDPS, is extracted by performing a template fit to the unfolded \DS distribution, obtained from the original inclusive four-jet sample.  The DPS signal and SPS background \DS distributions are both normalized to the integral of the \DS distribution obtained from the four-jet events in data.  
 The optimal value of the DPS fraction \fDPS is determined with a maximum likelihood technique using Poisson statistics~\cite{Brun:1997pa}:

\begin{align}
\sigma^\mathrm{data}(\DS) = \fDPS  \sigma^\mathrm{data}_\mathrm{DPS}(\DS) + (1 - \fDPS)  \sigma^\mathrm{MC}_\mathrm{SPS}(\DS). \label{eq:template}
\end{align}

The cross section $\sigmaDPS$, needed for the extraction of \sigmaeff, is then given by the integral of the \DS distribution measured in data, scaled with the DPS fraction \fDPS:
	
\begin{align}
\sigmaDPS = \fDPS \int \sigma^\mathrm{data}(\DS) {\rm d}(\DS) \label{eq:sigmaDPS_template}.
\end{align}

Because of the overlapping \PT ranges, the Processes A and B are not always distinguishable and the cross section for Process B has therefore to be rewritten as the sum of the cross section for Process A ($\sigma_{\rm A}$) and the difference between the cross sections for Processes B and A ($\sigma_{\rm B} - \sigma_{\rm A}$). Taking into account the correct combinatorial factor along with the four-jet efficiency, the DPS formula~\eqref{eq:DPSpocketformula} can be reformulated as:

\begin{align}
\sigmaDPS &= \frac{\ej}{\sigmaeff} \left( \frac{1}{2} \sigmaA^2 + \sigmaA  (\sigmaB - \sigma_{\rm A}) \right) \nonumber \\
&= \frac{\ej \sigmaA \sigmaB}{\sigmaeff} \left( 1 - \frac{1}{2} \frac{\sigmaA}{\sigmaB} \right). \label{eq:pocketformula_rewritten}
\end{align}

The cross sections $\sigmaA$ and $\sigmaB$ are determined by integrating the $\eta$ spectra for jets with $\PT$ above 50 and 30\GeV in data, which are unfolded in exactly the same manner as all other distributions.  

\section{Systematic uncertainties}
\label{sec:systematics}

Different sources of systematic uncertainties occurring in the data analysis are studied.  A summary is given in Tables\@~\ref{tab:systematics1} and \@~\ref{tab:systematics2}.

\begin{itemize}

\item[\textbf{Jet energy scale (JES) uncertainty:}] 
The low-pileup, low jet-\PT data sample used in this analysis necessitates a dedicated JES calibration. The methods discussed in Ref.~\cite{Khachatryan:2016kdb} are applied, scaling the four-momentum vectors of the jets by a series of sequential corrections.  The JES uncertainty depends on jet \PT and $\eta$ and is smaller than 10\% over the whole \PT spectrum and detector acceptance.  The jet momenta are  scaled downwards and upwards by the JES uncertainty to estimate its effect on the measured distributions.  The JES uncertainty is the dominant contribution to the total uncertainty for the observables in terms of the absolute cross section and results in a maximal upward (downward) uncertainty of 39 (33)\%. It largely cancels in the normalized distributions,  never exceeding 16\%.

\item[\textbf{Jet energy resolution (JER) uncertainty:}] The JER obtained from MC simulation differs from the one estimated for data, which would lead to a wrong estimation of the bin-to-bin migrations.  An additional smearing of the jet \PT at detector level is therefore applied to both MC samples used in the unfolding.  To estimate the JER uncertainty, the data-to-simulation smearing factor is varied up and down with its own uncertainty, resulting in migration matrices that differ from the nominal ones. The newly obtained migration matrices are used to unfold the distributions, which are then compared with the nominal distributions for all observables. The JER uncertainty is less than 9\%, except for the \PT spectrum of the leading jet where it reaches a maximum of 26\%. 

\item[\textbf{Trigger uncertainty:}] An event weight as a function of jet \PT is applied to data to correct for the efficiency of the trigger with the lowest threshold.   These weights are obtained by fitting the trigger efficiency curve determined in data using a least-squares minimization.   Varying the fit parameters by their uncertainty leads to a trigger uncertainty that never exceeds 1\%.

\item[\textbf{Model uncertainty:}] The data distributions are unfolded using migration matrices from the \PYTHIA{}8 and \HERWIGpp models. The averages of the two unfolded distributions are taken as the nominal unfolded distributions and the systematic error is estimated as half of the difference.  The model uncertainty varies between  1\% and 16\%, depending on the observable.

\item[\textbf{Pileup:}] Events with two $\Pp\Pp$ collisions in the same bunch crossing may be reconstructed with only one vertex if the collision points are separated by less than 0.12\unit{cm} along the beam axis.  Taking into account the spread of vertices and the relative yields of events with 1 and 2 collisions, the pileup contamination in the data sample is estimated to be 0.28\%.  No further correction is applied, and a systematic uncertainty is included by reproducing all distributions with a sample of events containing two vertices, normalizing these to 0.28\% of the nominal distributions, and estimating the effect of such pileup correction on the data.  The systematic uncertainty is smaller than 1\% in all bins for all observables.

\item[\textbf{Integrated luminosity uncertainty:}]	 The uncertainty in the integrated luminosity for data collected in 2016 is $1.2\%$~\cite{CMS:2021xjt}.

\end{itemize}

Because the four-jet efficiency is determined using uncorrected data, it has neither JER or model systematic uncertainty.  However, an additional systematic uncertainty is included to cover a possible difference with respect to the true efficiency to be applied on the corrected cross sections. This uncertainty is determined by examining the four-jet efficiency obtained with \PYTHIA{}8 and \HERWIGpp at both the detector and generator levels, after a 2-dimensional reweighting as a function of leading-jet \PT and  jet multiplicity to obtain a better description of the data.  A four-jet efficiency of $0.404$ and  $0.412$ is found with \PYTHIA{}8 on detector and generator level, respectively,  whereas values of $0.403$ and $0.392$ are obtained with  \HERWIGpp, with negligible statistical uncertainty.  The systematic uncertainty is therefore conservatively estimated to be smaller than 2\%. 

\begin{table}[h!]
\centering
\topcaption{Systematic uncertainties, along with the statistical and the total uncertainties for the \PT spectra, the $\eta$ spectra, and the DPS sensitive observables, in percent.
The JES uncertainty leads to asymmetric uncertainties (an upper and a lower error), while all other systematic uncertainties, as well as the statistical uncertainty, are symmetric.}

\label{tab:systematics1}
\resizebox{\textwidth}{!}{
\begin{tabular}{lllllllllll}
Observable	& \multicolumn{2}{c}{JES} & JER & Model & Trigger & Vertex & Lum. & Stat & \multicolumn{2}{c}{Total}	\\ 
& Upper & Lower		   & 	 & 	 & 	   & 	    &      & 	   & Upper & Lower		\\ \hline 
\multicolumn{11}{c}{Absolute cross section (\%)} \\ [\cmsTabSkip]
$p_\mathrm{T,1}$          & \multicolumn{1}{r}{11--39}	& \multicolumn{1}{r}{9--30}		& \multicolumn{1}{r}{2--26}	& \multicolumn{1}{r}{0--16}	& \multicolumn{1}{r}{$<1$}	& \multicolumn{1}{r}{$<1$}		& \multicolumn{1}{r}{1.2}		& \multicolumn{1}{r}{1--10} & \multicolumn{1}{r}{11--51}	& \multicolumn{1}{r}{10--44}	\\ 

$p_\mathrm{T,2}$		& \multicolumn{1}{r}{11--31}	& \multicolumn{1}{r}{10--24}	& \multicolumn{1}{r}{0--2}		& \multicolumn{1}{r}{0--7}		& \multicolumn{1}{r}{$<1$} & \multicolumn{1}{r}{$<1$}		& \multicolumn{1}{r}{1.2}		& \multicolumn{1}{r}{1--8}	& \multicolumn{1}{r}{14--33}	& \multicolumn{1}{r}{11--26}	\\  

$p_\mathrm{T,3}$          & \multicolumn{1}{r}{1--31}		& \multicolumn{1}{r}{7--24}		& \multicolumn{1}{r}{1--3}		& \multicolumn{1}{r}{2--7}		& \multicolumn{1}{r}{$<1$}	& \multicolumn{1}{r}{$<1$}		& \multicolumn{1}{r}{1.2}		& \multicolumn{1}{r}{2--15}	& \multicolumn{1}{r}{13--33}	& \multicolumn{1}{r}{13--25}	\\  

$p_\mathrm{T,4}$		& \multicolumn{1}{r}{10--25} & \multicolumn{1}{r}{0--21}  & \multicolumn{1}{r}{1--8} & \multicolumn{1}{r}{2--7} & \multicolumn{1}{r}{$<1$} & \multicolumn{1}{r}{$<1$} & \multicolumn{1}{r}{1.2} & \multicolumn{1}{r}{4--31} & \multicolumn{1}{r}{14--34} & \multicolumn{1}{r}{13--32} \\ [\cmsTabSkip] 

$\eta_1$		& \multicolumn{1}{r}{22--33} & \multicolumn{1}{r}{18--28} & \multicolumn{1}{r}{$<1$} & \multicolumn{1}{r}{1--9} & \multicolumn{1}{r}{$<1$} & \multicolumn{1}{r}{$<1$} & \multicolumn{1}{r}{1.2} & \multicolumn{1}{r}{3--5}  & \multicolumn{1}{r}{22--34} & \multicolumn{1}{r}{19--29} \\  

$\eta_2$		& \multicolumn{1}{r}{22--30} & \multicolumn{1}{r}{18--26} & \multicolumn{1}{r}{$<1$} & \multicolumn{1}{r}{0--6} & \multicolumn{1}{r}{$<1$} & \multicolumn{1}{r}{$<1$} & \multicolumn{1}{r}{1.2} & \multicolumn{1}{r}{3--6}  & \multicolumn{1}{r}{23--31} & \multicolumn{1}{r}{18--26} \\  

$\eta_3$		& \multicolumn{1}{r}{21--29} & \multicolumn{1}{r}{18--24} & \multicolumn{1}{r}{$<1$} & \multicolumn{1}{r}{0--7} & \multicolumn{1}{r}{$<1$} & \multicolumn{1}{r}{$<1$} & \multicolumn{1}{r}{1.2} & \multicolumn{1}{r}{3--5}  & \multicolumn{1}{r}{22--30} & \multicolumn{1}{r}{19--25} \\  

$\eta_4$		& \multicolumn{1}{r}{19--29} & \multicolumn{1}{r}{16--24} & \multicolumn{1}{r}{$<1$} & \multicolumn{1}{r}{1--8} & \multicolumn{1}{r}{$<1$} & \multicolumn{1}{r}{$<1$} & \multicolumn{1}{r}{1.2} & \multicolumn{1}{r}{3--4}  & \multicolumn{1}{r}{19--30} & \multicolumn{1}{r}{17--25} \\ [\cmsTabSkip]

\DphiS			& \multicolumn{1}{r}{21--24} & \multicolumn{1}{r}{19--20} & \multicolumn{1}{r}{$<1$} & \multicolumn{1}{r}{1--7} & \multicolumn{1}{r}{$<1$} & \multicolumn{1}{r}{$<1$} & \multicolumn{1}{r}{1.2} & \multicolumn{1}{r}{3--4}  & \multicolumn{1}{r}{22--25} & \multicolumn{1}{r}{20--22} \\  

\Dphimin		& \multicolumn{1}{r}{21--28} & \multicolumn{1}{r}{18--24} & \multicolumn{1}{r}{$<1$} & \multicolumn{1}{r}{1--6} & \multicolumn{1}{r}{$<1$} & \multicolumn{1}{r}{$<1$} & \multicolumn{1}{r}{1.2} & \multicolumn{1}{r}{3--7}  & \multicolumn{1}{r}{21--29} & \multicolumn{1}{r}{19--25} \\  

\DY			& \multicolumn{1}{r}{22--25} & \multicolumn{1}{r}{16--33} & \multicolumn{1}{r}{$<1$} & \multicolumn{1}{r}{0--6} & \multicolumn{1}{r}{$<1$} & \multicolumn{1}{r}{$<1$} & \multicolumn{1}{r}{1.2} & \multicolumn{1}{r}{3--6}  & \multicolumn{1}{r}{23--26} & \multicolumn{1}{r}{17--34} \\  

\phiij			& \multicolumn{1}{r}{23--26} & \multicolumn{1}{r}{19--22} & \multicolumn{1}{r}{$<1$} & \multicolumn{1}{r}{0--7} & \multicolumn{1}{r}{$<1$} & \multicolumn{1}{r}{$<1$} & \multicolumn{1}{r}{1.2} & \multicolumn{1}{r}{3--4}  & \multicolumn{1}{r}{24--27} & \multicolumn{1}{r}{19--22} \\  

\DptS			& \multicolumn{1}{r}{22--25} & \multicolumn{1}{r}{19--20} & \multicolumn{1}{r}{0--3} & \multicolumn{1}{r}{2--6} & \multicolumn{1}{r}{$<1$} & \multicolumn{1}{r}{$<1$} & \multicolumn{1}{r}{1.2} & \multicolumn{1}{r}{3--4}  & \multicolumn{1}{r}{23--26} & \multicolumn{1}{r}{19--21} \\ 

\DS			& \multicolumn{1}{r}{4--34}  & \multicolumn{1}{r}{13--20} & \multicolumn{1}{r}{$<1$} & \multicolumn{1}{r}{0--5} & \multicolumn{1}{r}{$<1$} & \multicolumn{1}{r}{$<1$} & \multicolumn{1}{r}{1.2} & \multicolumn{1}{r}{3--13} & \multicolumn{1}{r}{12--37} & \multicolumn{1}{r}{15--22} \\  [\cmsTabSkip]  

\multicolumn{11}{c}{Bin-normalized cross section (\%)} \\ [\cmsTabSkip]

\DphiS 				& \multicolumn{1}{r}{0--1}      & \multicolumn{1}{r}{0--1}		& \multicolumn{1}{r}{$<1$}		& \multicolumn{1}{r}{0--4}		& \multicolumn{1}{r}{$<1$}	& \multicolumn{1}{r}{$<1$}		& \multicolumn{1}{r}{\NA}	& \multicolumn{1}{r}{3--4}	& \multicolumn{1}{r}{3--6}		& \multicolumn{1}{r}{3--6}	\\  

\Dphimin		 	& \multicolumn{1}{r}{0--5}      & \multicolumn{1}{r}{0--4}		& \multicolumn{1}{r}{$<1$}		& \multicolumn{1}{r}{0--4}		& \multicolumn{1}{r}{$<1$}	& \multicolumn{1}{r}{$<1$}		& \multicolumn{1}{r}{\NA}	& \multicolumn{1}{r}{3--7}	& \multicolumn{1}{r}{4--8}		& \multicolumn{1}{r}{3--8}	\\  

\DY				& \multicolumn{1}{r}{0--2}      & \multicolumn{1}{r}{0--18}		& \multicolumn{1}{r}{$<1$}		& \multicolumn{1}{r}{0--5}		& \multicolumn{1}{r}{$<1$}	& \multicolumn{1}{r}{$<1$}		& \multicolumn{1}{r}{\NA}	& \multicolumn{1}{r}{3--6}	& \multicolumn{1}{r}{3--10}		& \multicolumn{1}{r}{3--21}	\\  

\phiij 				& \multicolumn{1}{r}{0--3}      & \multicolumn{1}{r}{0--2}		& \multicolumn{1}{r}{$<1$}		& \multicolumn{1}{r}{0--4}		& \multicolumn{1}{r}{$<1$}	& \multicolumn{1}{r}{$<1$}		& \multicolumn{1}{r}{\NA}	& \multicolumn{1}{r}{3--4}	& \multicolumn{1}{r}{3--6}		& \multicolumn{1}{r}{3--6}	\\  

\DptS 				& \multicolumn{1}{r}{0--2}      & \multicolumn{1}{r}{0--2}		& \multicolumn{1}{r}{0--2}		& \multicolumn{1}{r}{0--2}		& \multicolumn{1}{r}{$<1$}	& \multicolumn{1}{r}{$<1$}		& \multicolumn{1}{r}{\NA}	& \multicolumn{1}{r}{3--4}	& \multicolumn{1}{r}{3--5}		& \multicolumn{1}{r}{3--5}	\\  

\DS				& \multicolumn{1}{r}{0--16}	& \multicolumn{1}{r}{0--7}		& \multicolumn{1}{r}{$<1$}		& \multicolumn{1}{r}{0--7}		& \multicolumn{1}{r}{$<1$}	& \multicolumn{1}{r}{$<1$}		& \multicolumn{1}{r}{\NA}	& \multicolumn{1}{r}{3--13}	& \multicolumn{1}{r}{3--22}		& \multicolumn{1}{r}{3--15}	\\ 

\end{tabular}
}
\end{table}

\begin{table}[h!]
\centering
\topcaption{Systematic uncertainties, along with the statistical and the total uncertainties for the cross sections of the two phase space regions, 
along with the observables needed for the extraction of $\sigmaDPS$, in percent. The JES uncertainty leads to asymmetric uncertainties (an upper and a lower error): 
all other systematic uncertainties, as well as the statistical uncertainty, are symmetric. An additional uncertainty in \ej because of possible differences between 
generator- and detector-level events, is estimated to be 2\%.}

\label{tab:systematics2}
\resizebox{\textwidth}{!}{
\begin{tabular}{lllllllllll}
Observable	& \multicolumn{2}{c}{JES} & JER & Model & Trigger & Vertex & Lum. & Stat & \multicolumn{2}{c}{Total}	\\ 
& Upper & Lower		   & 	 & 	 & 	   & 	    &      & 	   & Upper & Lower		\\ \hline 

\multicolumn{11}{c}{Integrated cross section (\%)} \\ [\cmsTabSkip]

\sigmaI & \multicolumn{1}{r}{24} 	& \multicolumn{1}{r}{19}		& \multicolumn{1}{r}{$<1$}			& \multicolumn{1}{r}{4}			& \multicolumn{1}{r}{$<1$}	& \multicolumn{1}{r}{$<1$}		& \multicolumn{1}{r}{1.2}			& \multicolumn{1}{r}{1}	& \multicolumn{1}{r}{25}		& \multicolumn{1}{r}{20}	\\

\sigmaII & \multicolumn{1}{r}{17} 	& \multicolumn{1}{r}{13}		& \multicolumn{1}{r}{$<1$}			& \multicolumn{1}{r}{6}			& \multicolumn{1}{r}{$<1$}	& \multicolumn{1}{r}{$<1$}		& \multicolumn{1}{r}{1.2}			& \multicolumn{1}{r}{2}	& \multicolumn{1}{r}{20}		& \multicolumn{1}{r}{16}	\\ [\cmsTabSkip]

\multicolumn{11}{c}{\sigmaeff extraction (\%)} \\ [\cmsTabSkip]

\DSDPS				& \multicolumn{1}{r}{7--19}		& \multicolumn{1}{r}{15--24}	& \multicolumn{1}{r}{$<1$}		& \multicolumn{1}{r}{0--3}		& \multicolumn{1}{r}{$<1$}	& \multicolumn{1}{r}{$<1$}		& \multicolumn{1}{r}{1.2}			& \multicolumn{1}{r}{1--2}	& \multicolumn{1}{r}{7--19}		& \multicolumn{1}{r}{15--25} \\ 

\sigmaA & \multicolumn{1}{r}{10} 	& \multicolumn{1}{r}{9}		& \multicolumn{1}{r}{$<1$}			& \multicolumn{1}{r}{4}			& \multicolumn{1}{r}{$<1$}	& \multicolumn{1}{r}{$<1$}		& \multicolumn{1}{r}{1.2}			& \multicolumn{1}{r}{1}	& \multicolumn{1}{r}{11}		& \multicolumn{1}{r}{10}	\\	

\sigmaB & \multicolumn{1}{r}{7}		& \multicolumn{1}{r}{9}		& \multicolumn{1}{r}{$<1$}			& \multicolumn{1}{r}{4}		& \multicolumn{1}{r}{$<1$}	& \multicolumn{1}{r}{$<1$}		& \multicolumn{1}{r}{1.2}			& \multicolumn{1}{r}{1}	& \multicolumn{1}{r}{9}		& \multicolumn{1}{r}{10} 	\\ 

\ej	& \multicolumn{1}{r}{11}		& \multicolumn{1}{r}{20}		& \multicolumn{1}{r}{\NA}	& \multicolumn{1}{r}{\NA}	& \multicolumn{1}{r}{$<1$}	& \multicolumn{1}{r}{$<1$}		& \multicolumn{1}{r}{\NA}	& \multicolumn{1}{r}{$<1$}	& \multicolumn{1}{r}{11}		& \multicolumn{1}{r}{20}	\\  

\end{tabular}
}
\end{table}

\section{Results}
\label{sec:results}

The total cross sections in the two phase space regions defined by thresholds on the \PT of the four leading jets, \regioni and \regionii, are obtained by integrating the differential cross section as a  function of the leading jet $\eta$ and the \DS observable, respectively, yielding:

\begin{align}
\sigmaI \left(\Pp\Pp\mathrm{\to 4j+X}\right)& = 2.77\pm 0.02\,(\mathrm{stat})\,^{+0.68}_{-0.55}\,(\mathrm{syst})\,\mathrm{\mu b}, \\
\sigmaII \left(\Pp\Pp\mathrm{\to 4j+X}\right) &= 0.61 \pm 0.01\,(\mathrm{stat})\,^{+0.12}_{-0.10}\,(\mathrm{syst})\,\mathrm{\mu b}.
\end{align}

Tables~\ref{tab:PHTunes}--~\ref{tab:Total} compares the values measured in data with the ones obtained from MC event generators.
A discussion of the total and  differential cross sections for each of the sets of models introduced in Section\@~\ref{sec:models} is presented below.

The cross sections of the inclusive single-jet Processes A and B are determined by integrating the leading jet $\eta$ distribution for both processes, and are: 
\begin{align}
\sigmaA \left(\Pp\Pp\mathrm{\to 1j+X}\right)&= 15.9 \pm 0.1\,(\mathrm{stat})\,^{+1.8}_{-1.6}\,(\mathrm{syst})\,\mathrm{\mu b}, \label{eq:sigmaA} \\
\sigmaB \left(\Pp\Pp\mathrm{\to 1j+X}\right)&= 106 \pm 1\,(\mathrm{stat})\,^{+10}_{-11}\,(\mathrm{syst})\,\mathrm{\mu b}. \label{eq:sigmaB}
\end{align}	

A large increase in cross section is observed when lowering the \PT threshold from 50\GeV to 30\GeV, as expected. These cross sections will be used as input to Eq.\@~\eqref{eq:DPSpocketformula} for the determination of the DPS cross section and \sigmaeff along with the four-jet efficiency from Eq.\@~\eqref{eq:four-jet-eff}.

\subsection{The \texorpdfstring{\PYTHIA{}8}{PYTHIA8} and \texorpdfstring{\HERWIG}{HERWIG} models}
\label{subsec:PHmodels}

\begin{figure}[htp!] 
\centering
\subfloat{\includegraphics[width=0.48\textwidth]{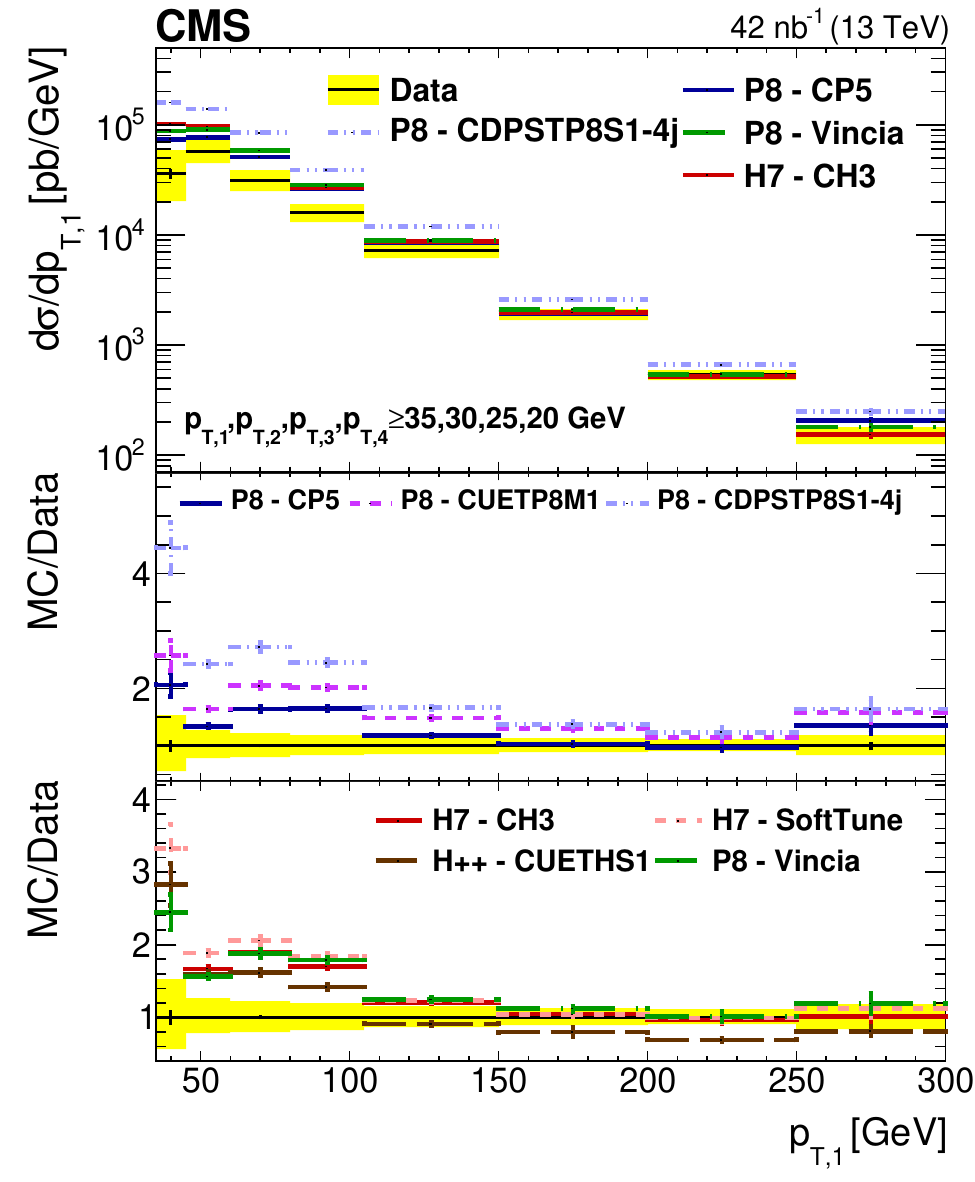}} \hspace*{\fill}
\subfloat{\includegraphics[width=0.48\textwidth]{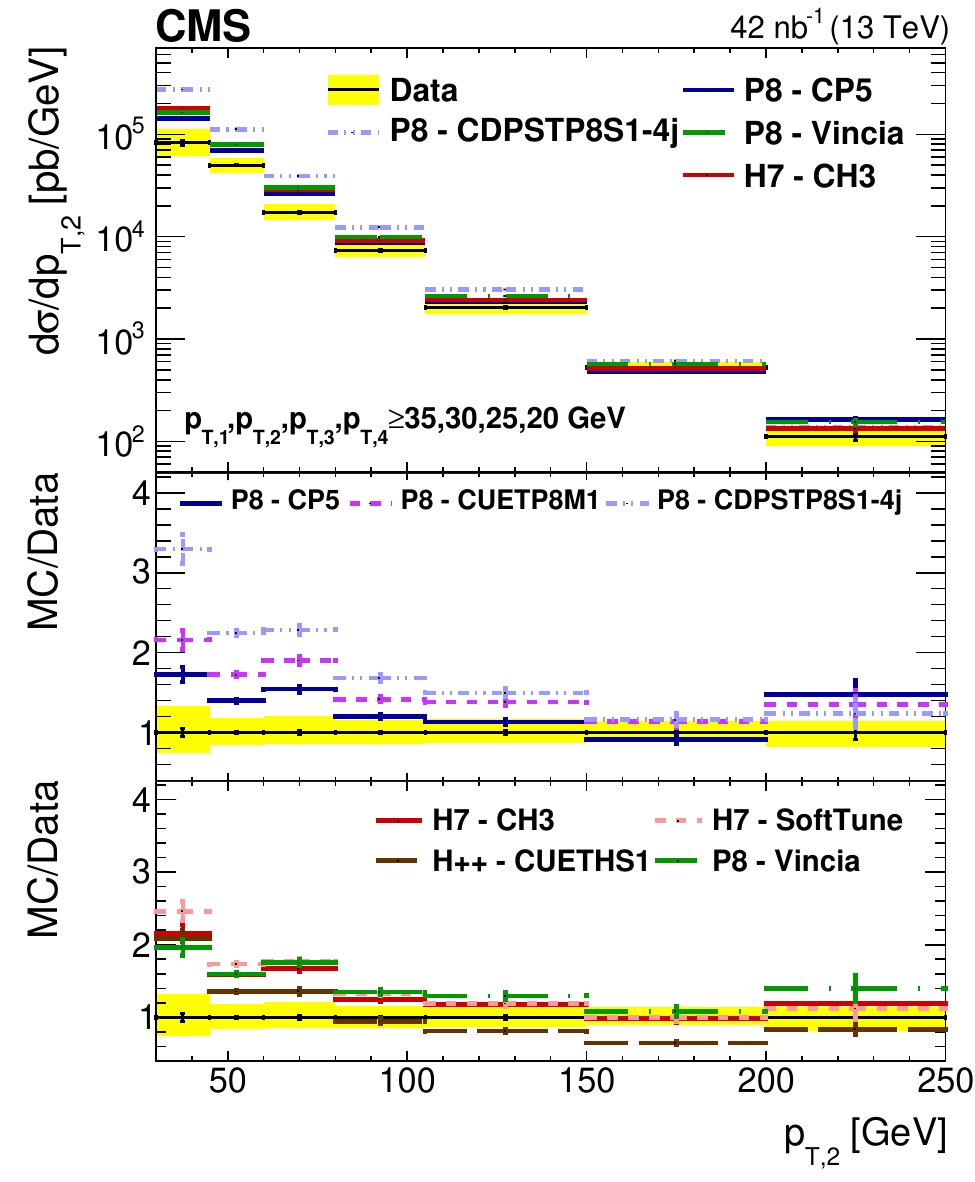}} 

\subfloat{\includegraphics[width=0.48\textwidth]{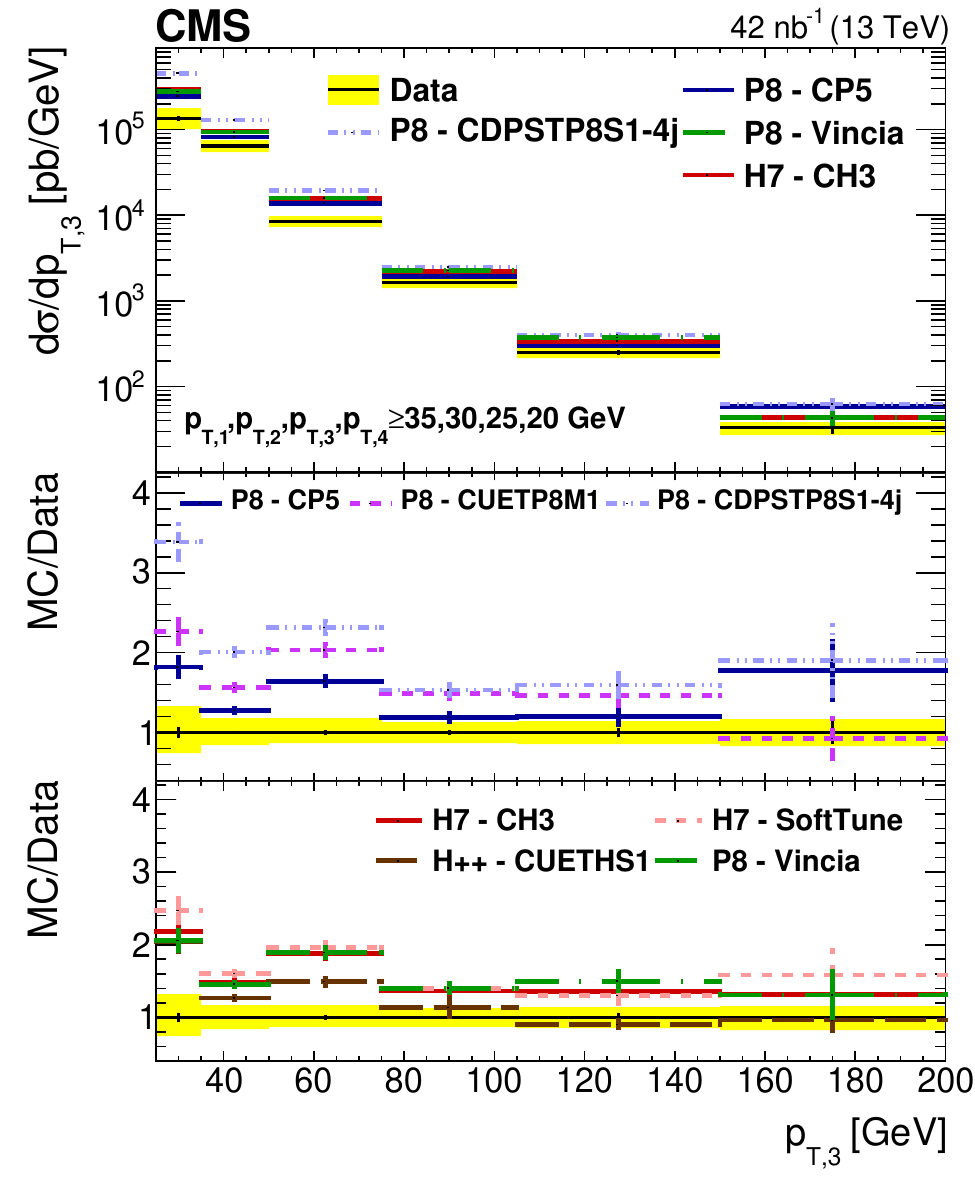}} \hspace*{\fill}
\subfloat{\includegraphics[width=0.48\textwidth]{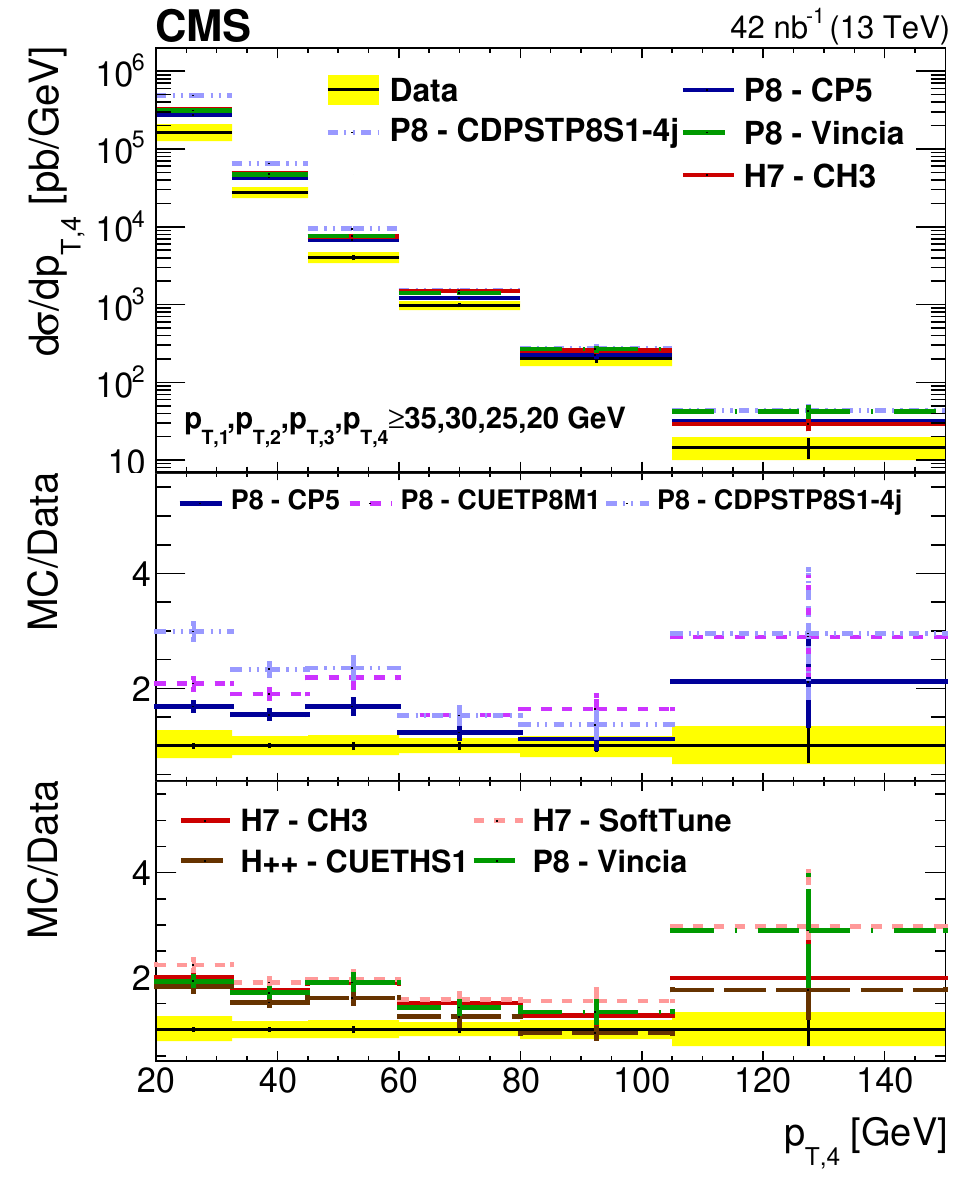}} 

\caption{Comparison of the \PT spectra from data to different \PYTHIA{}8 (P8), \HERWIGpp (H++), and \HERWIG{}7 (H7) tunes, for the leading (upper left), subleading (upper right), third leading (lower left), and fourth leading (lower right) jet in \regioni. The error bars represent the statistical uncertainty, and the yellow band indicates the total (statistical+systematic) uncertainty in the measurement.}
\label{fig:PHtunes_pt}
\end{figure}

\begin{figure}[htp!] 
\centering
\subfloat{\includegraphics[width=0.48\textwidth]{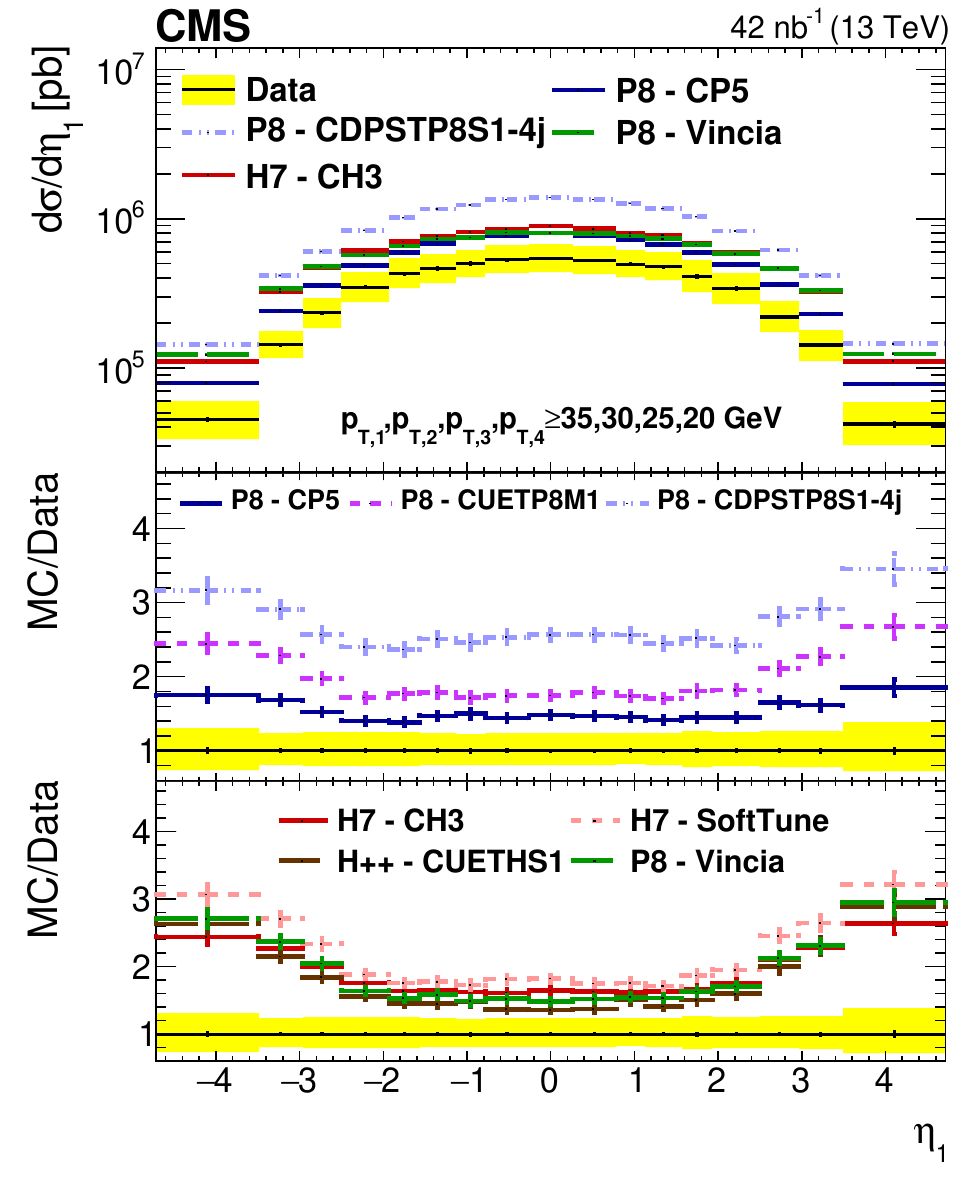}} \hspace*{\fill}
\subfloat{\includegraphics[width=0.48\textwidth]{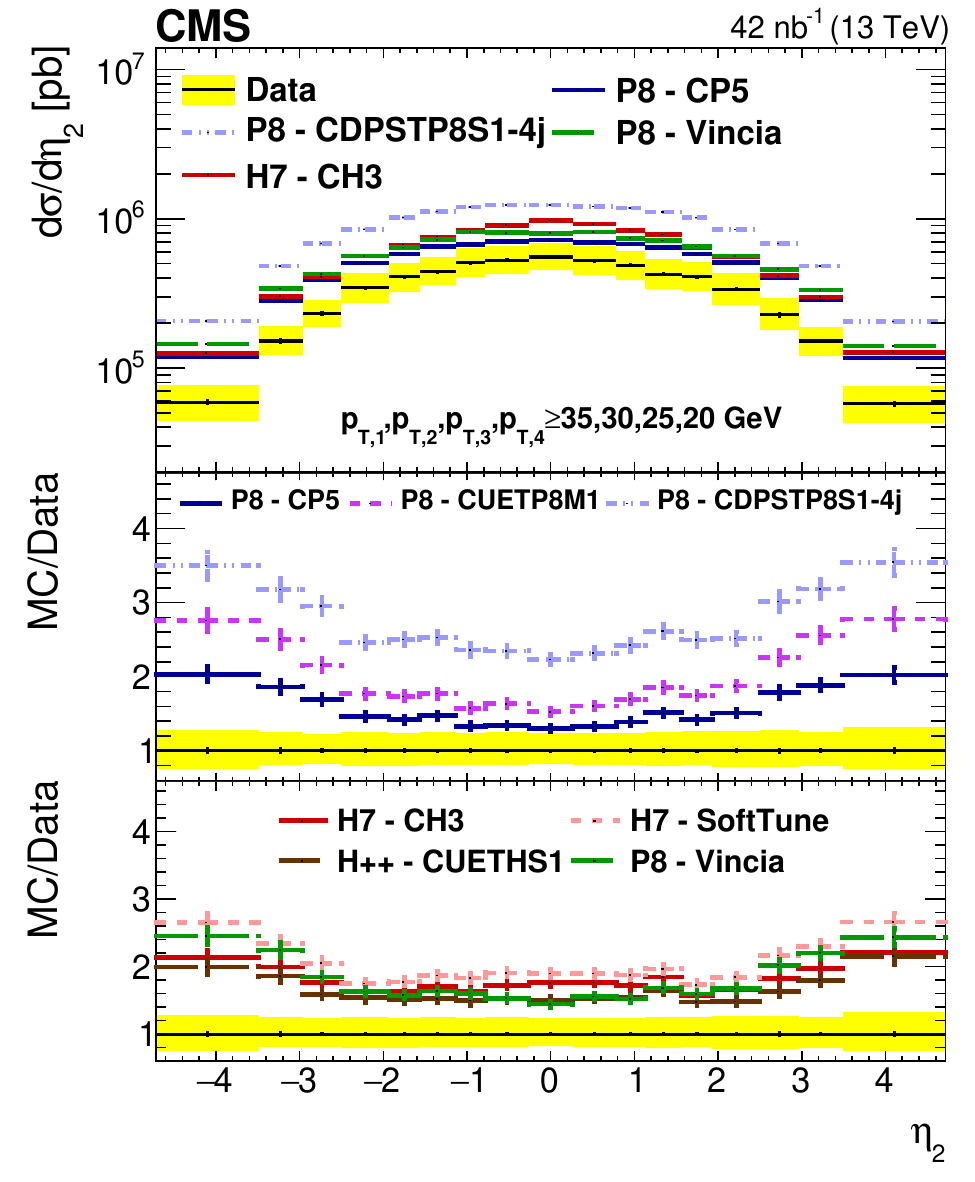}} 

\subfloat{\includegraphics[width=0.48\textwidth]{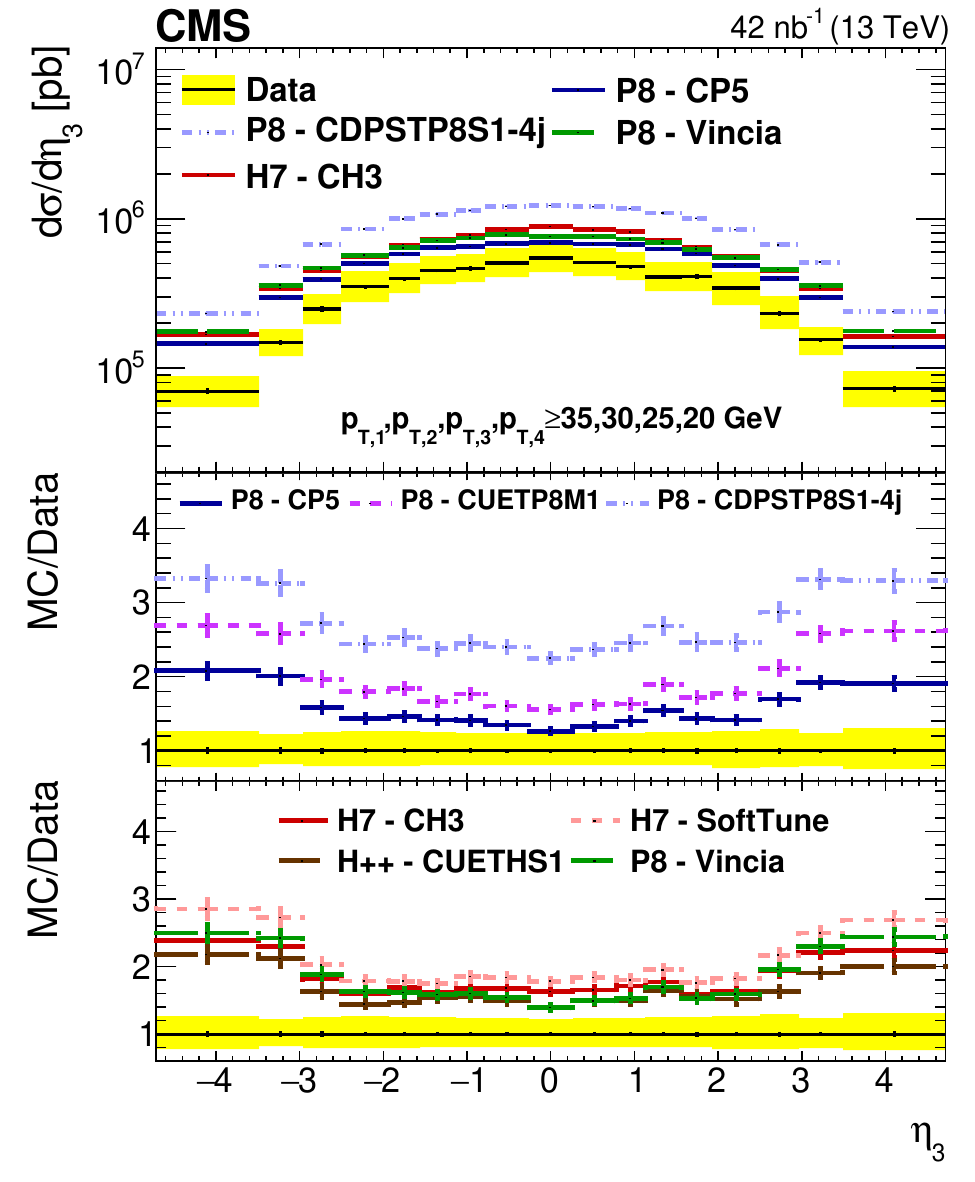}} \hspace*{\fill}
\subfloat{\includegraphics[width=0.48\textwidth]{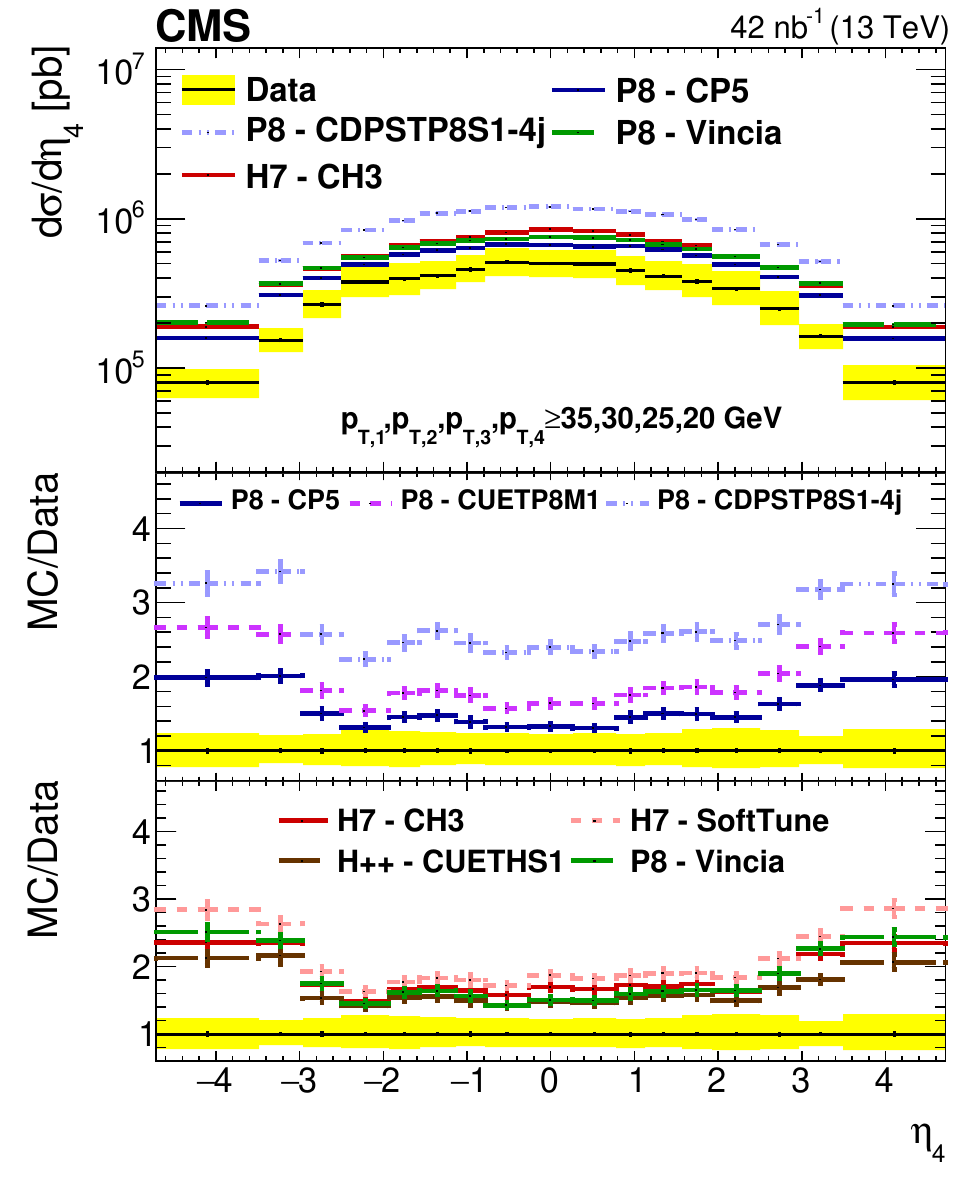}} 

\caption{Comparison of the  $\eta$ spectra from data to different \PYTHIA{}8 (P8), \HERWIGpp (H++), and \HERWIG{}7 (H7) tunes, for the leading (upper left), subleading (upper right), third leading (lower left), and fourth leading (lower right) jet in \regioni. The error bars represent the statistical uncertainty, and the yellow band indicates the total (statistical+systematic) uncertainty in the measurement.}
\label{fig:PHtunes_rap}
\end{figure}

\begin{figure}[htp!] 
\centering
\subfloat{\includegraphics[width=0.48\textwidth]{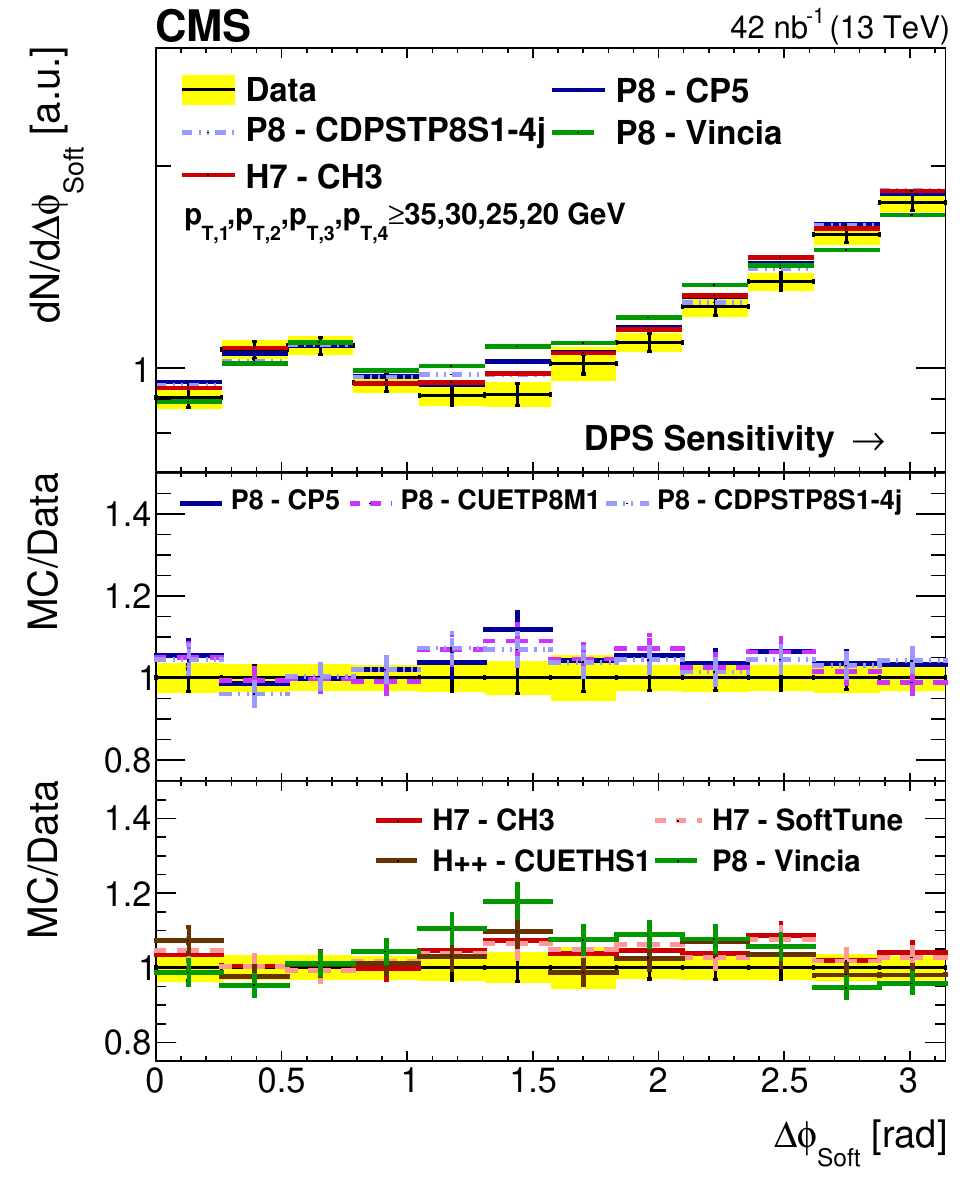}} \hspace*{\fill}
\subfloat{\includegraphics[width=0.48\textwidth]{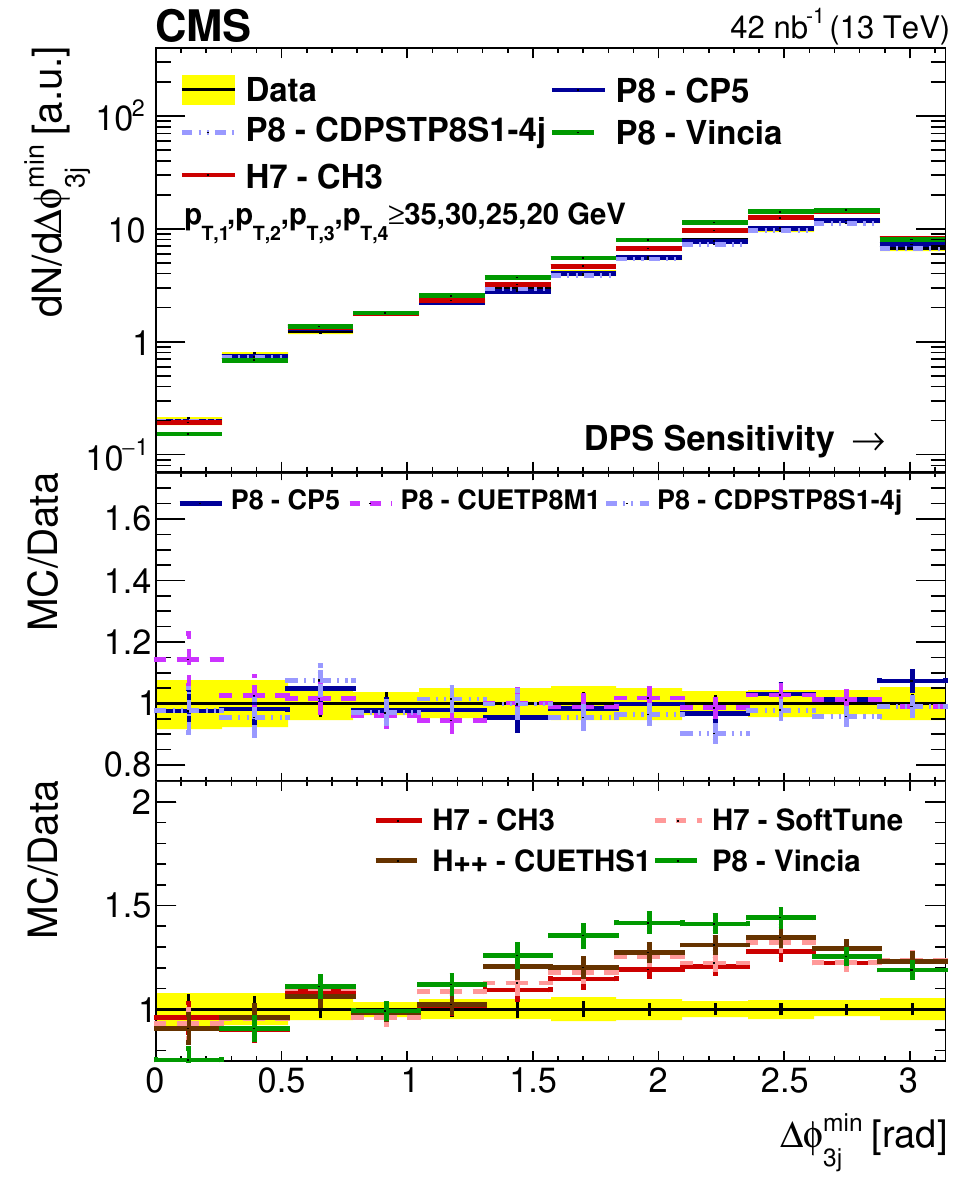}} 

\subfloat{\includegraphics[width=0.48\textwidth]{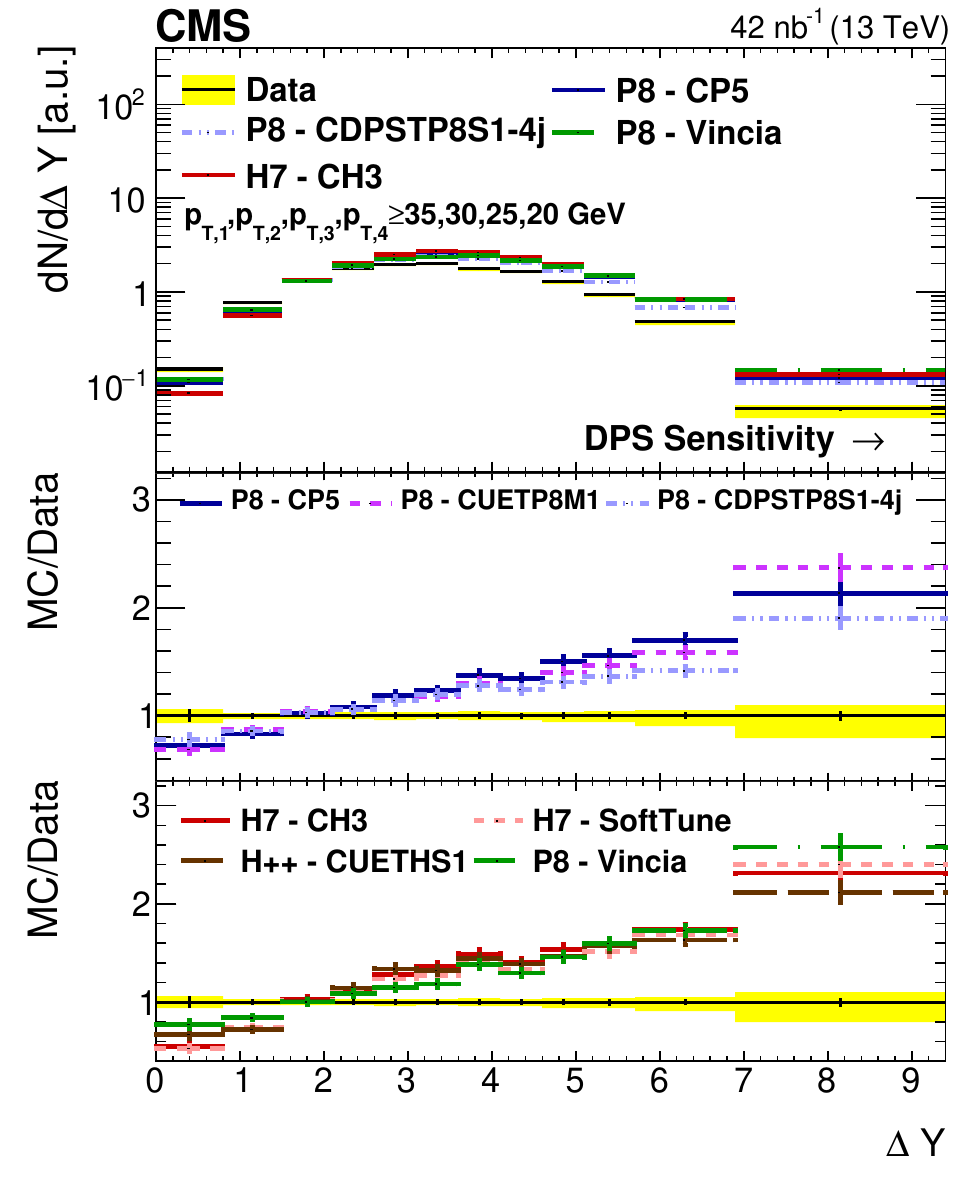}} \hspace*{\fill}
\subfloat{\includegraphics[width=0.48\textwidth]{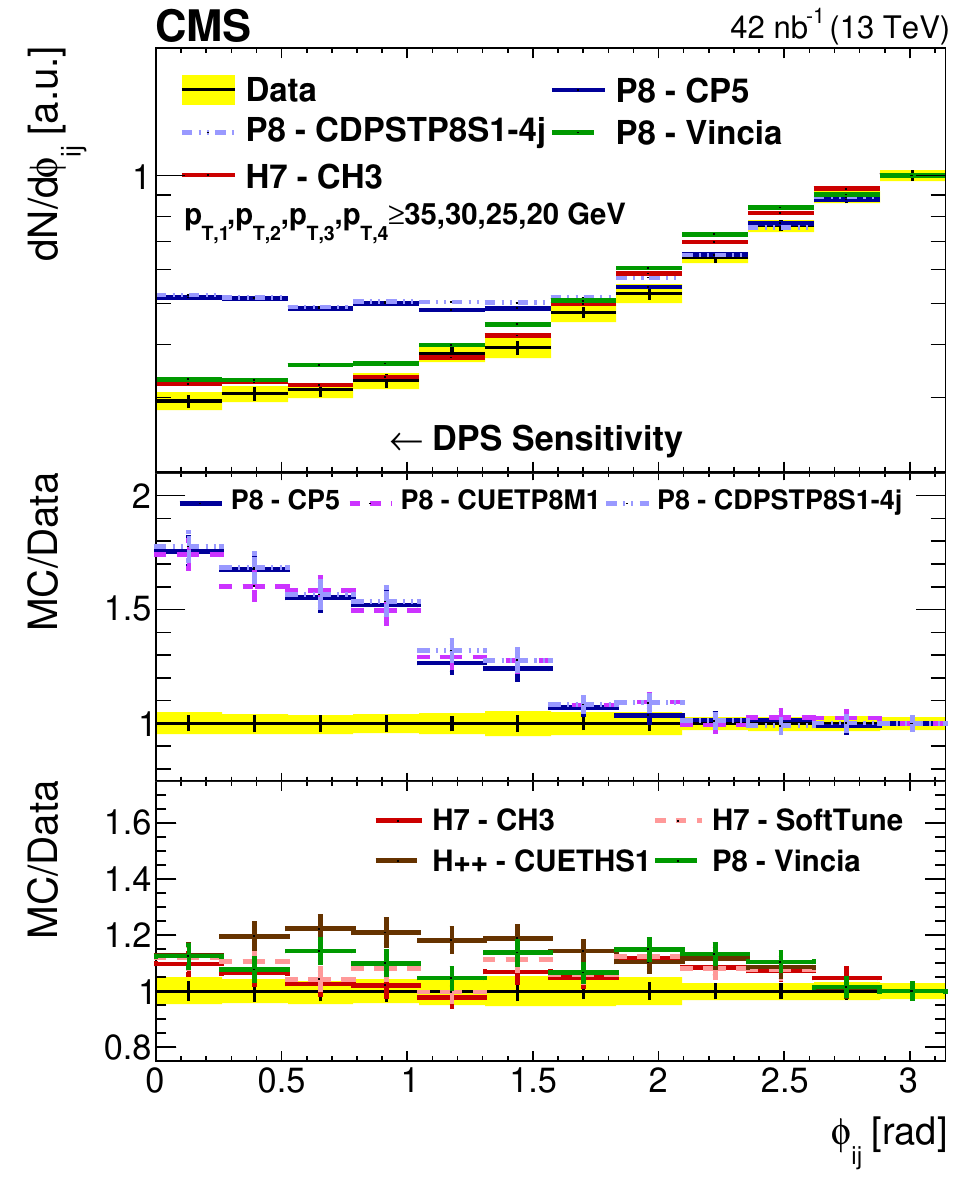}} 

\caption{Comparison of the  \DphiS, \Dphimin, \DY, and \phiij distributions from data to different \PYTHIA{}8 (P8), \HERWIGpp (H++), and \HERWIG{}7 (H7) tunes  in \regioni. All distributions have been normalized to  regions where a reduced DPS contribution is expected. The error bars represent the statistical uncertainty, and the yellow band indicates the total (statistical+systematic) uncertainty in the measurement.}
\label{fig:PHtunes_var1}
\end{figure}

\begin{figure}[htp!] 
\centering
\subfloat{\includegraphics[width=0.48\textwidth]{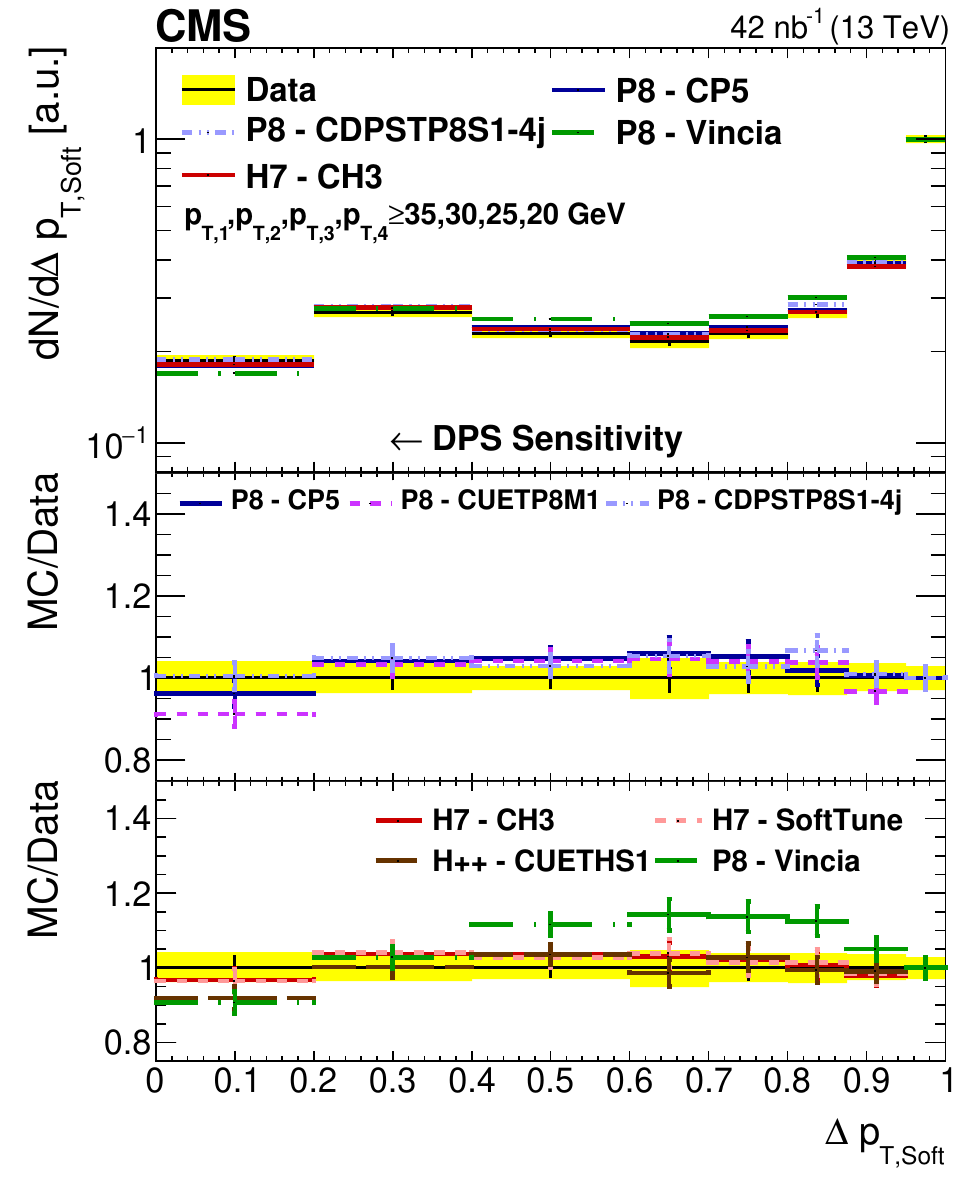}} \hspace*{\fill}
\subfloat{\includegraphics[width=0.48\textwidth]{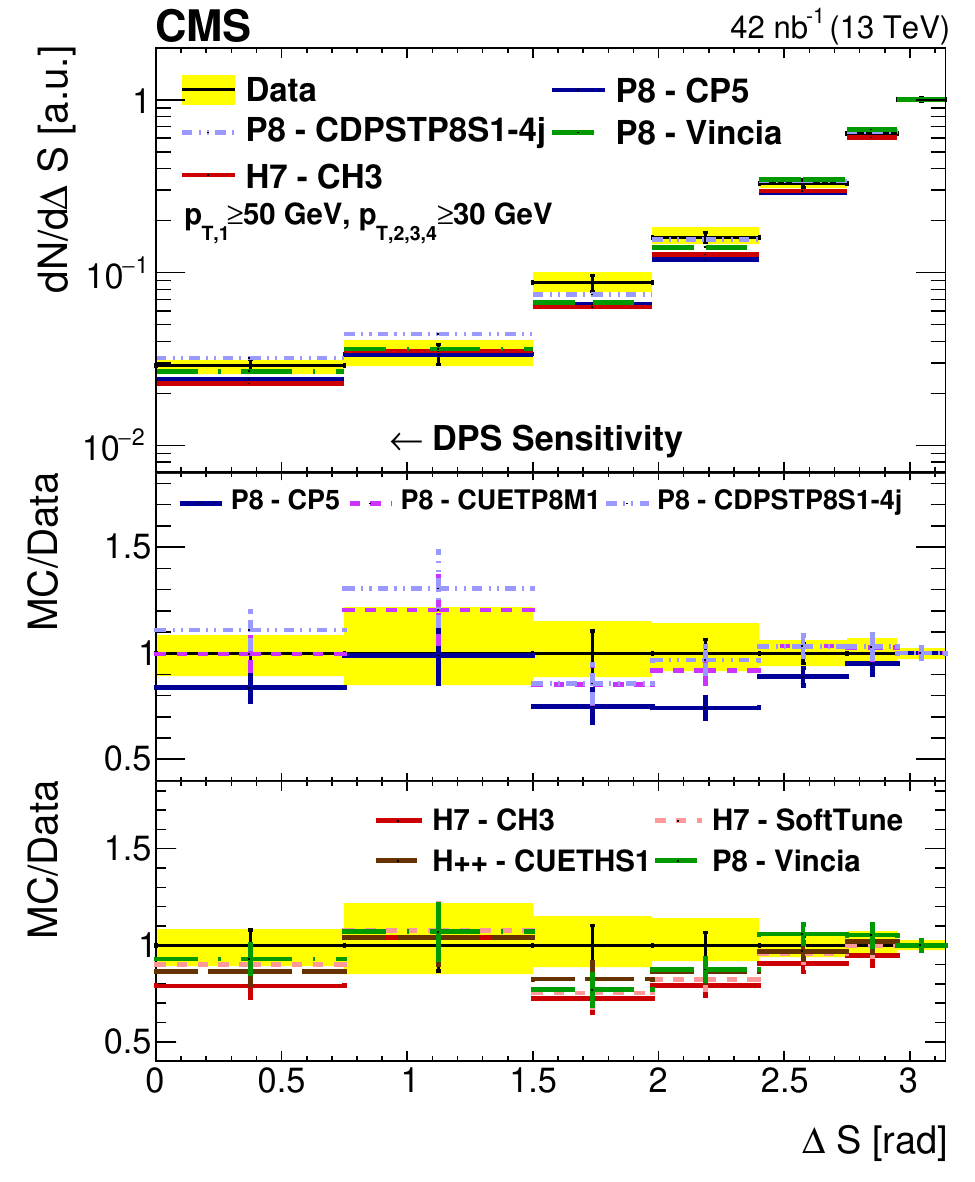}}  

\caption{Comparison of the  \DptS and \DS distributions from data to different \PYTHIA{}8 (P8), \HERWIGpp (H++), and \HERWIG{}7 (H7) tunes in \regioni and \regionii, respectively. All distributions have been normalized to  regions where a  reduced DPS contribution is expected. The error bars represent the statistical uncertainty, and the yellow band indicates the total (statistical+systematic) uncertainty in the measurement.}
\label{fig:PHtunes_var2}
\end{figure}

The models based on LO \MEtwo matrix elements, \PYTHIA{}8, \HERWIGpp, and \HERWIG{}7, respectively labeled as P8, H++, and H7 in the figures, are compared with data. Table~\ref{tab:PHTunes} gives an overview of the event generators, tunes, and PDF sets, along with their respective cross sections. All LO \MEtwo models predict cross sections that are much larger than the measured ones; especially \PYTHIA{}8 with the CDPSTP8S1-4j tune predicts a cross section that is roughly 2.5 times larger than the one observed in data. Figures~\ref{fig:PHtunes_pt}--~\ref{fig:PHtunes_var2} show a comparison of the data to various MC models as a function of \PT, $\eta$, and the DPS-sensitive observables. Three \PYTHIA{}8 models and one \HERWIG{}7 model are shown in direct comparison with the data. These models employ the most recent CP5 and CH3 tunes, the dedicated DPS tune or combine \PYTHIA{}8 with a dipole-antenna shower provided by \VINCIA, while all of the models are represented in the ratio plots.

The \PT spectra in Fig.~\ref{fig:PHtunes_pt}, obtained for \regioni, show that the much larger integrated cross section of the MC models, compared with the data, comes from an abundance of low-\PT jets,  whereas for $\PT \gtrsim 100\GeV$ the models show agreement within 50\% of the data; for \HERWIG{}7 it is even within the total uncertainty. Fig.~\ref{fig:PHtunes_rap}, the $\eta$ spectra, shows that a large part of the excess of low-\PT jets is located in the forward $\eta$ regions. 

\begin{table}[htp!]
\centering
\topcaption{Cross sections obtained from data and from the \PYTHIA{}8, \HERWIGpp, and \HERWIG{}7 models in \regioni and \regionii of the phase space, where ME stands for matrix element.}
\label{tab:PHTunes}
\resizebox{\textwidth}{!}{
\begin{tabular}{p{17.5mm}lllll}
Sample & ME & Tune & PDF & \sigmaI (\unit{$\mu$b}) & \sigmaII (\unit{$\mu$b})	\\ \hline

Data & \NA & \NA & \NA & \multicolumn{1}{r}{$2.77\pm 0.02\,^{+0.68}_{-0.55}$}	& \multicolumn{1}{l}{$0.61$ $\pm 0.01\,^{+0.12}_{-0.10}$}	\\ [\cmsTabSkip]

\PYTHIA{}8   & LO \MEtwo	& {CUETP8M1} & {NNPDF2.3\_LO}		& \multicolumn{1}{l}{5.03}		& \multicolumn{1}{l}{1.07}	   \\  

\PYTHIA{}8 	& LO \MEtwo	& {CP5}	& {NNPDF2.3\_NNLO}		& \multicolumn{1}{l}{4.07} 		& \multicolumn{1}{l}{0.84}	   \\  

\PYTHIA{}8 	& LO \MEtwo	& {CDPSTP8S1-4j} & {CTEQ6L1}		& \multicolumn{1}{l}{7.06} 		& \multicolumn{1}{l}{1.28}	   \\  

\PYTHIA{}8 +\VINCIA   & LO \MEtwo	& Default  & {NNPDF2.3\_LO}		& \multicolumn{1}{l}{4.66} 		& \multicolumn{1}{l}{0.97}	   \\  [\cmsTabSkip]

\HERWIGpp 	& LO \MEtwo	& {CUETHS1}  & {CTEQ6L1}		& \multicolumn{1}{l}{4.35}		& \multicolumn{1}{l}{0.83}	   \\  

\HERWIG{}7 	& LO \MEtwo	& {CH3}	& {NNPDF2.3\_NNLO}		& \multicolumn{1}{l}{4.82}		& \multicolumn{1}{l}{0.98}	   \\  

\HERWIG{}7 	& LO \MEtwo	& {SoftTune} & {MMHT2014lo68cl}		& \multicolumn{1}{l}{5.34}		& \multicolumn{1}{l}{1.07}	   \\  

\end{tabular}
}	
\end{table}

The distributions in the DPS-sensitive observables are shown in Figs.\@~\ref{fig:PHtunes_var1} and~\ref{fig:PHtunes_var2}, for \regionii in the case of \DS and \regioni for all other observables.  To make qualitative statements about the shape, the fully corrected distributions have been normalized to one or more bins where a much reduced DPS contribution is expected. The distribution in \DphiS is normalized to the average of the first five bins, covering the tail of the distribution, which is determined by the jet cone size. The distributions in \Dphimin and \DY are normalized to the average of their first four bins. Normalizing to the average of multiple bins reduces the effect of statistical fluctuations. The distributions in \phiij, \DptS, and \DS  are all normalized to their last bin, since these bins already have a small relative statistical uncertainty.

The \DphiS and \DptS distributions are relatively well described by all LO \MEtwo models. Deviations from data never exceed 20\%, albeit being larger than the total uncertainty in the data points in certain bins for some of the models. Similar results are observed for the predictions of the \DptS observable from the DPS tune in~\cite{CMS:2021wfx}. Deviations of 10--20\% occur between a model employing a similar DPS tune (CDPSTP8S1-WJ) and the data in the Z+jets final state.

The shape of the \DY distribution predicted by all the LO \MEtwo models differs significantly from data, and the MC-to-data ratio increases towards higher values of \DY. The overshoot at large \DY is consistent with the excess of low-\PT forward jets.

A distinction between two classes of models becomes apparent in the \Dphimin and \phiij  distributions. The models implementing a \PT-ordered parton shower describe the distribution in \Dphimin well, although yielding a distribution in \phiij that is more uncorrelated than observed in data. The slope of the \Dphimin distribution obtained from the models with a \PT-ordered shower algorithm going to zero, overshoots the slope of the distribution obtained from data. For models that use an angular-ordered or dipole-antenna parton shower,  the \Dphimin correlation is too strong. The slope of the distributions \Dphimin distributions obtained from the models with an angular-ordered shower overshoot the distribution obtained from data when going to $\pi$, whereas the shape of the data is described more accurately by \phiij. The \PYTHIA{}8+\VINCIA model confirms that the parton shower algorithm is responsible for the different tendencies observed for the two classes of models.  This observation makes these observables less suitable to untangle SPS and DPS contributions to the cross section.

The \DS distribution is less affected by different parton shower implementations.  The DPS tune {CDPSTP8S1-4j} agrees very well with the shape of the data, but lies slightly above the data  at low values of \DS, pointing to a potential overestimation of the DPS contribution. All other models underestimate the data at low \DS, indicating a possible need for more DPS to obtain a proper description of the shape of the \DS distribution.

\subsection{Multijet models}
\label{subsec:multijetmodels}

\begin{figure}[htp!] 
\centering
\subfloat{\includegraphics[width=0.48\textwidth]{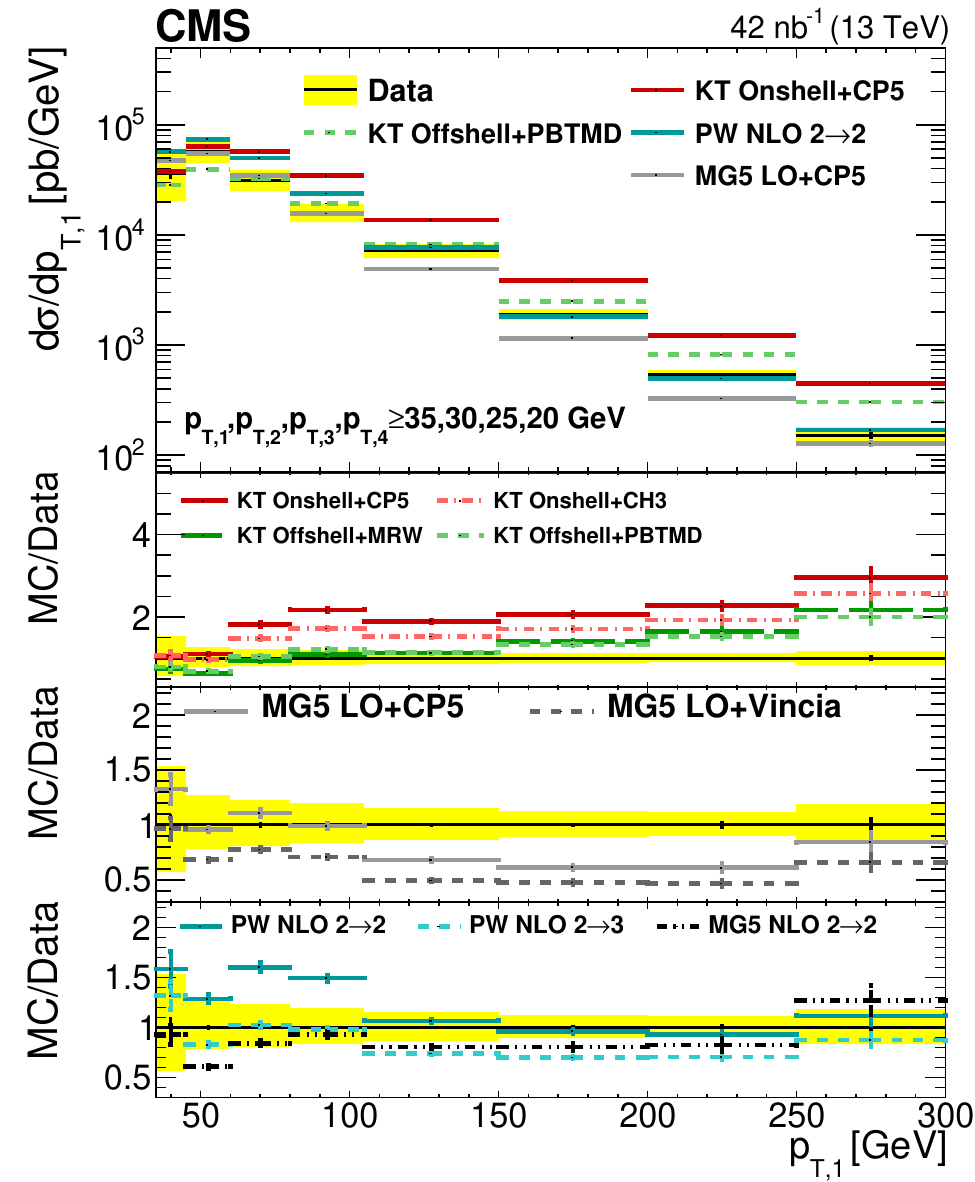}} \hspace*{\fill}
\subfloat{\includegraphics[width=0.48\textwidth]{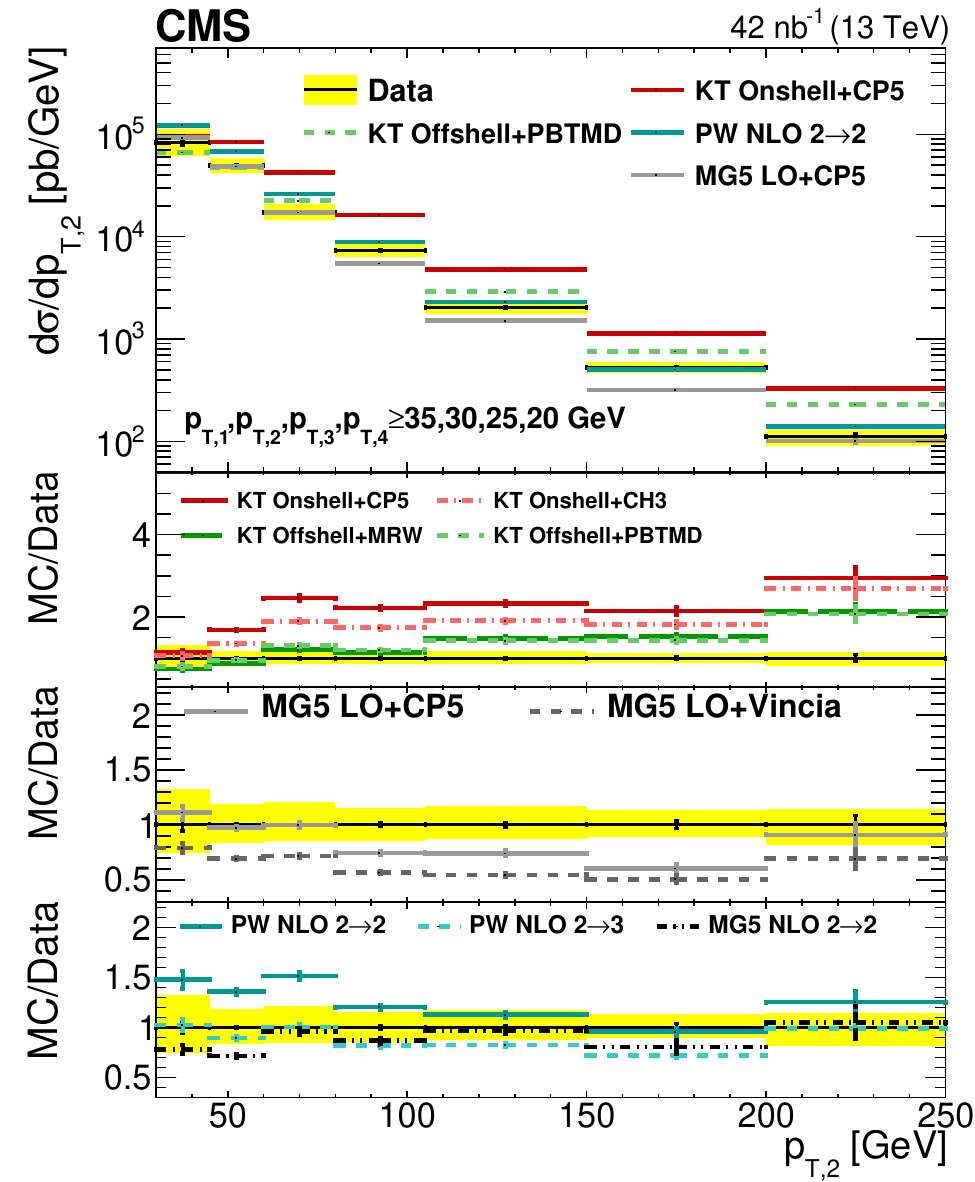}} 

\subfloat{\includegraphics[width=0.48\textwidth]{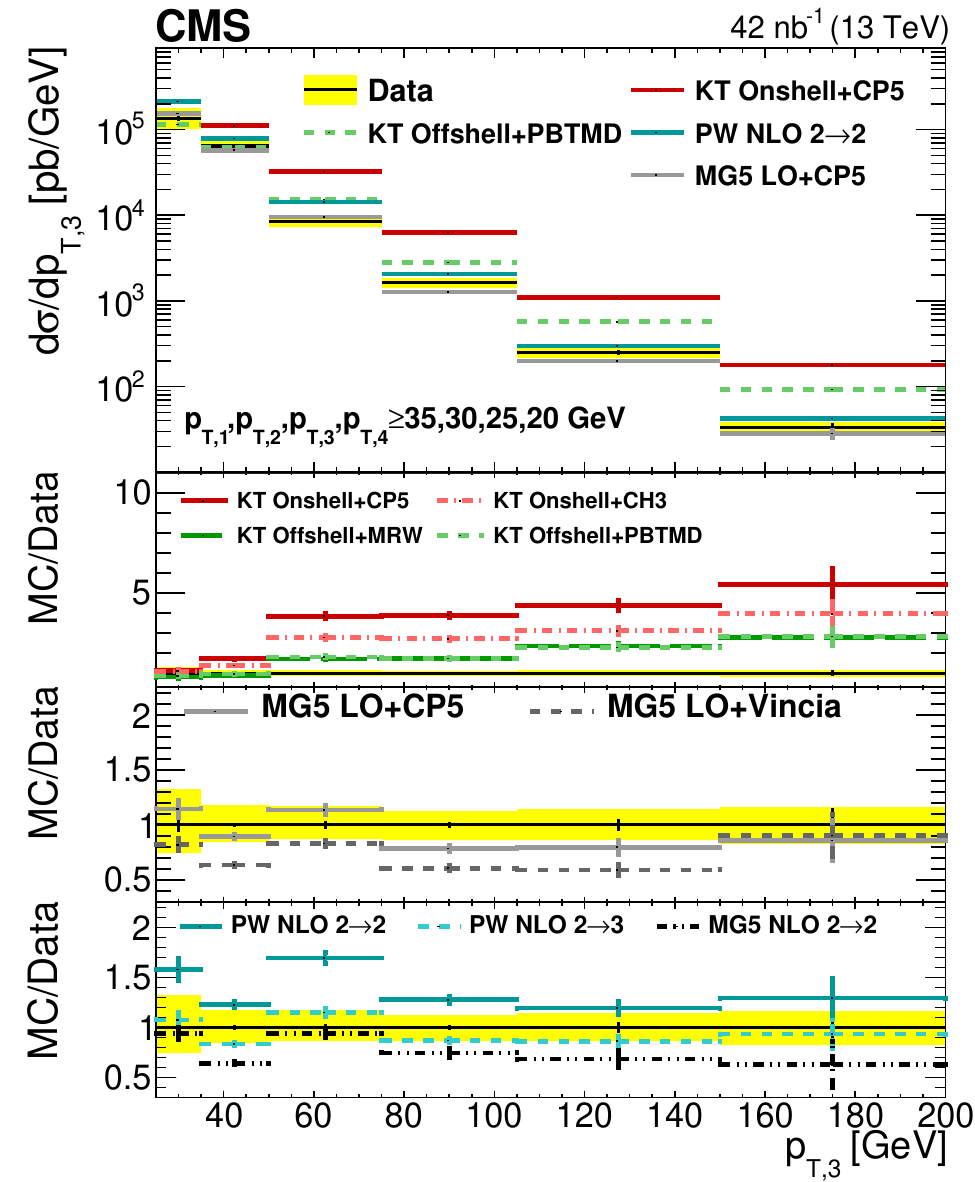}} \hspace*{\fill}
\subfloat{\includegraphics[width=0.48\textwidth]{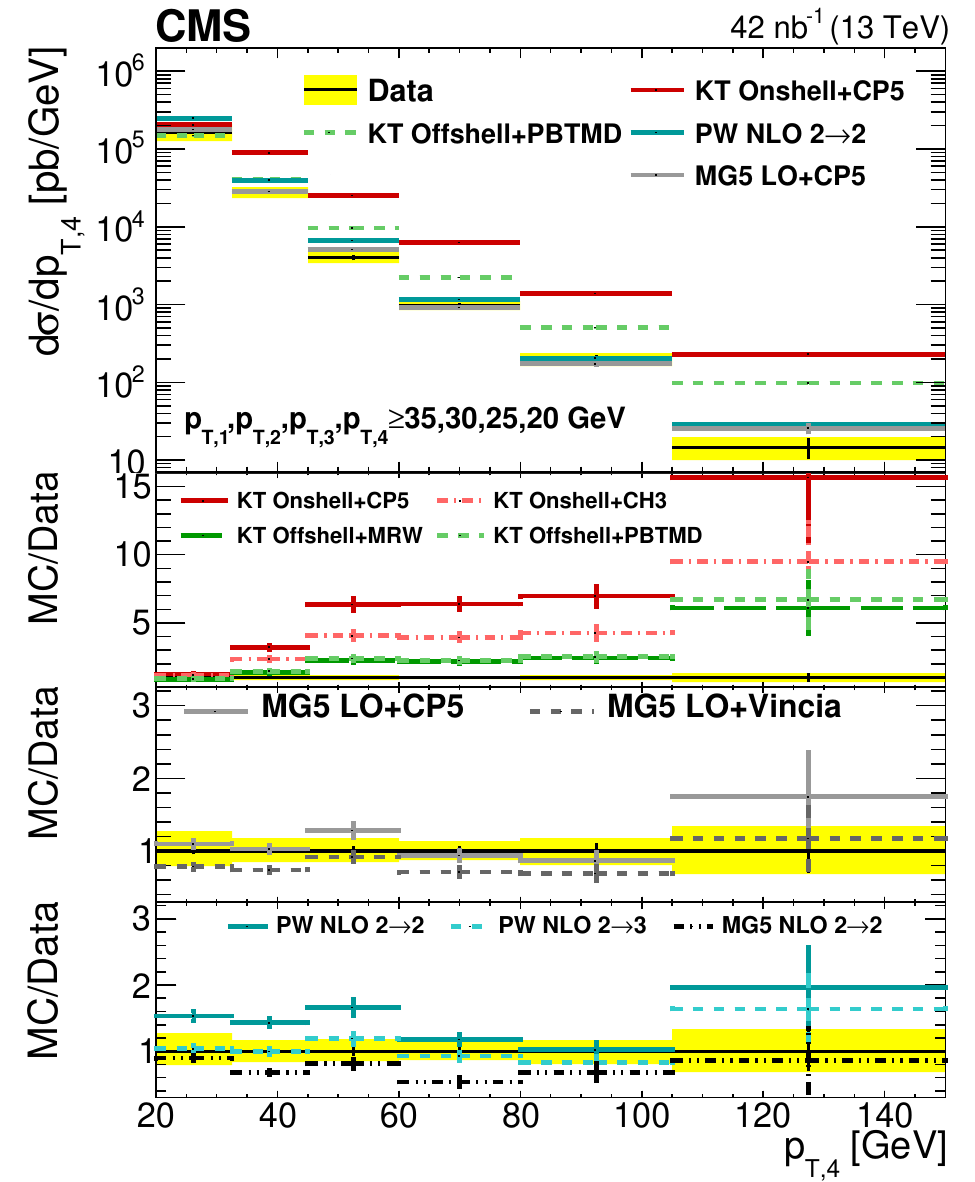}} 

\caption{Comparison of the unfolded \PT spectra of data with different \KATIE (KT), \MGvATNLO (MG5), and \POWHEG (PW) models, for the leading (upper left), subleading (upper right), third leading (lower left), and fourth leading (lower right) jet in \regioni. The error bars represent the statistical uncertainty, and the yellow band indicates the total (statistical+systematic) uncertainty in the measurement.}
\label{fig:multijet_pt}
\end{figure}

\begin{figure}[htp!] 
\centering
\subfloat{\includegraphics[width=0.48\textwidth]{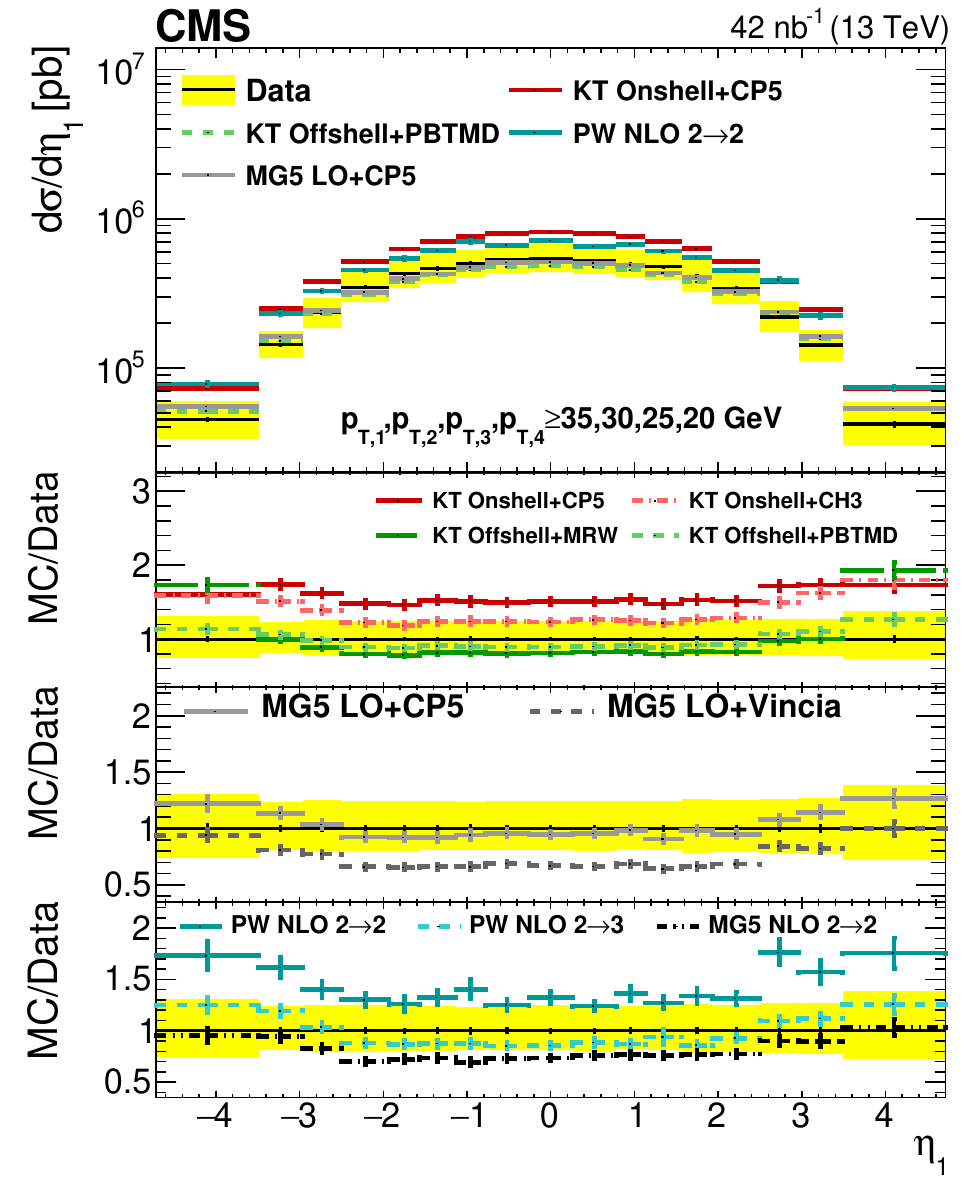}} \hspace*{\fill}
\subfloat{\includegraphics[width=0.48\textwidth]{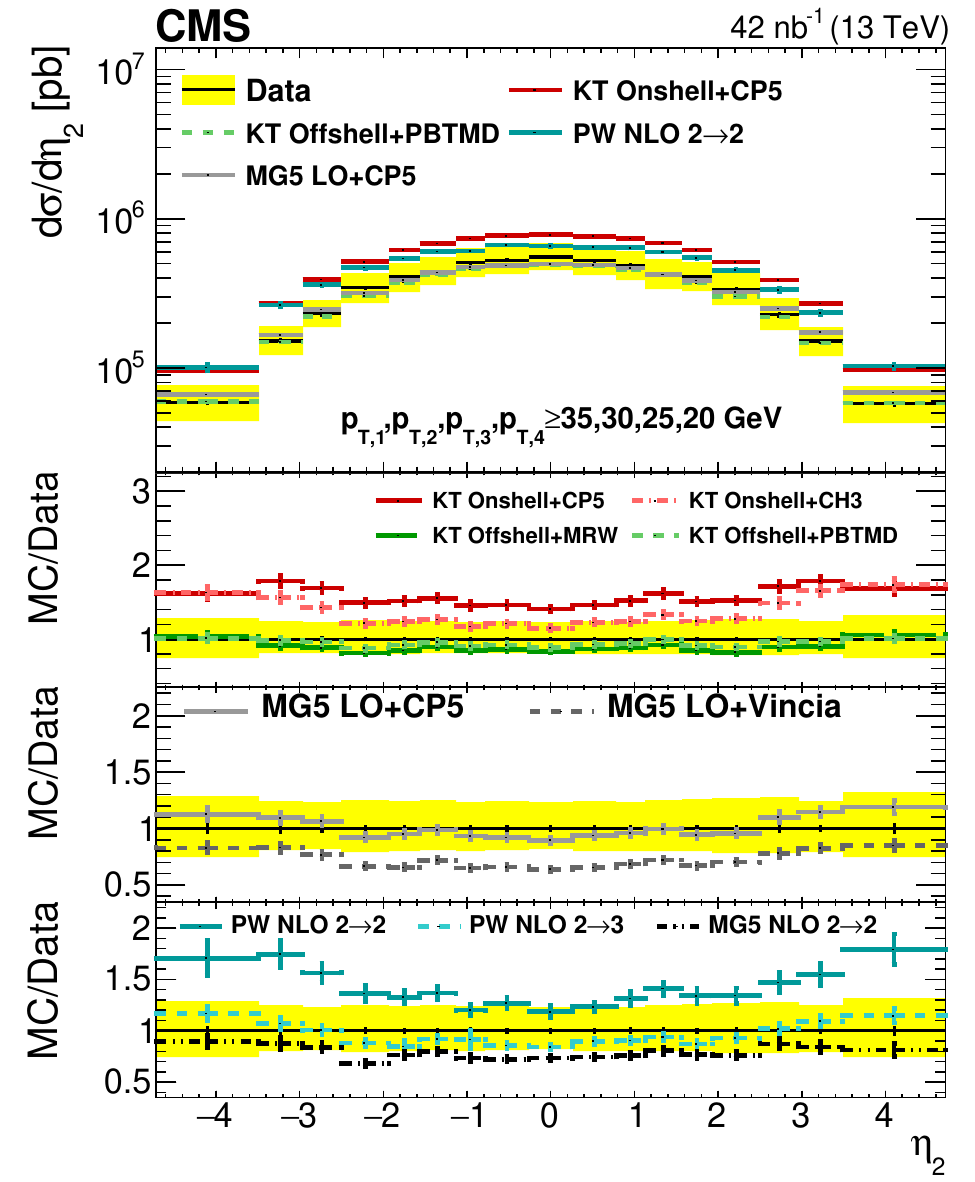}} 

\subfloat{\includegraphics[width=0.48\textwidth]{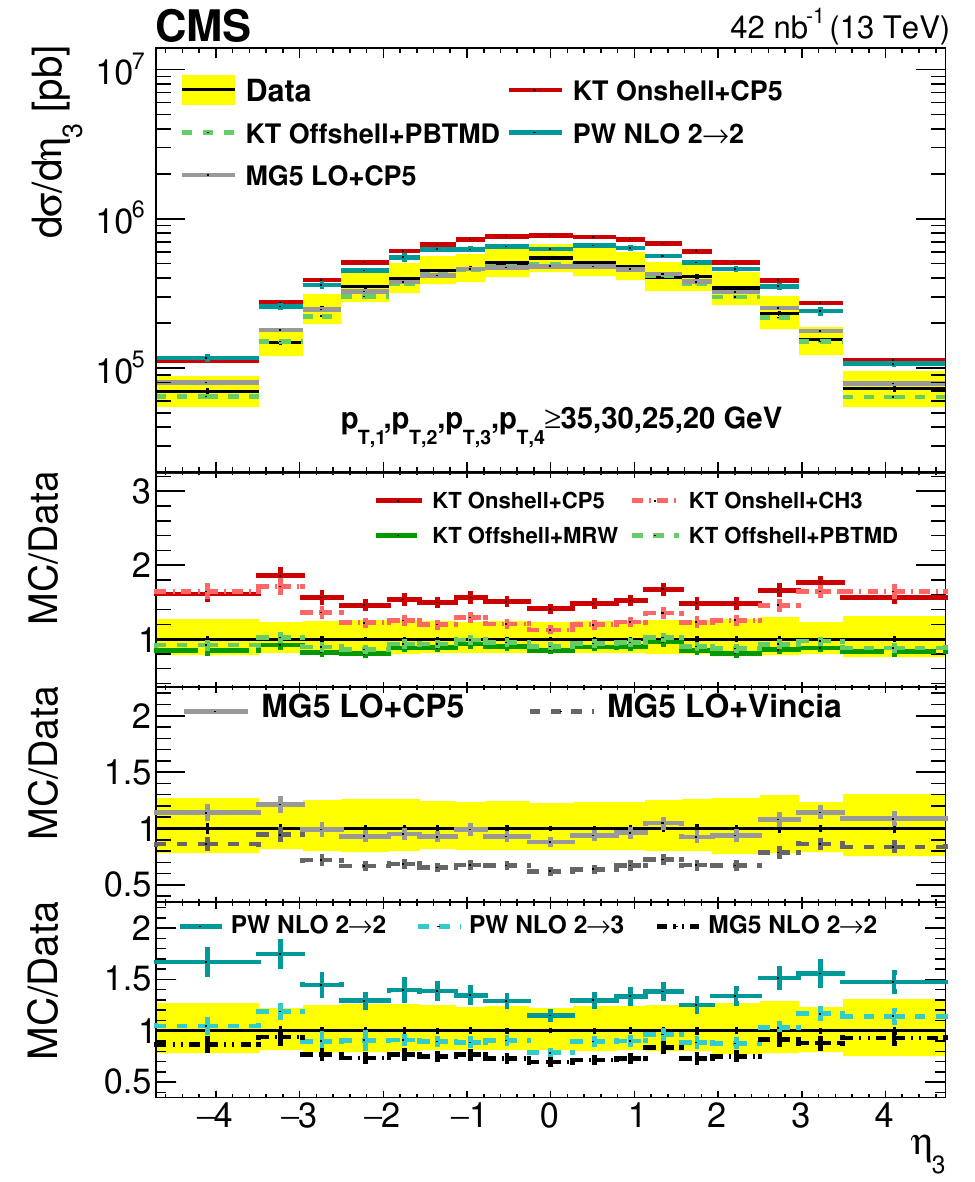}} \hspace*{\fill}
\subfloat{\includegraphics[width=0.48\textwidth]{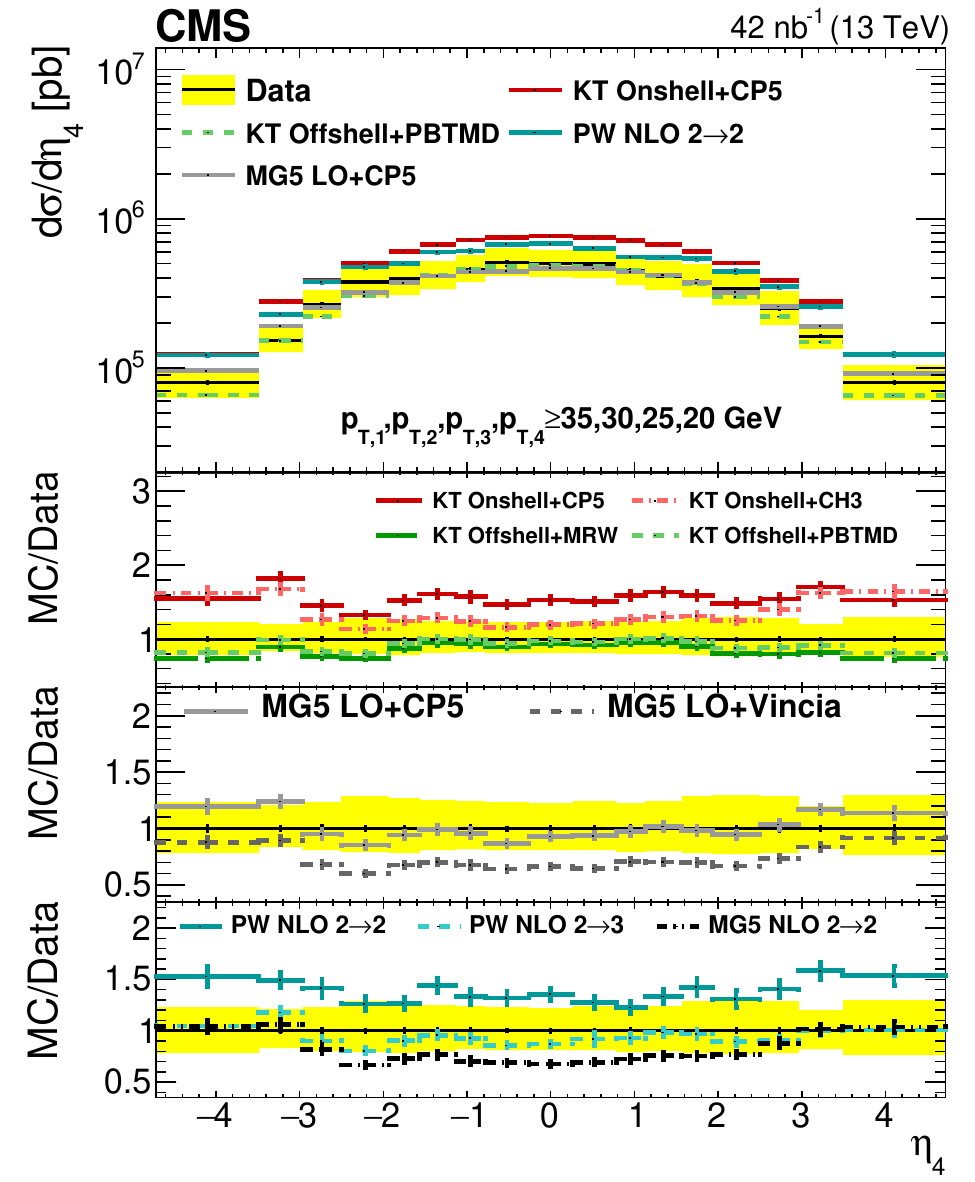}} 

\caption{Comparison of the unfolded $\eta$ spectra of data with different \KATIE (KT), \MGvATNLO (MG5), and \POWHEG (PW) models, for the leading (upper left), subleading (upper right), third leading (lower left), and fourth leading (lower right) jet in \regioni. The error bars represent the statistical uncertainty, and the yellow band indicates the total (statistical+systematic) uncertainty in the measurement.}
\label{fig:multijet_rap}
\end{figure}

\begin{figure}[htp!] 
\centering
\subfloat{\includegraphics[width=0.48\textwidth]{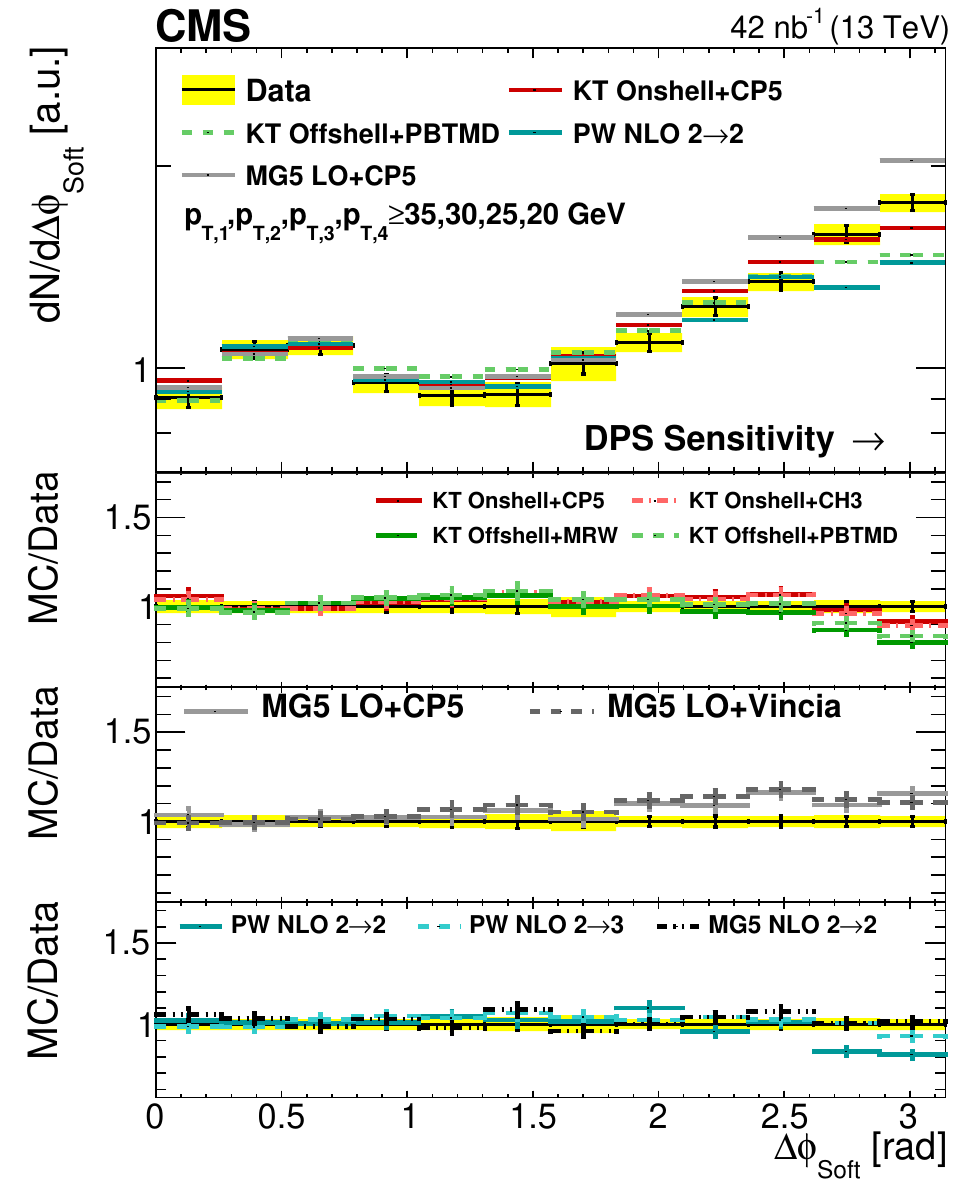}} \hspace*{\fill}
\subfloat{\includegraphics[width=0.48\textwidth]{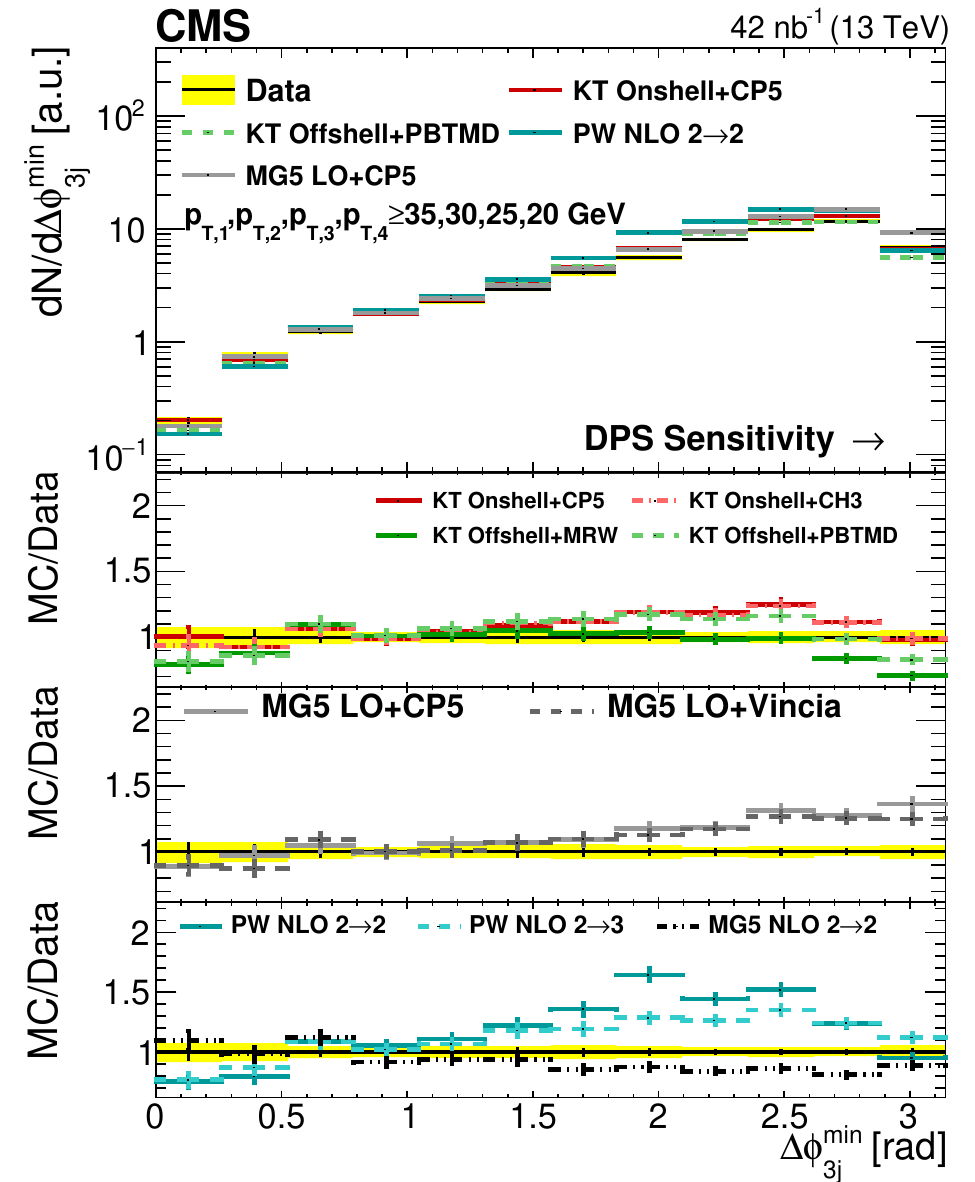}} 

\subfloat{\includegraphics[width=0.48\textwidth]{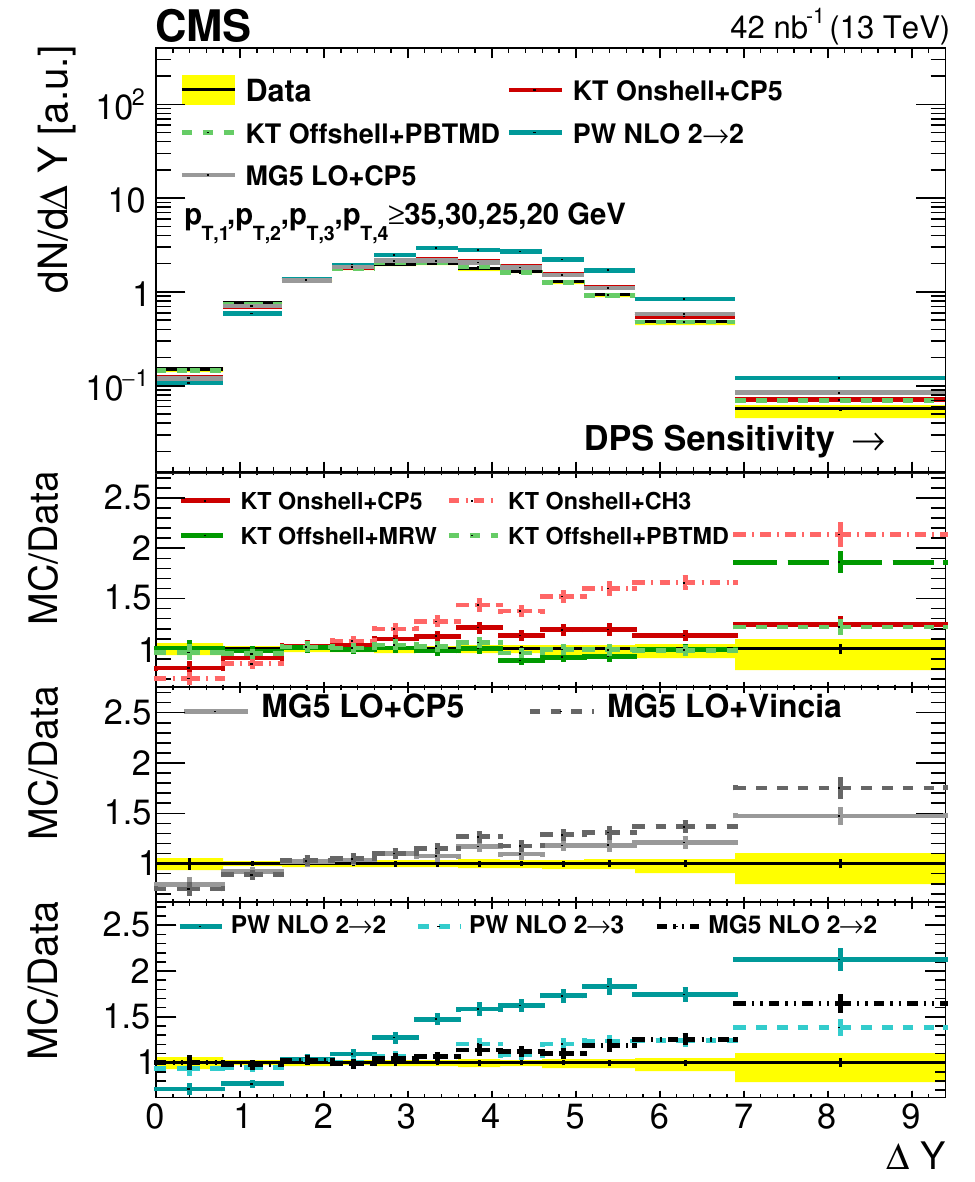}} \hspace*{\fill}
\subfloat{\includegraphics[width=0.48\textwidth]{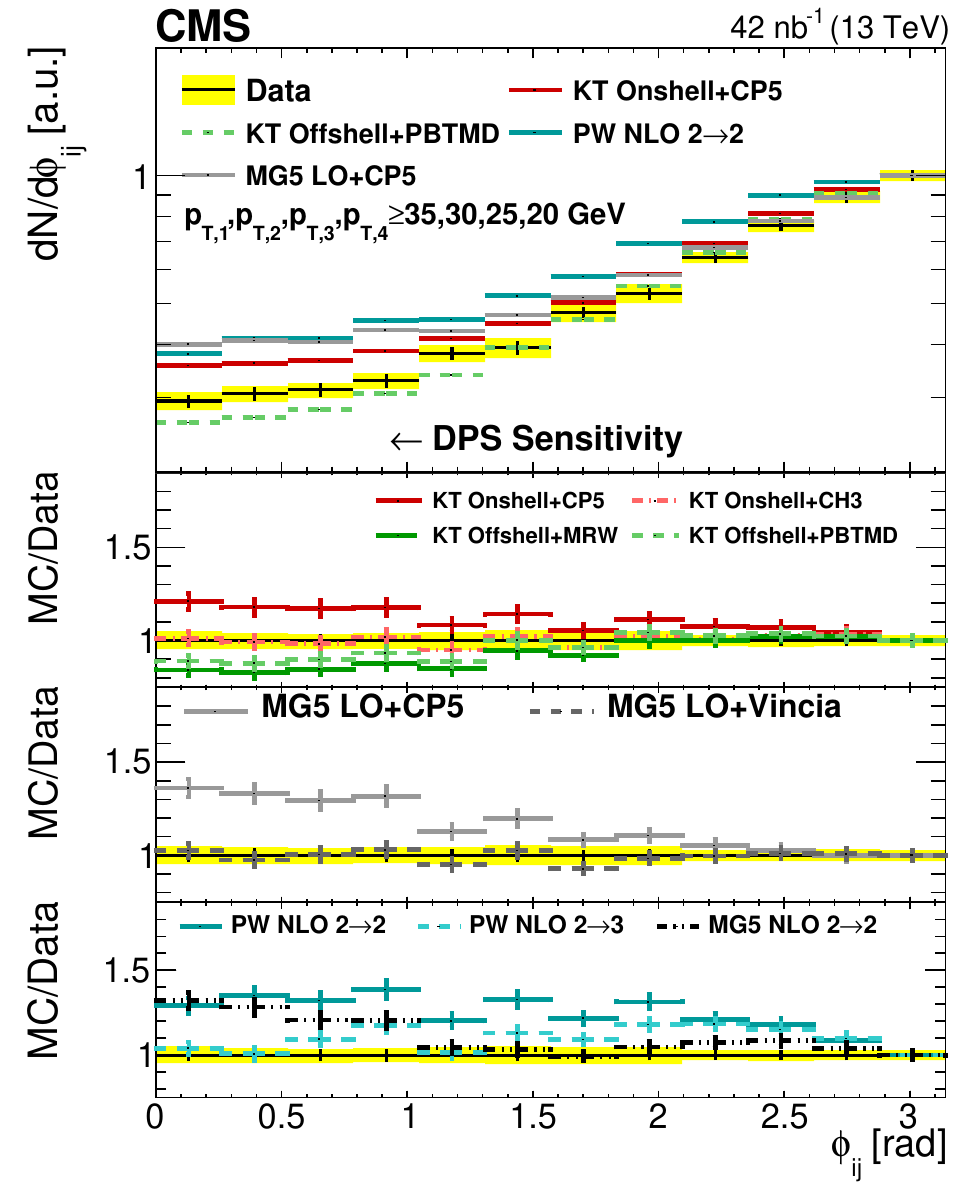}}  

\caption{Comparison of the  \DphiS, \Dphimin, \DY, and \phiij distributions from data to different \KATIE (KT), \MGvATNLO (MG5), and \POWHEG (PW) implementations in \regioni. All distributions have been normalized to regions where a reduced DPS sensitivity is expected. The error bars represent the statistical uncertainty, and the yellow band indicates the total (statistical+systematic) uncertainty on the measurement.}
\label{fig:multijet_var1}
\end{figure}

\begin{figure}[htp!] 
\centering
\subfloat{\includegraphics[width=0.48\textwidth]{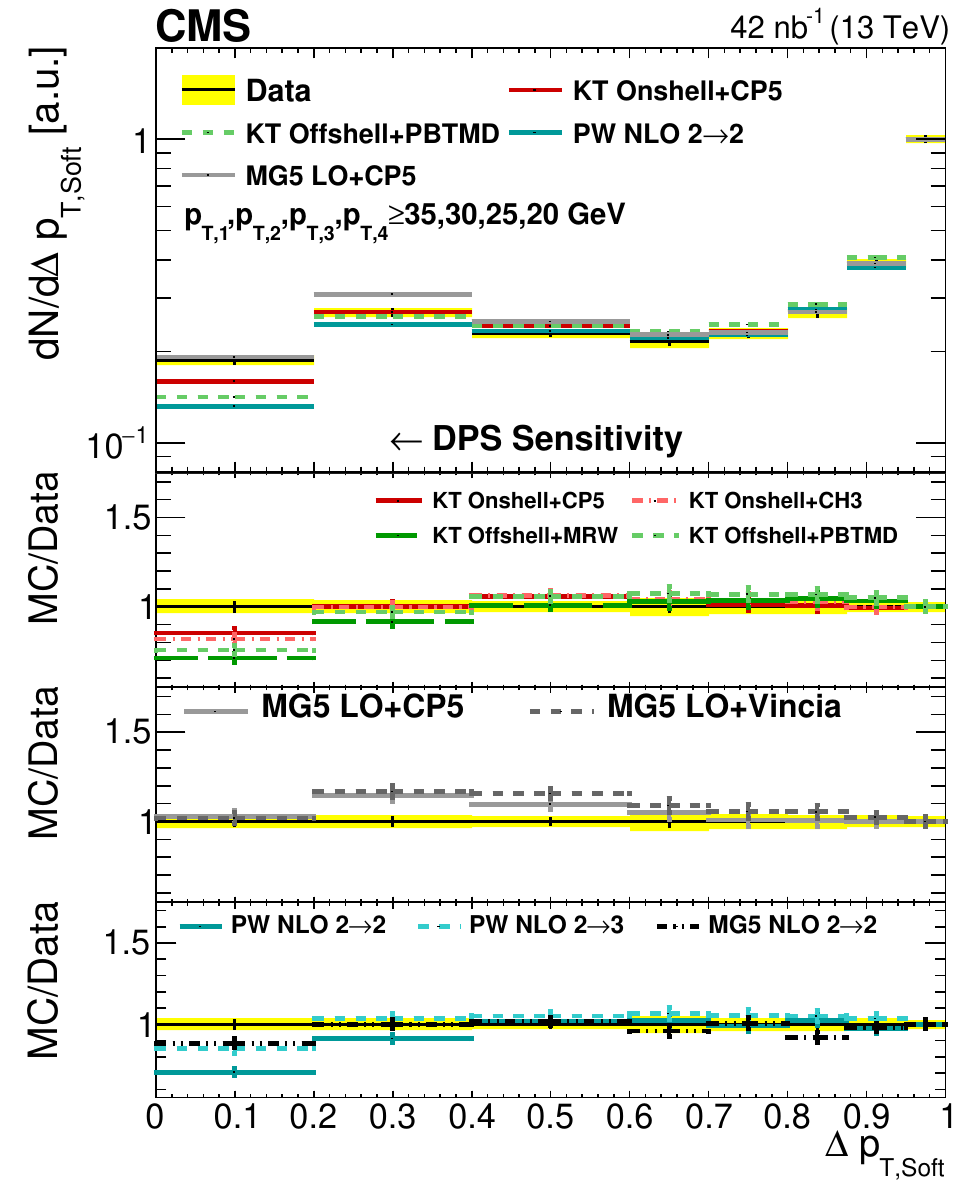}} \hspace*{\fill}
\subfloat{\includegraphics[width=0.48\textwidth]{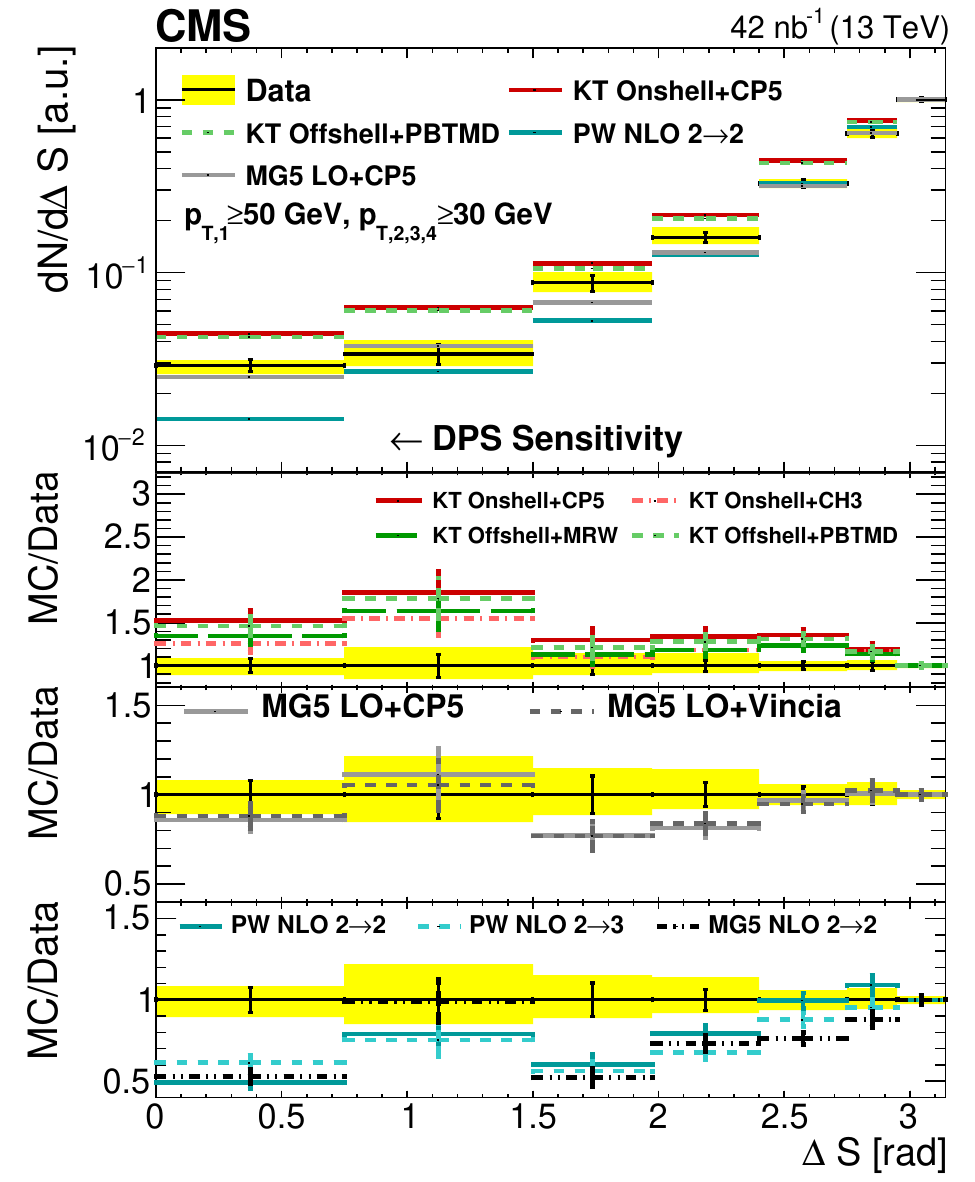}} 

\caption{Comparison of the \DptS and \DS distributions from data to different \KATIE (KT), \MGvATNLO (MG5), and \POWHEG (PW) implementations in \regioni and \regionii, respectively. All distributions have been normalized to  regions where a reduced DPS sensitivity is expected. The error bars represent the statistical uncertainty, and the yellow band indicates the total (statistical+systematic) uncertainty on the measurement.}
\label{fig:multijet_var2}
\end{figure}

Data distributions are also compared with the multijet samples that are obtained from models based on  LO $2 \to n (n\geq 2)$ and NLO \MEtwo and \MEthree matrix elements.  The group of multijet samples includes \MEfour on-shell and off-shell predictions made by \KATIE, two \MGvATNLO LO samples for which \MEmixed matrix element are all included, a \MGvATNLO NLO \MEtwo sample, and two \POWHEG NLO samples that use a \MEtwo and a \MEthree matrix element. Table~\ref{tab:MultiJet} gives a complete overview of all models, tunes, and PDF sets, along with their respective cross sections. In the figures, the labels KT, PW, and MG5 are used for \KATIE, \POWHEG, and \MGvATNLO, respectively. Figures~\ref{fig:multijet_pt}--~\ref{fig:multijet_var2} show a comparison of the data to various MC models as a function of \PT, $\eta$, and the DPS-sensitive observables. Four models are shown in direct comparison with the data. These models include one of the two on- and off-shell \KATIE models, the \MGvATNLO LO sample interfaced with the CP5 tune and the \POWHEG NLO \MEtwo sample, while all of the models are represented in the ratio plots.

\begin{table}[htp!]
\centering
\topcaption{Cross sections obtained from data and from \KATIE, \MGvATNLO, and \POWHEG in region \regioni and \regionii of the phase space, where ME stands for matrix element.}
\label{tab:MultiJet}
\resizebox{\textwidth}{!}{
\begin{tabular}{p{40mm}lp{20mm}lll}
Sample & ME & Tune & PDF/TMD & \sigmaI (\unit{$\mu$b}) & \sigmaII (\unit{$\mu$b})	\\ \hline

Data & \NA & \NA & \NA & \multicolumn{1}{l}{$2.77\pm 0.02\,^{+0.68}_{-0.55}$}	& \multicolumn{1}{l}{$0.61 \pm 0.01\,^{+0.12}_{-0.10}$}	\\ [\cmsTabSkip]

\KATIE on-shell + \PYTHIA{}8  & LO \MEfour 	& {CP5} & {NNPDF2.3\_NNLO}	& \multicolumn{1}{l}{4.23} 	& \multicolumn{1}{l}{2.87}	\\ 

\KATIE on-shell + \HERWIG{}7 & LO \MEfour 		& {CH3} & {NNPDF2.3\_NNLO}	& \multicolumn{1}{l}{3.56} 	& \multicolumn{1}{l}{2.25} \\ 

\KATIE off-shell + \CASCADE	& LO \MEfour    & \NA	    & {MRW}	& \multicolumn{1}{l}{2.40}	& \multicolumn{1}{l}{1.46}	\\ 
	
\KATIE off-shell + \CASCADE	& LO \MEfour    & \NA    & {PBTMD}& \multicolumn{1}{l}{2.57}	& \multicolumn{1}{l}{1.56}	\\ [\cmsTabSkip]

\MGvATNLO + \PYTHIA{}8 & LO \MEmixed 	& {CP5}  & {NNPDF2.3\_NNLO}	& \multicolumn{1}{l}{2.69}	& \multicolumn{1}{l}{1.26}	\\ 

\MGvATNLO + \PYTHIA{}8+\VINCIA &  LO \MEmixed 	& Default  & {NNPDF2.3\_LO}	& \multicolumn{1}{l}{1.93}	& \multicolumn{1}{l}{0.90}	\\ 

\MGvATNLO + \PYTHIA{}8 & NLO \MEtwo 	& {CP5}  & {NNPDF2.3\_NNLO} 	& \multicolumn{1}{l}{2.12}	& \multicolumn{1}{l}{1.03}	\\  [\cmsTabSkip]

\POWHEG + \PYTHIA{}8 & NLO \MEtwo 	& {CP5}  & {NNPDF2.3\_NNLO}	& \multicolumn{1}{l}{3.50}	& \multicolumn{1}{l}{1.62}	\\ 

\POWHEG + \PYTHIA{}8 & NLO \MEthree 	& {CP5}  & {NNPDF2.3\_NNLO}	& \multicolumn{1}{l}{2.55}	& \multicolumn{1}{l}{1.22}	\\

\end{tabular}
}	
\end{table}

The predicted cross sections obtained from the on-shell \KATIE samples interfaced with \PYTHIA{}8 and \HERWIG{}7 are larger than the cross sections obtained from data. They sharply decrease when an off-shell matrix element is used,  showing agreement within the data uncertainty for \regioni. The \MGvATNLO LO samples and all the NLO samples predict cross sections that are roughly in agreement with the cross sections obtained from data in \regioni, but are larger than those from data in \regionii, as are all the \KATIE cross sections in the same region.

Fig.~\ref{fig:multijet_pt} compares the \PT spectra of the various models with the data.
The on-shell \KATIE predictions agree with the data in the first bin of each of the \PT spectra, but are above the data at higher \PT. 
This may be explained because most jets originate from the \MEfour matrix element and not from the parton shower. 
The same effect, but less pronounced, is observed for the off-shell \KATIE curves. 
The different PDF sets used with off-shell \KATIE result in small variations. A better description of the \PT spectra is given by the \MGvATNLO LO sample, with a \PT-ordered parton shower.  In this case, some jets must originate from the parton shower, yielding a softer spectrum.  The combination of the \MGvATNLO LO \MEmixed sample with the dipole-antenna showering from \PYTHIA{}8+\VINCIA, results in a lowering of the total cross section.  All NLO models give a similar description as the \MGvATNLO sample; the higher-order matrix element, including virtual corrections, contributes to a lower cross section. A comparison of the multi jet data samples to the standard \PYTHIA{}8 and \HERWIG curves from the previous section demonstrates that NLO corrections and the inclusion of multi-leg matrix elements improve the description of the \PT spectra.

The $\eta$ spectra are shown in Fig.~\ref{fig:multijet_rap}. The central region is consistently described by all models, even the on-shell \KATIE models; the overall cross section is too large but the ratio remains flat for $\vert \eta \vert \leq 3.0$. An excess of jets is observed in the forward region, although this is less pronounced than in the case of the \PYTHIA{}8 and \HERWIG models. The excess is also strongest for the leading jet and diminishes for the second, third and fourth leading jet, yielding a good description of the shape for the latter.

Differential cross sections for all other observables are shown in Figs.~\ref{fig:multijet_var1} and~\ref{fig:multijet_var2}.  As before, these distributions have been normalized to a region with a much reduced DPS contribution.

The distributions in \DphiS and \DptS demonstrate that most multijet models leave room for an additional DPS contribution in the regions where this can be expected. The exceptions are the \MGvATNLO LO \MEmixed distributions that describe the shape of the distributions from data reasonably well.  This last model contains a LO \MEtwo contribution, and, as observed in Sec 8.1, the LO \MEtwo models describe the \DphiS and \DptS distributions well.

The \DY distributions show that some multijet models still have an excess of jets at large rapidity separation.  

The distributions in \phiij confirm the strong dependence on the parton shower implementation observed with the LO \MEtwo models (as shown, \eg, the \MGvATNLO LO \MEmixed predictions for \phiij).  Angular-ordered parton showers, reproduced by \VINCIA and \KATIE, describe the shape of the data distributions better.  The off-shell \KATIE predictions show a depletion in the region where additional DPS is expected.  The effect of the parton shower on the \Dphimin distributions is less pronounced in the multijet models compared to the LO \MEtwo models.

The \DS distributions again show a more robust behavior with respect to the parton shower implementation. The LO \MGvATNLO samples leave less room for an additional DPS contribution compared with the NLO models. Due to the sole use of a \MEfour matrix element, the \PT spectra are far too hard, resulting in a large overestimation of the slope for all the \KATIE models.

\subsection{SPS+DPS Models}
\label{subsec:totaldpsmodels}

The last group of model predictions that are compared with data are those including an explicit DPS contribution, as described in Section~\ref{sec:total_samples}. The \PYTHIA{}8 {CDPSTP8S1-4j} predictions presented in this section are the same as in previous sections; there is no need to add a DPS contribution explicitly,  because the underlying event parameters are specifically tuned to include a DPS contribution.  All models are labeled as before in the figures. 

A comparison of the integrated cross sections obtained from the SPS+DPS samples  in Table~\ref{tab:Total} with those from pure SPS samples, shows the expected increase. Only the values of the off-shell \KATIE models agree with the measured cross section in \regioni within the uncertainty. The cross sections predicted for \regionii are all too high.

\begin{table}[htp!]
\centering
\topcaption{Cross sections obtained from data and from models with an explicit DPS contribution in \regioni and \regionii of the phase space, where ME stands for matrix element.}
\label{tab:Total}
\resizebox{\textwidth}{!}{
\begin{tabular}{p{40mm}lllll}
Sample & ME & Tune & PDF/TMD & \sigmaI (\unit{$\mu$b}) & \sigmaII (\unit{$\mu$b})	\\ \hline

Data & \NA & \NA & \NA & \multicolumn{1}{l}{$2.77\pm 0.02\,^{+0.68}_{-0.55}$}	& \multicolumn{1}{l}{$0.61 \pm 0.01\,^{+0.12}_{-0.10}$}	\\ [\cmsTabSkip]

\PYTHIA{}8  & LO \MEtwo	& {CDPSTP8S1-4j}	& {CTEQ6L1} & \multicolumn{1}{l}{7.06} 	& \multicolumn{1}{l}{1.28}	  \\

SPS+DPS \PYTHIA{}8 & LO \MEtwo & {CP5}	& {NNPDF2.3\_NNLO}	& \multicolumn{1}{l}{4.76}	& \multicolumn{1}{l}{0.94}	\\ [\cmsTabSkip]

SPS+DPS \KATIE on-shell + \PYTHIA{}8 & LO \MEfour	& {CP5}	& {NNPDF2.3\_NNLO}	& \multicolumn{1}{l}{5.04} 	& \multicolumn{1}{l}{2.14}	\\ 

SPS+DPS \KATIE off-shell + \CASCADE & LO \MEfour & \NA	& {MRW}		& \multicolumn{1}{l}{3.11}	& \multicolumn{1}{l}{0.95}	\\ 

SPS+DPS \KATIE off-shell + \CASCADE & LO \MEfour & \NA	& {PBTMD}	& \multicolumn{1}{l}{3.12}	& \multicolumn{1}{l}{0.99}	\\

\end{tabular}
}	
\end{table}

\begin{figure}[htp!] 
\centering
\subfloat{\includegraphics[width=0.48\textwidth]{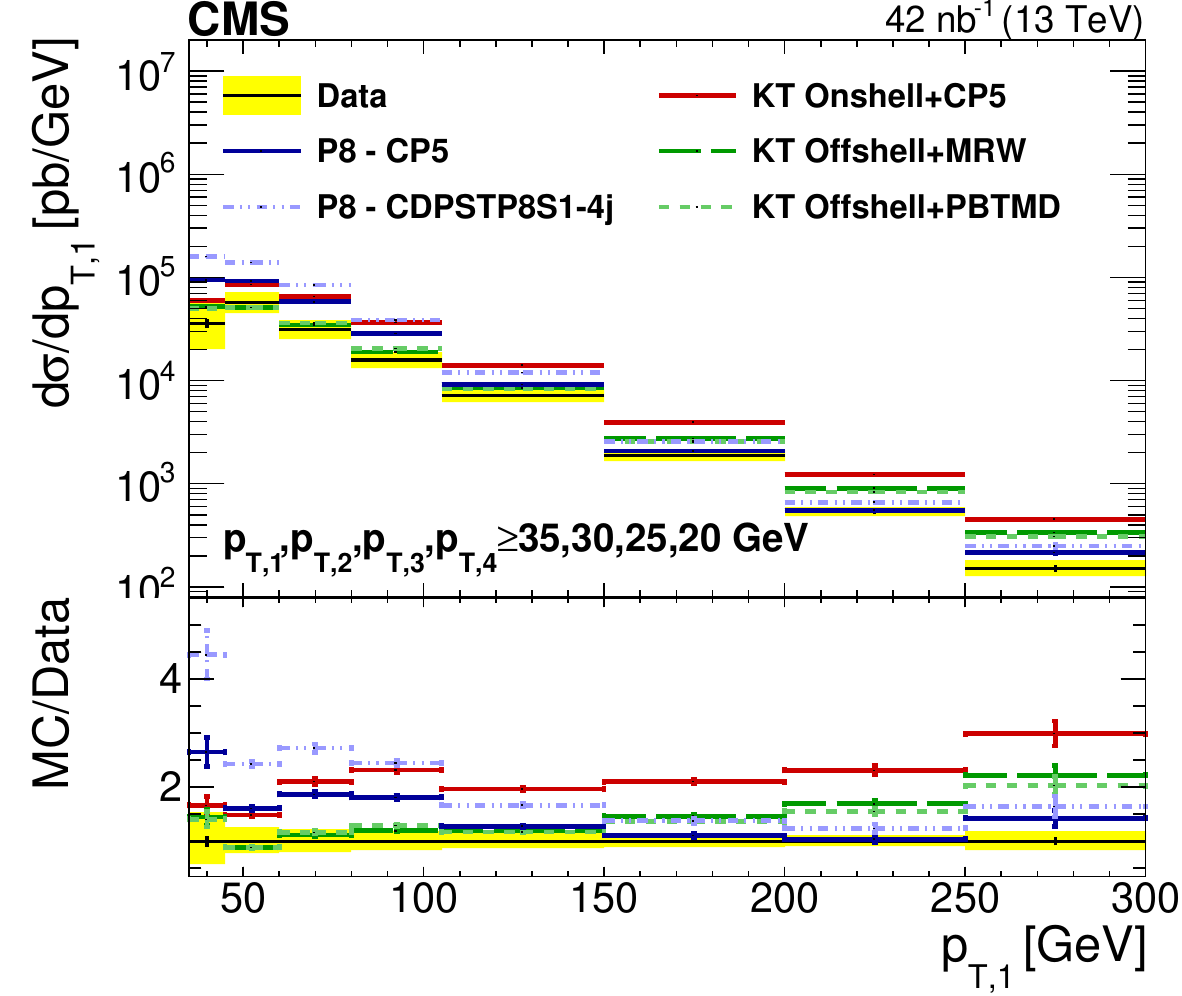}} \hspace*{\fill}
\subfloat{\includegraphics[width=0.48\textwidth]{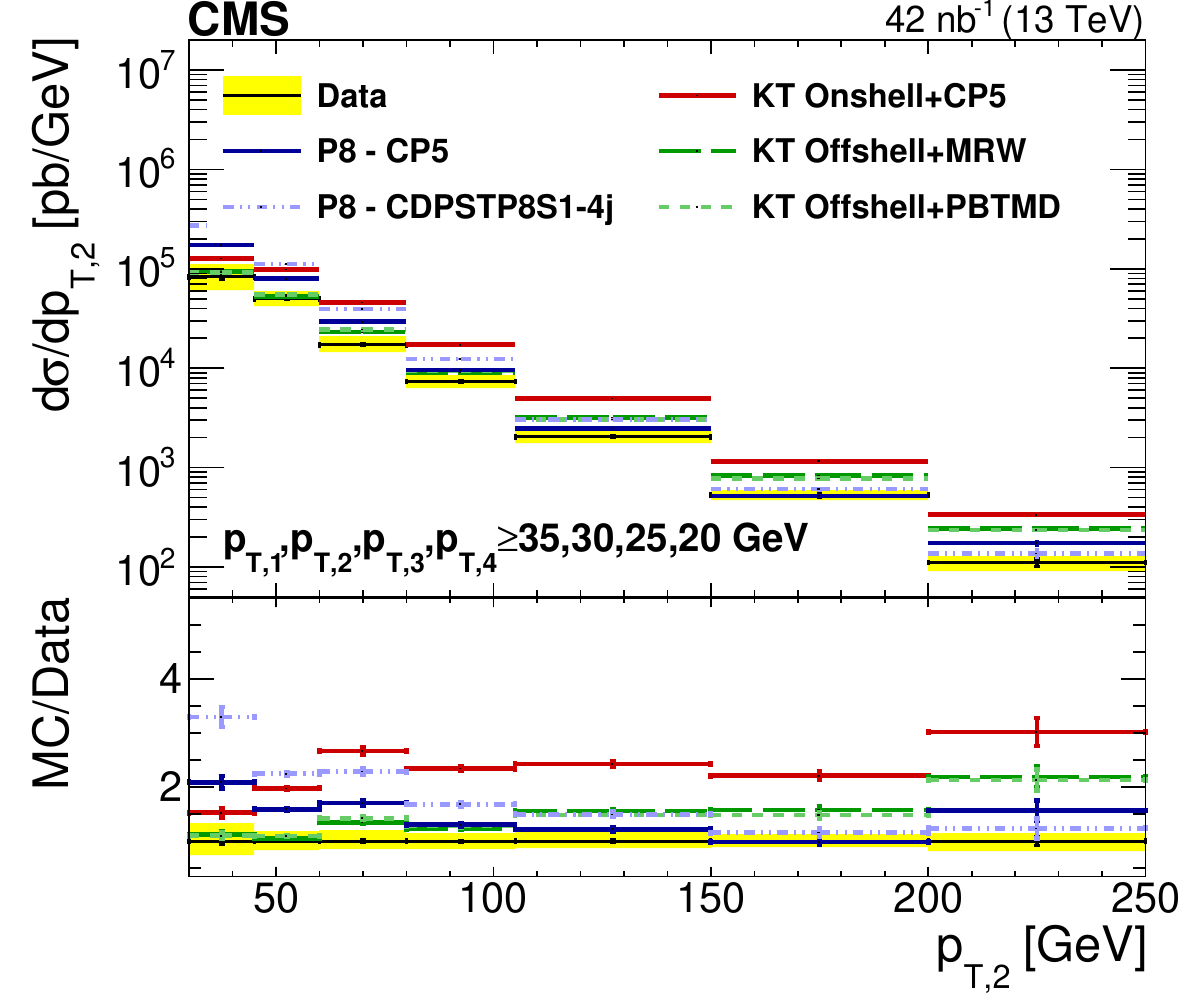}} 

\subfloat{\includegraphics[width=0.48\textwidth]{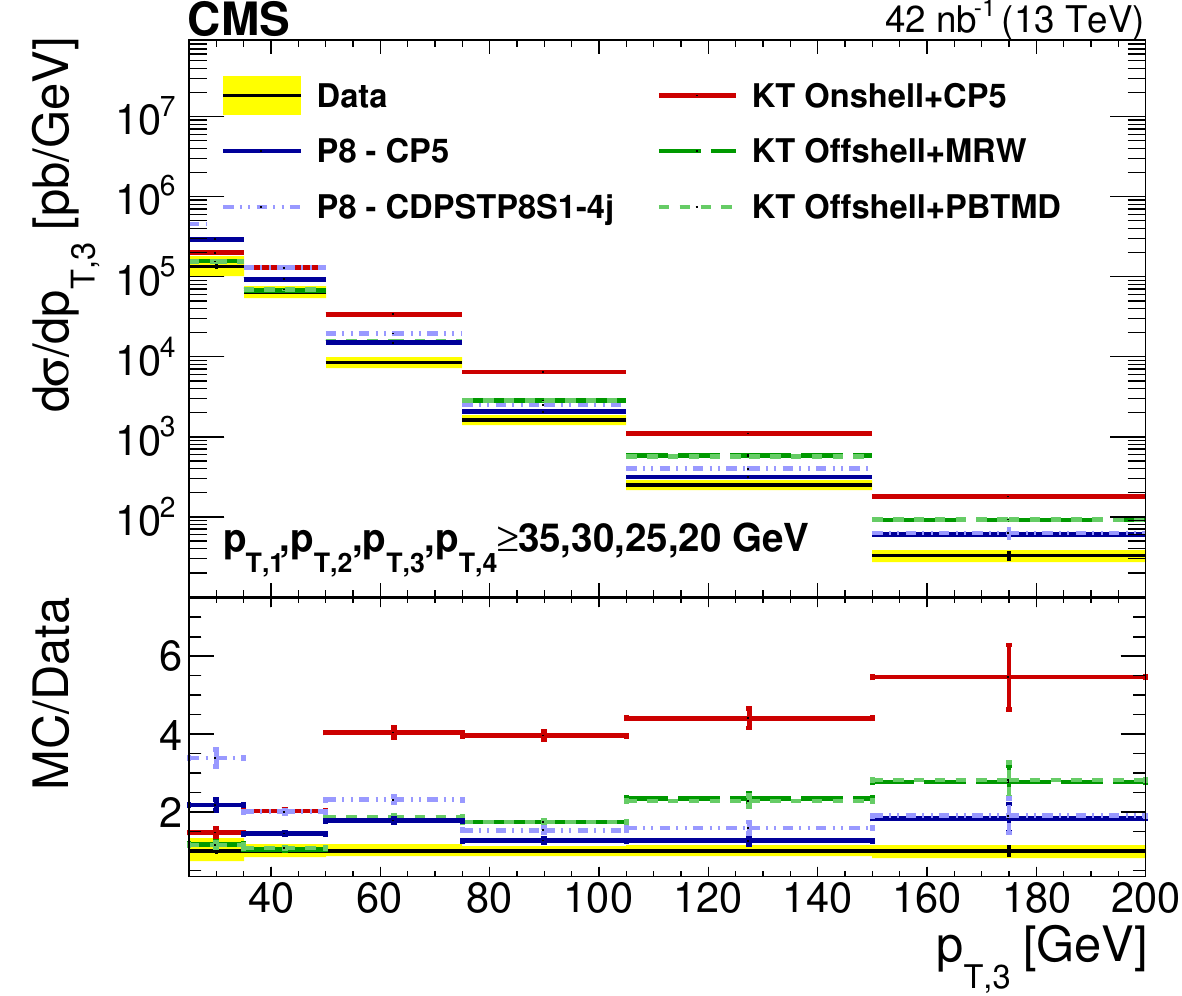}} \hspace*{\fill}
\subfloat{\includegraphics[width=0.48\textwidth]{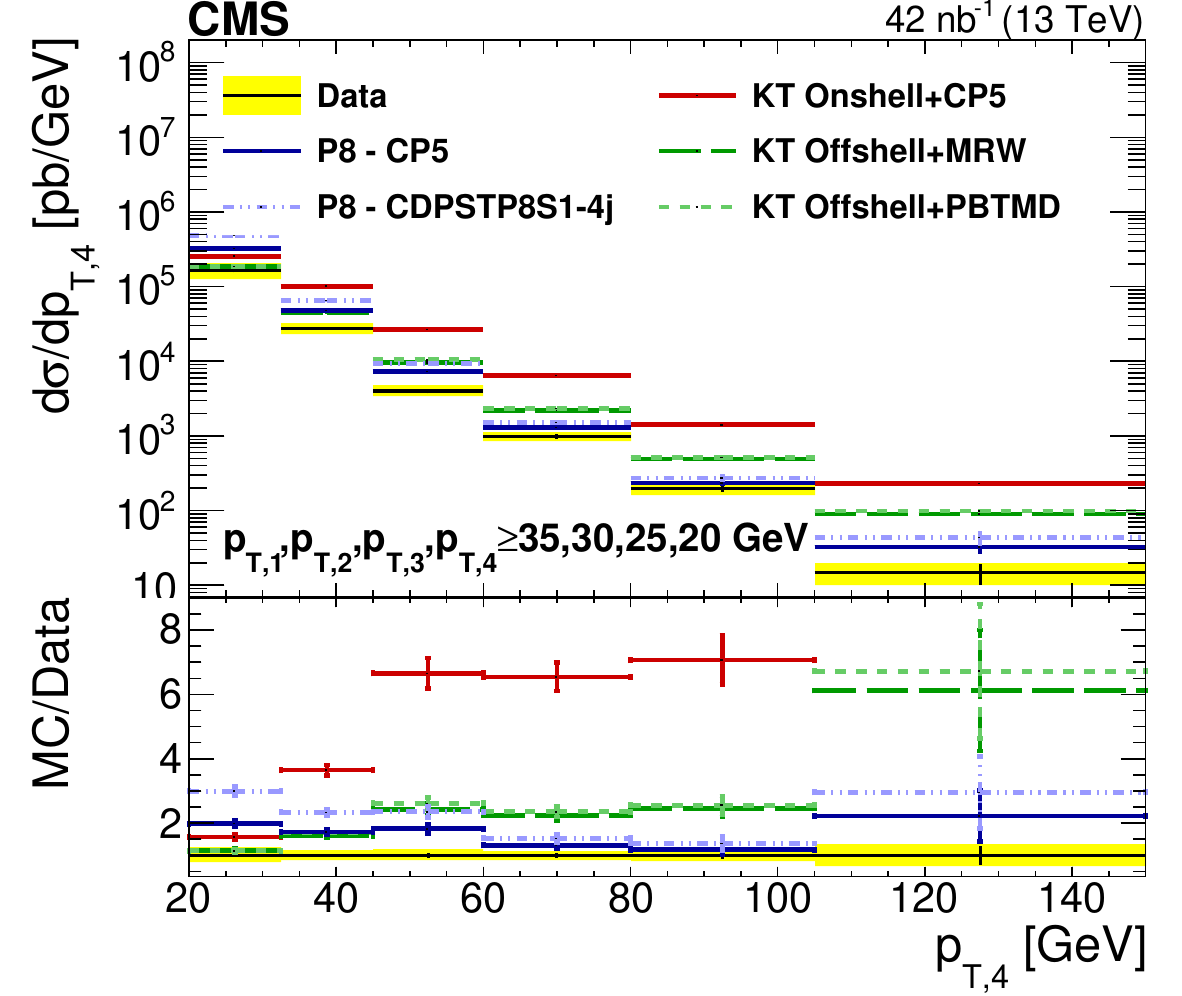}} 

\caption{Comparison of the unfolded \PT spectra of data with different SPS+DPS \KATIE (KT) and \PYTHIA{}8 (P8) models, for the leading (upper left), subleading (upper right), third leading (lower left), and fourth leading (lower right) jet in \regioni. The error bars represent the statistical uncertainty, and the yellow band indicates the total (statistical+systematic) uncertainty in the measurement.}
\label{fig:total_pt}
\end{figure}

\begin{figure}[htp!] 
\centering
\subfloat{\includegraphics[width=0.48\textwidth]{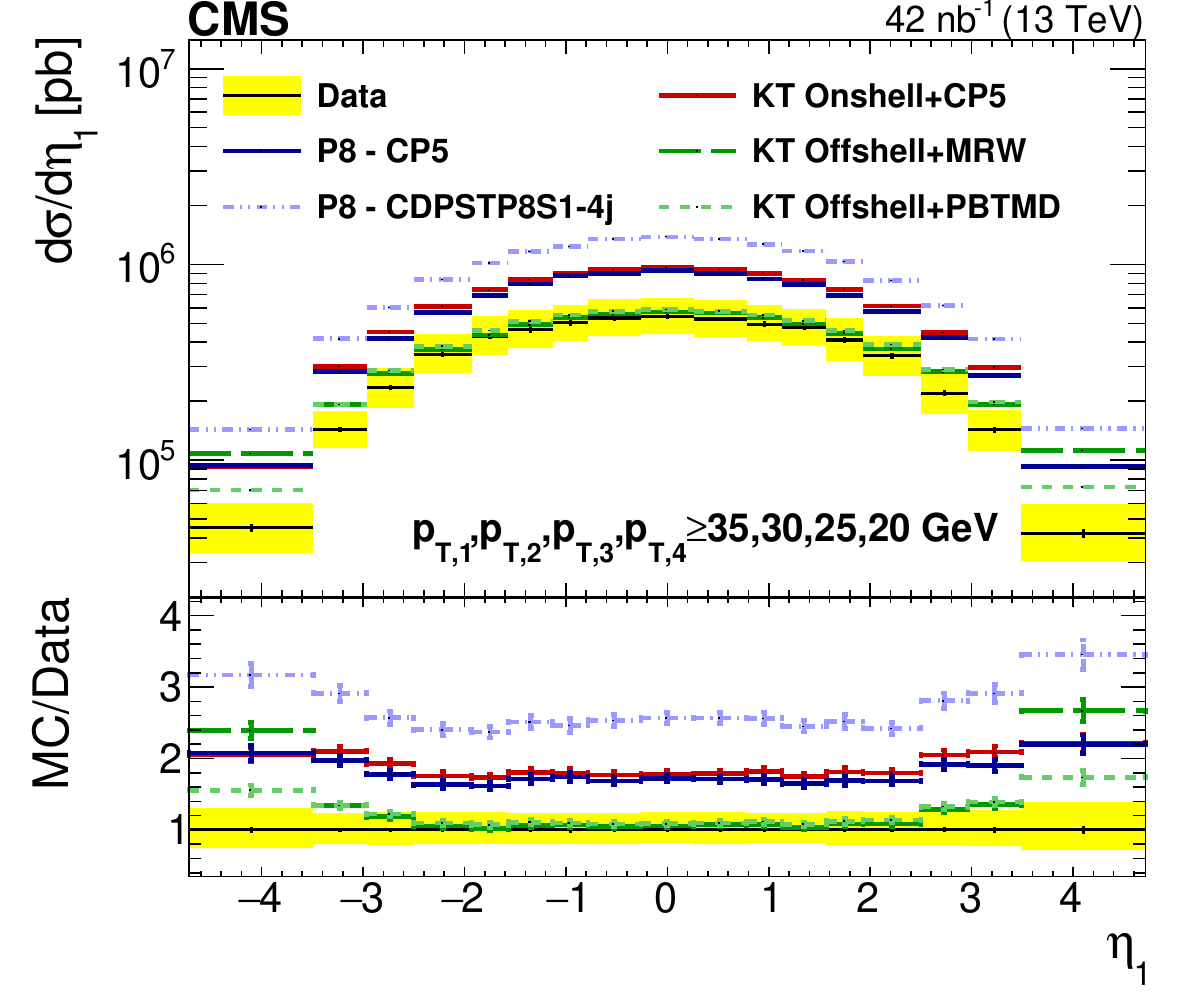}} \hspace*{\fill}
\subfloat{\includegraphics[width=0.48\textwidth]{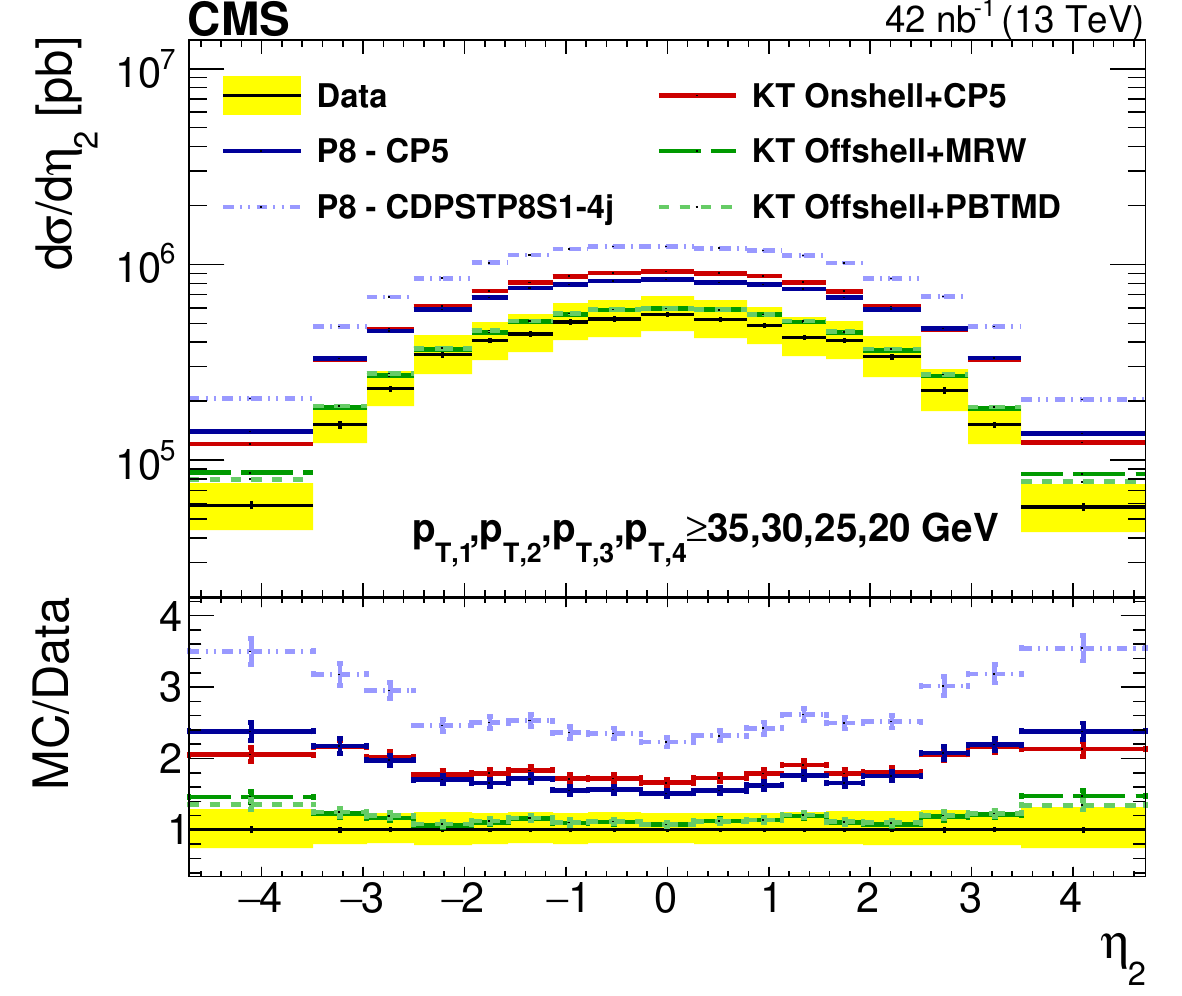}} 

\subfloat{\includegraphics[width=0.48\textwidth]{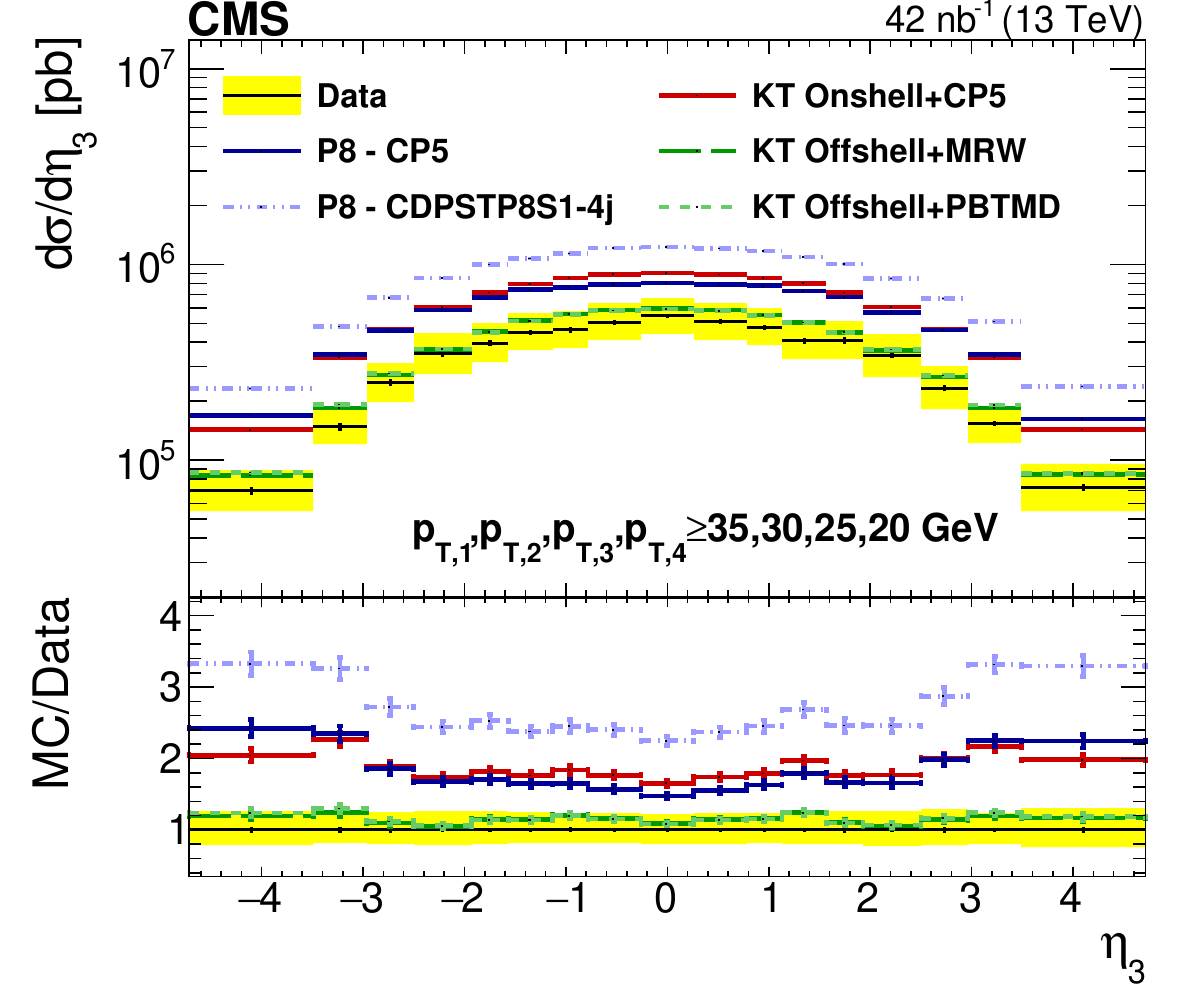}} \hspace*{\fill}
\subfloat{\includegraphics[width=0.48\textwidth]{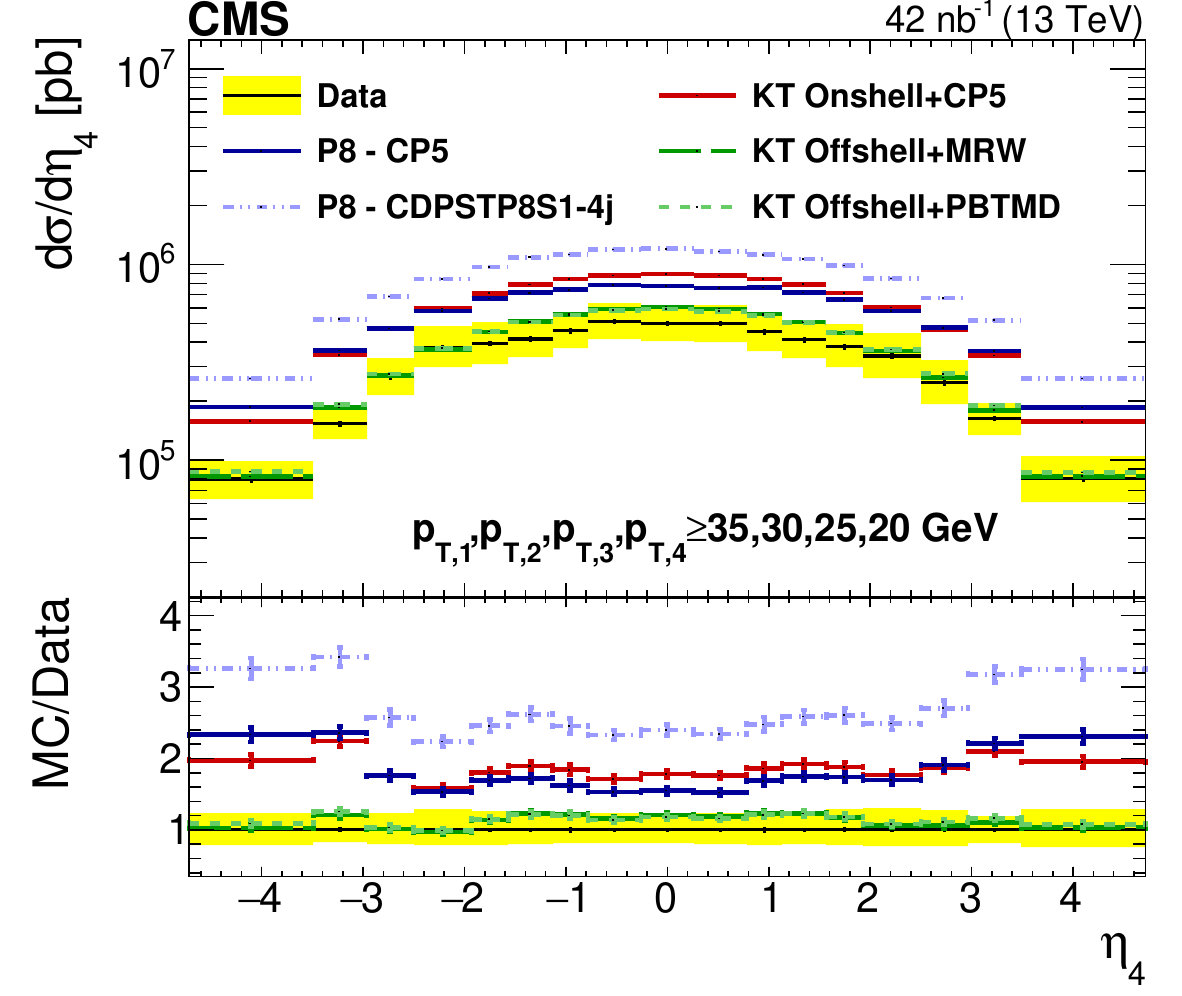}}

\caption{Comparison of the unfolded $\eta$ spectra of data with different SPS+DPS \KATIE (KT) and \PYTHIA{}8 (P8) models, for the leading (upper left), subleading (upper right), third leading (lower left), and fourth leading (lower right) jet in \regioni. The error bars represent the statistical uncertainty, and the yellow band indicates the total (statistical+systematic) uncertainty in the measurement.}
\label{fig:total_rap}
\end{figure}

Figures~\ref{fig:total_pt} and~\ref{fig:total_rap} show the \PT and $\eta$ spectra of the four leading jets, respectively. A comparison with the spectra of the pure SPS samples from the previous sections shows that the DPS samples contribute in the low-\PT region, and mostly in the forward regions of the $\eta$ spectra.	

\begin{figure}[htp!] 
\centering
\subfloat{\includegraphics[width=0.48\textwidth]{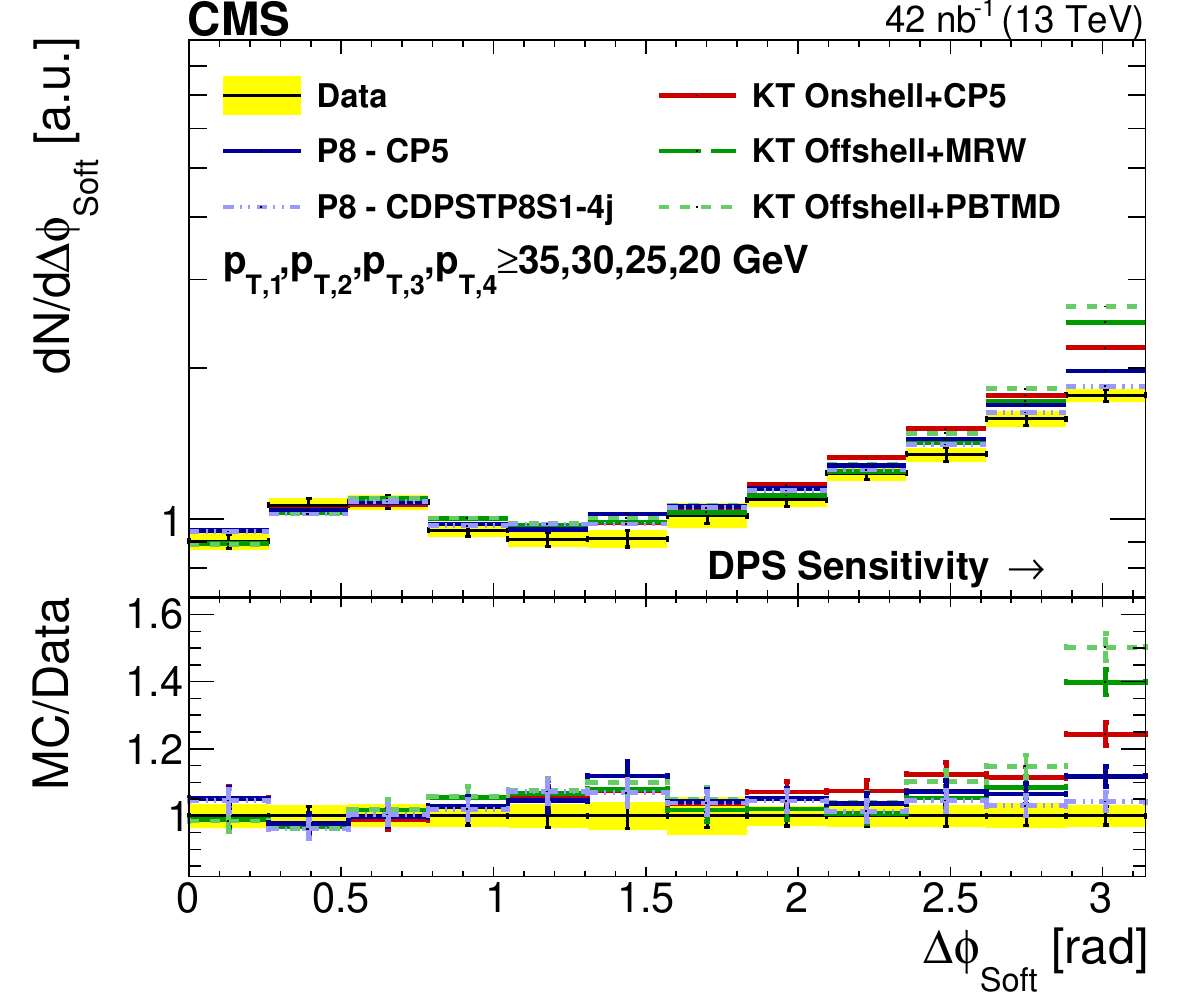}} \hspace*{\fill}
\subfloat{\includegraphics[width=0.48\textwidth]{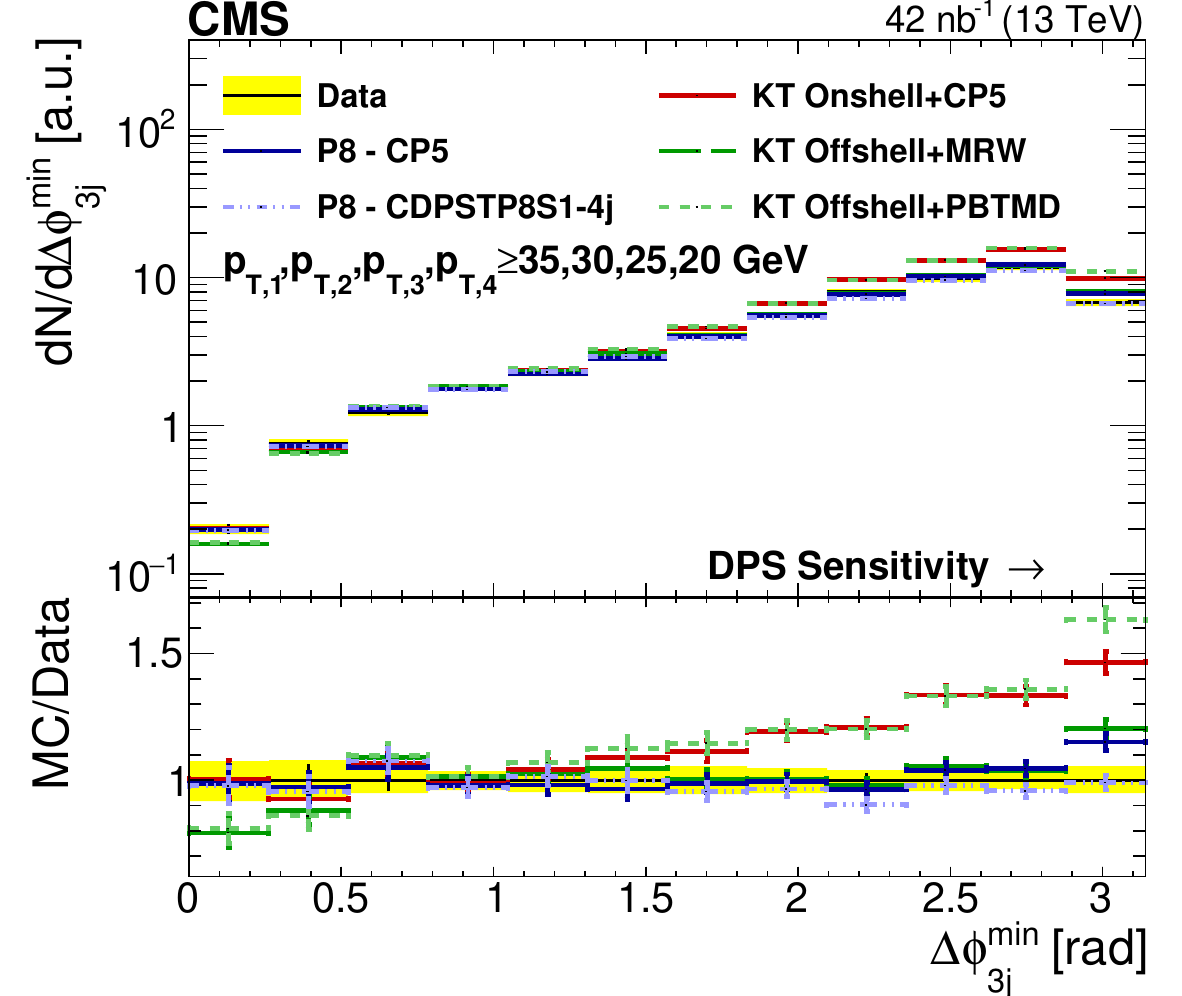}} 

\subfloat{\includegraphics[width=0.48\textwidth]{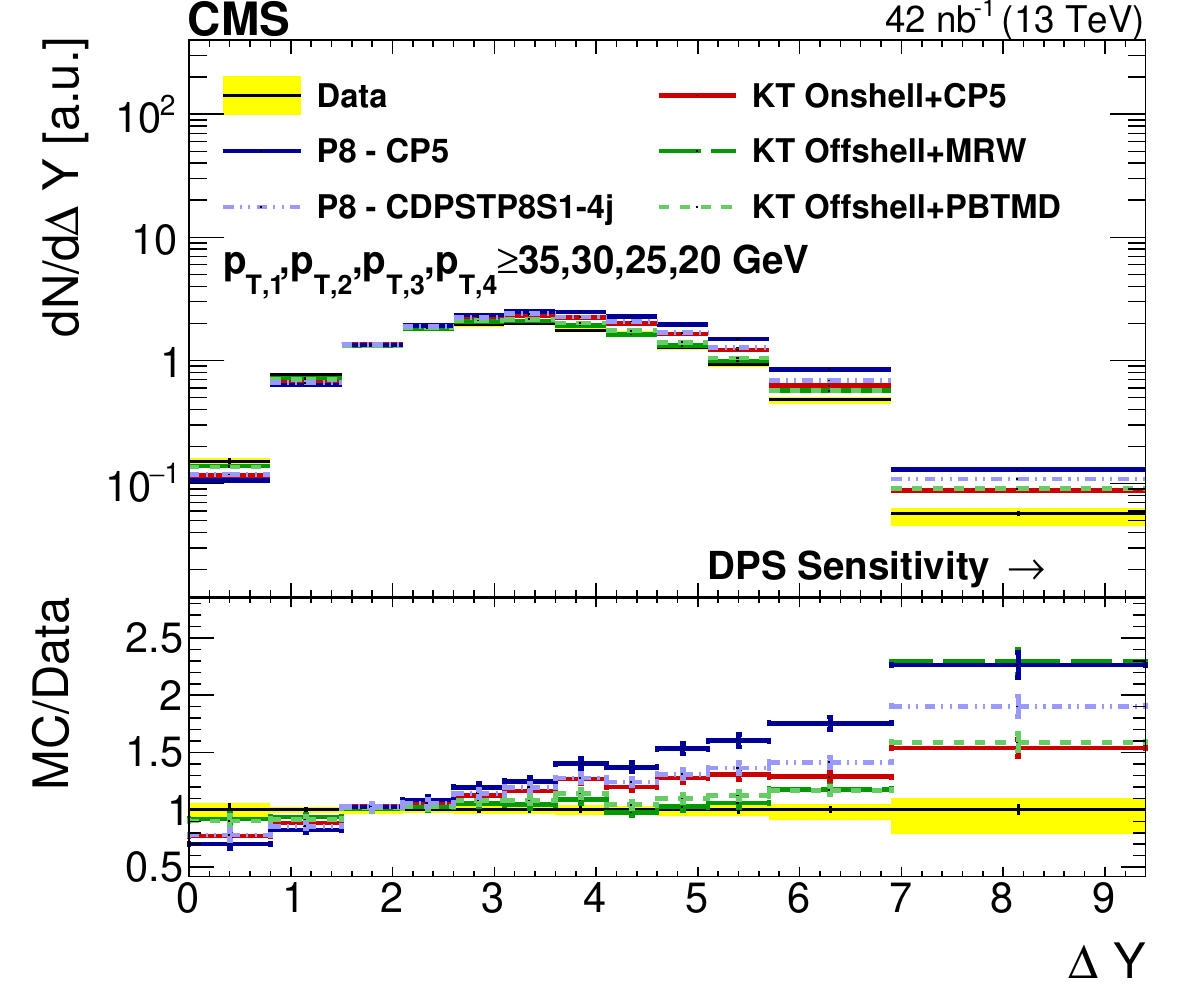}} \hspace*{\fill}
\subfloat{\includegraphics[width=0.48\textwidth]{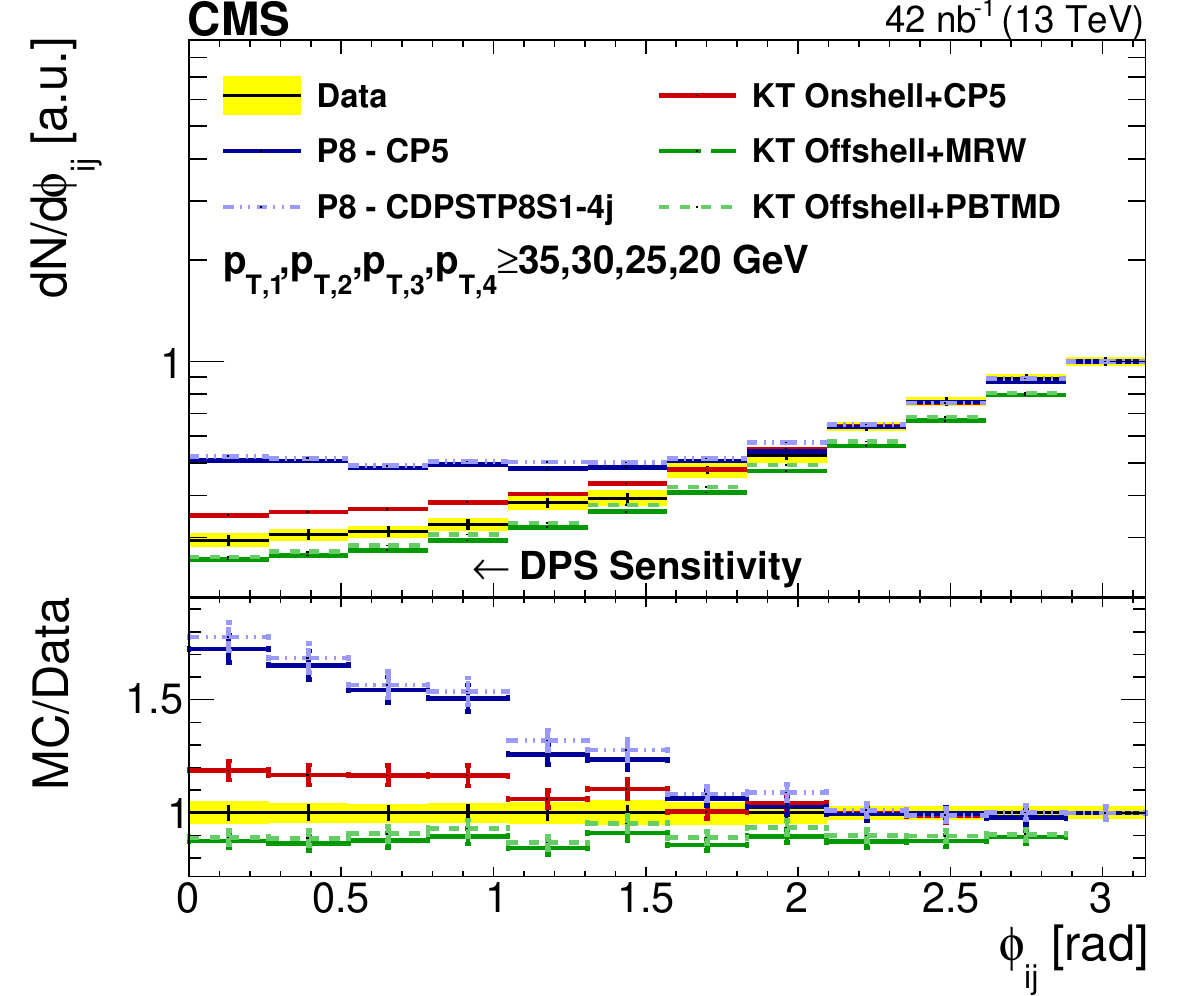}}

\subfloat{\includegraphics[width=0.48\textwidth]{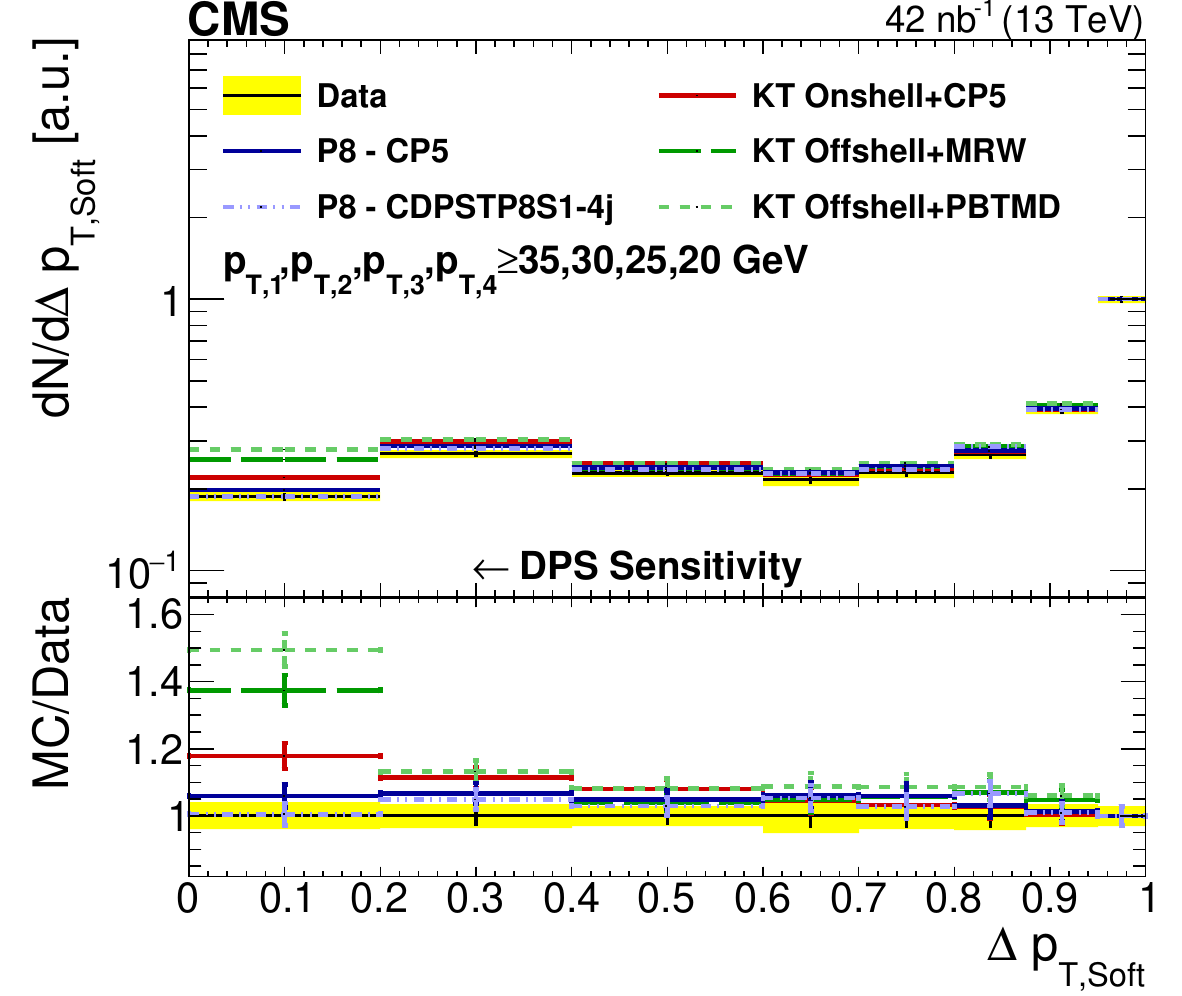}} \hspace*{\fill}
\subfloat{\includegraphics[width=0.48\textwidth]{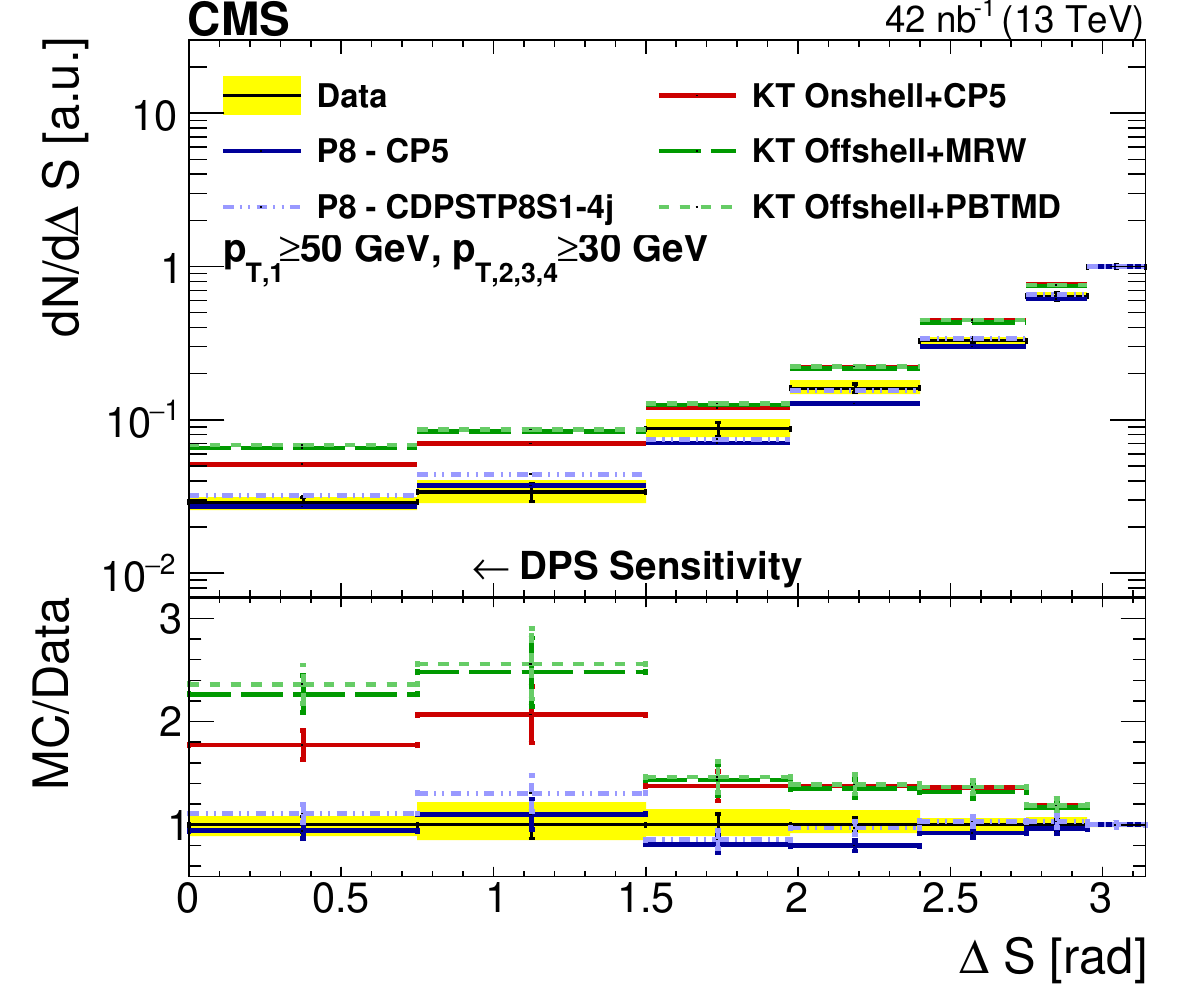}}

\caption{Comparison of the distributions in DPS-sensitive observables obtained from data to different SPS+DPS \KATIE (KT) and \PYTHIA{}8 (P8) models. All distributions have been determined in \regioni, except for the \DS distribution which has been measured in \regionii. All distributions have been normalized to the region where a reduced DPS contribution is expected. The error bars represent the statistical uncertainty, and the yellow band indicates the total (statistical+systematic) uncertainty on the measurement.}
\label{fig:total_var1}
\end{figure}

Normalized distributions in the DPS-sensitive observables are shown in Fig.~\ref{fig:total_var1}.   Small differences in shape, typically $\sim5$\%, occur in the DPS-sensitive regions when comparing the SPS+DPS \PYTHIA{}8 sample, interfaced with the {CP5} tune, with its pure SPS counterpart in Figs.~\ref{fig:PHtunes_var1} and~\ref{fig:PHtunes_var2}. The \PYTHIA{}8 samples give a good description of shape of the distributions in the observables \DphiS, \Dphimin, \DptS and \DS, with the {CDPSTP8S1-4j} tune performing slightly better than the \PYTHIA{}8 SPS+DPS sample with the {CP5} tune. This might be expected since the {CDPSTP8S1-4j} tune was obtained by fitting the \PYTHIA{}8 underlying event parameters to distributions in \DphiS, \DptS, and \DS measured with data at a center-of-mass energy of 7\TeV, as detailed in~\cite{Khachatryan:2015pea}.
 
The \KATIE curves show a more noticeable increase where a DPS contribution is expected, especially in the case of the off-shell samples. This is in contrast with the performance of the SPS-only predictions (see Figs.~\ref{fig:multijet_var1} and~\ref{fig:multijet_var2}) where the off-shell \KATIE models underestimate the data. This shows that a DPS contribution is needed in the \KATIE model to improve the description of the data, but that the value of $\sigmaeff = 21.3$\unit{mb} (taken from Ref.~\cite{Khachatryan:2015pea}), is too small for this model based on a \MEfour matrix element.
			
All models fail to describe the shape of the \DY observable, and the \phiij observable shows the same tendencies found in the previous sections. The LO \MEtwo \PYTHIA{}8 models using a \PT-ordered parton shower show too large a decorrelation, whereas predictions using higher-order matrix element calculations perform better, demonstrated by the \KATIE distributions.

\subsection{Extraction of the effective cross section}

As demonstrated in the above sections, the \DS observable exhibits the most robust sensitivity to DPS.  Other observables are either less sensitive to DPS or suffer from large variations induced by different parton shower models.  Therefore, the \DS distribution is used in the extraction of \sigmaeff. The DPS cross section is determined and the effective cross section, \sigmaeff, is extracted using different SPS MC event samples with and without the hard MPI removed following the template method laid out in Section~\ref{sec:extraction}. The \PYTHIA{}8 sample with the \text{CUETP8M1} tune yields a DPS fraction lower compared to all other tunes, while the \HERWIGpp sample with the {CUETHS1} tune gives similar results as when \HERWIG{}7 with the {CH3} tune is used, and therefore both are omitted below. The \PYTHIA{}8 sample interfaced with the {CDPSTP8S1-4j} tune without the hard MPI removed already contains a DPS contribution, scaled with an effective cross section equal to 21.3\unit{mb}~\cite{Khachatryan:2015pea}. Therefore, the extracted DPS fraction for this model is expected to be close to zero.  The extraction is still performed as a consistency check and to test the performance of the tune at a center-of-mass energy of $\sqrt{s}=13\TeV$. The multijet \KATIE models are not considered; they significantly overshoot the DPS-sensitive slope of the \DS observable and using them would result in a negative DPS contribution.

\begin{figure}[h!] 
\centering
\includegraphics[width=0.70\textwidth]{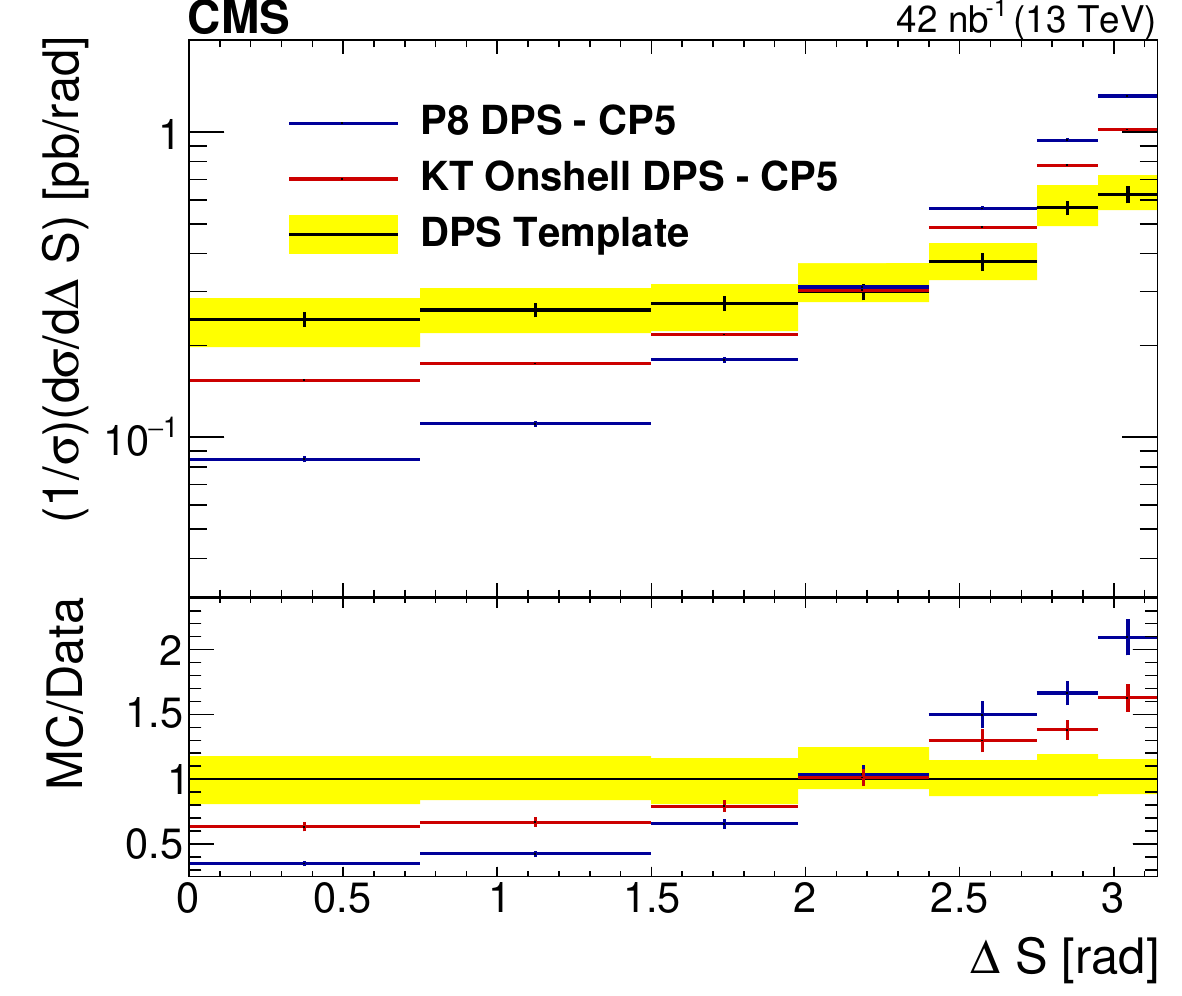}
\caption{The \DS distribution obtained from the mixed data sample compared to predictions from the pure DPS sample in \PYTHIA{}8 (P8) and \KATIE (KT). The distributions are normalized to unity. The error bars represent the statistical uncertainty, and the yellow band indicates the total (statistical+systematic) uncertainty on the data.}
\label{fig:DSDPS}
\end{figure}

The mixed event sample obtained from data, as described in Section\@~\ref{sec:extraction}, is used to get the \DS distributions for pure DPS events and is corrected by means of unfolding in the same way as the other distributions.   This corrected \DS distribution is shown in Fig.~\ref{fig:DSDPS}, along with the \DS distributions obtained from the DPS component of the \PYTHIA{}8 and on-shell \KATIE generators, both interfaced with the {CP5} tune. All distributions have been normalized to unity. The distributions show a maximum at $\DS = \pi$ because of jet pairs with overlapping \PT values such that the two softest and the two hardest jets do not coincide with the jet pairs from separate events or parton collisions.
The DPS data sample exhibits a larger decorrelation as compared to the DPS MC samples, which can be attributed to disparities in the \PT spectra observed in data and for MC events.

\begin{figure}[h!] 
\centering
\includegraphics[width=0.70\textwidth]{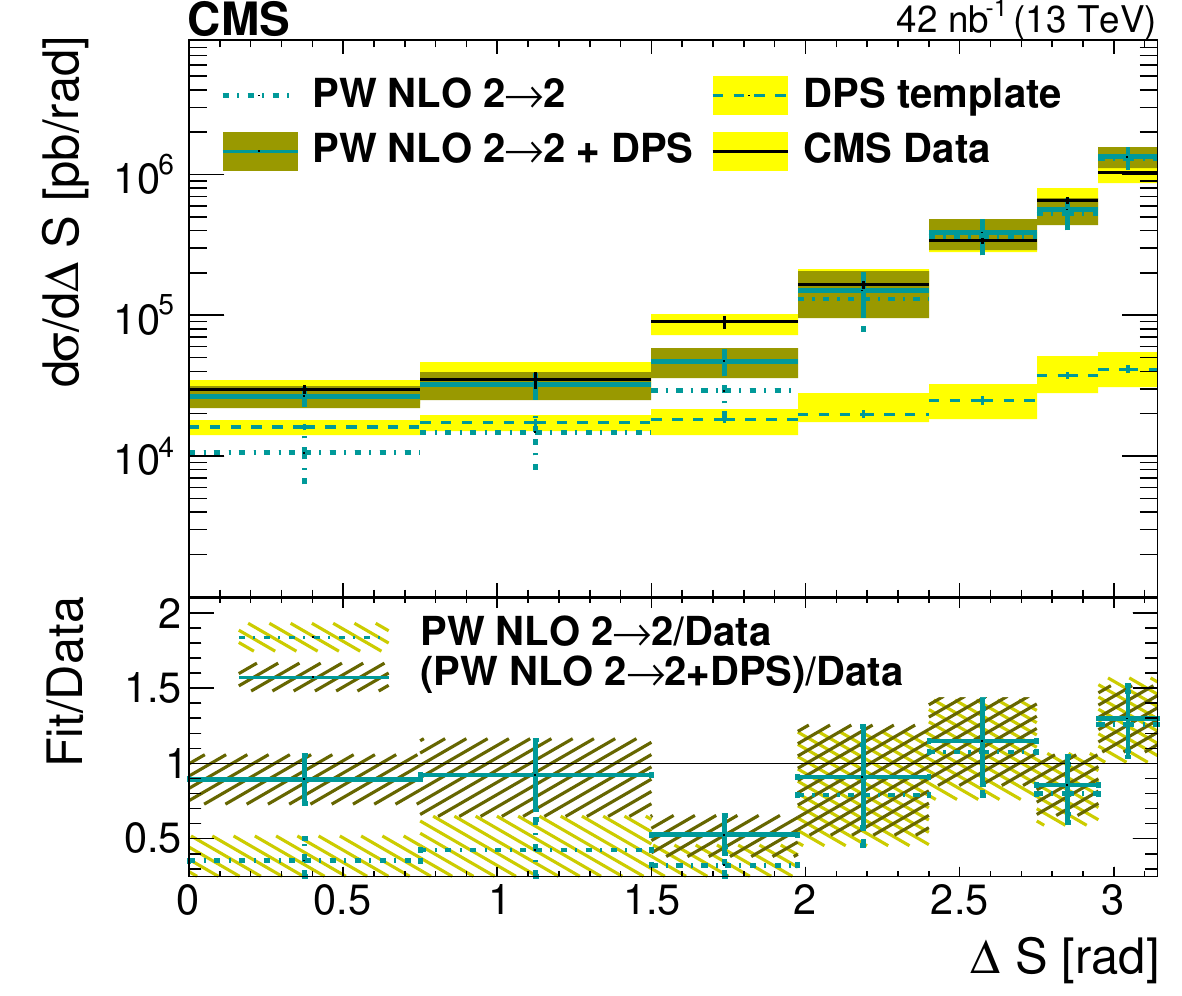}

\caption{The results of the template fit for the \POWHEG (PW) NLO \MEtwo model without the hard MPI removed. The yellow bands represent the total uncertainty of the distribution. The ratio of the scaled MC model and of the total fitted result over the data are shown in the bottom plot. Since the \DS distribution obtained from the mixed data sample carries a statistical and systematic uncertainty, so does the total fitted sample. The total uncertainty in the ratio is shown on the plot.}
\label{fig:multijet_temp}
\end{figure}

The template fitter program~\cite{Brun:1997pa} takes the SPS MC distributions along with the DPS data sample as input, and uses the template method to determine the fraction of DPS events, \fDPS. The results for the extracted values of \fDPS are shown in Table~\ref{tab:fdps_temp}. Equations \eqref{eq:pocketformula_rewritten} and \eqref{eq:sigmaDPS_template} are used to determine the DPS cross sections and values for \sigmaeff shown in Tables~\ref{tab:sigmadps_temp} and~\ref{tab:sigmaeff_temp}, respectively. The results for both sets of samples (with and without the hard MPI removed) are shown, along with the net difference between both cross sections since this can be interpreted as the amount of DPS inherent to the tune. An example of the template fit is presented in Fig.~\ref{fig:multijet_temp}, where the fitted distributions using the \POWHEG NLO \MEtwo model without the hard MPI removed along with its statistical and systematic uncertainties are shown.

The DPS cross section obtained for all models with and without the hard MPI removed range from 14.6 to 70\unit{mb}, yielding values for \sigmaeff between 7.7 and 34.8\unit{mb}. The exception is \PYTHIA{}8 with the {CDPSTP8S1-4j} tune without the hard MPI removed where an excess of DPS events is found, resulting in a negative DPS cross section. Therefore, the effective cross section for this model has not been calculated because it would yield a nonphysical result. 

In the case of the LO \MEtwo models, 2 final state jets originate from the ME, and 2 additional jets stem from the parton shower. A distinction between two groups of models using a LO \MEtwo matrix element can be made. On the one hand, the \PYTHIA{}8 and \HERWIG{}7 models using the {CP5} and {CH3} tunes, respectively, use the latest underlying event tune and up-to-date PDFs. On the other hand, both \PYTHIA{}8+\VINCIA and \HERWIG{}7 with the {SoftTune} rely on an older underlying event tune and older PDFs. The extracted values for the DPS cross section for the former two models are 38.5 and 38.2\unit{mb}, whereas for the latter two models values of 23.3 and 29.5\unit{mb} were obtained where events containing one or more hard partons originating from MPI have been removed. The results indicate that it might be the different underlying event tunes and the usage of older PDFs that are responsible for the more DPS-like topology in both \PYTHIA{}8+\VINCIA and \HERWIG{}7 with the {SoftTune}. 

Introducing higher multiplicity MEs reduces the effect of the underlying event tune, parton showers, and the PDFs since more jets will originate from the ME.  The dominating contribution to the event sample comes from \MEfour ME, where the 4 final state jets all stem from the ME.  The extracted DPS cross section with hard MPI removed is 31\unit{mb} for both \PYTHIA{}8 with the {CP5} tune and \PYTHIA{}8+\VINCIA, confirming that the effect of the different underlying event tunes, parton showers, and PDFs is suppressed.   In the case of \PYTHIA{}8 with the {CP5} tune, the higher multiplicity MEs reduce the need for additional DPS, whereas in the case of \PYTHIA{}8+\VINCIA more room for additional DPS exists.  For both models, about half of the DPS cross section that is needed to describe the data can be covered by the MPI that are intrinsic to the tune.

The NLO \MEtwo models contain a combination of exclusive NLO \MEtwo and LO \MEthree ME.  One or more parton-shower jets are required to produce a 4-jet final state, with a ME/parton-shower matching algorithm applied to avoid double counting.  Large DPS cross sections of up to 70\unit{mb} are needed to describe the data if events containing hard MPI partons originating from the underlying event description, provided by the {CP5} tune, are removed. Compared with the other models, the NLO \MEtwo models are outliers.  If events with hard MPI partons from the underlying event description are included, they can again account for about half of the DPS cross section needed to describe the data.

For the \POWHEG NLO \MEthree sample, NLO \MEthree and LO \MEfour MEs are combined, leaving room for zero or one additional jets from the parton shower.  The extracted DPS cross section approaches the one obtained for the models with a LO \MEmixed ME.  Contrary to the LO \MEmixed ME models, however, the NLO \MEthree models do not allow events containing hard MPI partons to contribute as much to the DPS cross section.
 
\begin{figure}[h!] 
\centering
\includegraphics[width=0.70\textwidth]{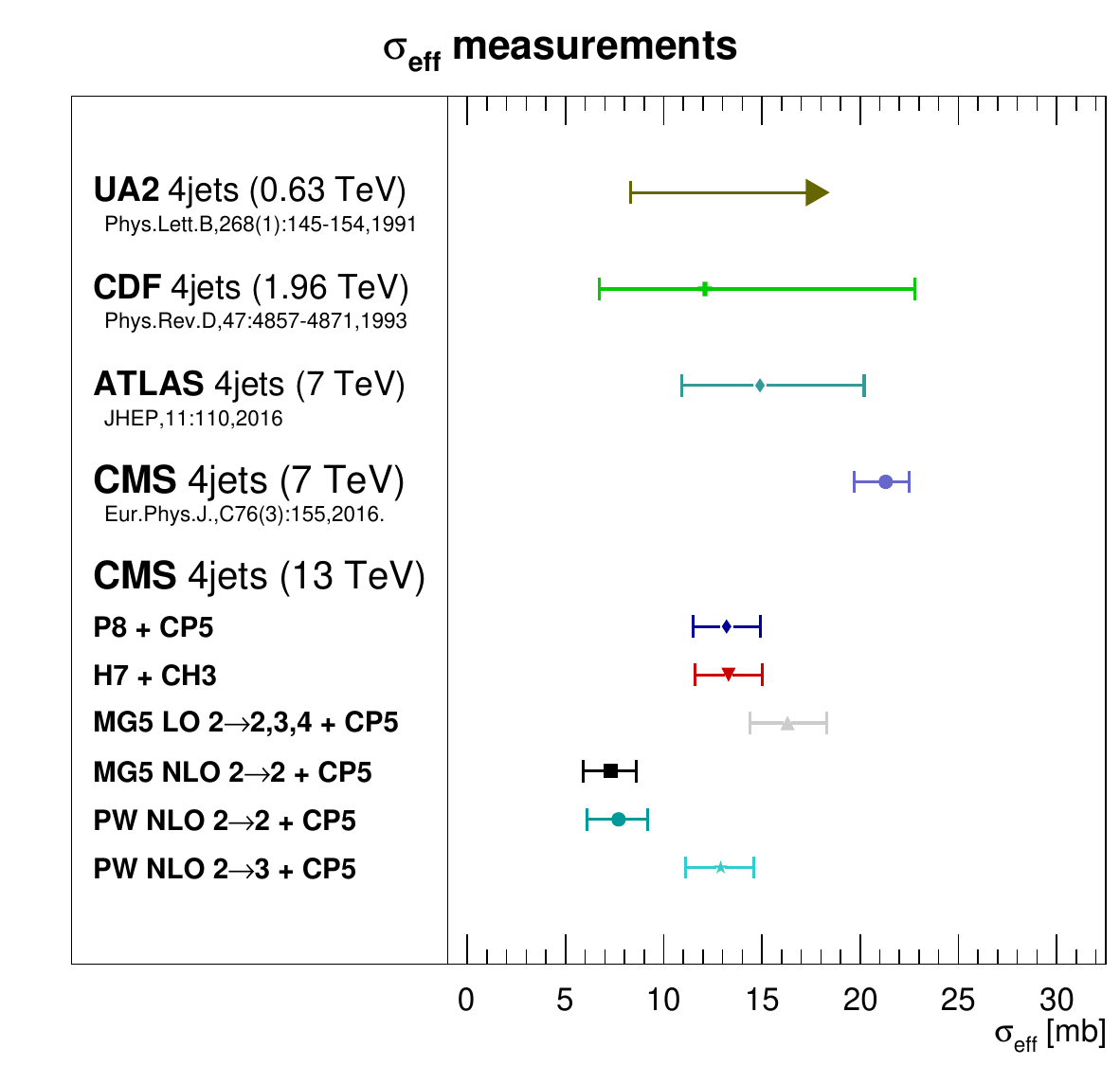}

\caption{Comparison of the values for \sigmaeff extracted from data using different SPS models where events that have generated one or more hard MPI partons with $\PT^\text{parton} \geq 20$\GeV, have been removed. The results from four-jet measurements performed at lower center-of-mass energies \cite{ALITTI1991145,PhysRevD.56.3811,Aaboud:2016dea,Khachatryan:2015pea} are shown alongside the newly extracted values. The error bars in each of the values of \sigmaeff represent the total (statistical+systematic) uncertainties.}
\label{fig:sigmaeff_overview}
\end{figure}

Fig.~\ref{fig:sigmaeff_overview} shows the results for \sigmaeff extracted with the models that are based on the recent CP5 and CH3 tunes where the hard MPI have been removed.
The extracted \sigmaeff values show agreement with the results from the UA2 and CDF experiments, which set a lower bound on $\sigmaeff >  8.3$\unit{mb}~\cite{ALITTI1991145} and found a \sigmaeff value of $12.1^{+10.7}_{-5.4}$\unit{mb}, respectively. All results, except for the values obtained with the NLO \MEtwo models, agree with the measurement performed by the ATLAS collaboration~\cite{Aaboud:2016dea} at a center-of-mass energy of 7\TeV~\cite{Abe:1993rv}, where a \sigmaeff equal to $14.9^{+1.2}_{-1.0}\mathrm{(stat)}^{+5.1}_{-3.8}\mathrm{(syst)}$\unit{mb} was reported, whereas none agree with the value of $21.3^{+1.2}_{-1.6}$\unit{mb} from the CMS measurement at a center-of-mass energy of 7\TeV~\cite{Khachatryan:2015pea}, which is more in line with the results obtained with some of the models based on older UE tunes.

Two other DPS measurements have been performed at a center-of-mass energy $\sqrt{s}=13\TeV$. A value of $\sigma_\mathrm{eff}$ equal to $12.7^{+5.0}_{-2.9}$\unit{mb} has been extracted from the same-sign WW measurement in Ref.~\cite{Sirunyan:2019zox}. A $\sigma_\mathrm{eff}$ of $7.3 \pm 0.5\,(\text{stat}) \pm 1.0\,(\text{syst})$\unit{mb} has been obtained from the \cPJgy pair production measurement in Ref.~\cite{Aaij:2016bqq}. 
It has been shown that \sigmaeff is expected to be process independent for inclusive final states~\cite{Seymour:2013sya}, therefore, it is noteworthy that the result the \cPJgy meson pair production measurement shows only agreement with the NLO \MEtwo models. The extracted \sigmaeff of the same-sign WW measurement shows agreement with all models due to the size of the errors in the measurement, indicating that further measurements in this channel are desirable before any further conclusions can be made.

\begin{table}[htp!]
\centering
\caption{The values of the DPS fraction $\fDPS$ extracted from data using different SPS models, along with their statistical and systematic uncertainties. The results are shown for the model where the full tune is used, and for the same models where the hard MPI have been removed. The last column shows the net difference between the two first columns, and is interpreted as the fraction of DPS inherent to the tune.}
\label{tab:fdps_temp}
\resizebox{\textwidth}{!}{
\begin{tabular}{p{40mm}p{28mm}lll} 
MC Model & Tune & \multicolumn{3}{c}{$\fDPS\pm(\text{stat})\pm\text{(syst)}$ (\unit{\%})}  \\
\NA & \NA & Full tune & Hard MPI removed & Inherent DPS \\ \hline
\PYTHIA{}8 										& {CP5}					& \multicolumn{1}{r}{$3.77\pm0.08\,^{+0.45}_{-0.68}$}	& \multicolumn{1}{r}{$6.34\pm0.07\,^{+0.32}_{-0.57}$} 	& \multicolumn{1}{r}{$2.57\pm0.11\,^{+0.36}_{-0.62}$}	\\  	   							
\PYTHIA{}8+\VINCIA								& Standard \PYTHIA{}8.3	& \multicolumn{1}{r}{$2.40\pm0.07\,^{+0.41}_{-0.68}$}	& \multicolumn{1}{r}{$3.84\pm0.07\,^{+0.34}_{-0.63}$}	& \multicolumn{1}{r}{$1.44\pm0.10\,^{+0.37}_{-0.65}$}		\\  
\PYTHIA{}8										& {CDPSTP8S1-4j}		& \multicolumn{1}{r}{$-1.30\pm0.08\,^{+0.39}_{-0.69}$} 	& \multicolumn{1}{r}{$3.06\pm0.07\,^{+0.28}_{-0.62}$} 	& \multicolumn{1}{r}{$4.36\pm0.11\,^{+0.34}_{-0.66}$} 	\\
\HERWIG{}7 										& {CH3}					& \multicolumn{1}{r}{$3.72\pm0.07\,^{+0.38}_{-0.68}$} 	& \multicolumn{1}{r}{$6.28\pm0.08\,^{+0.29}_{-0.58}$}	& \multicolumn{1}{r}{$2.56\pm0.11\,^{+0.33}_{-0.63}$}	\\   
\HERWIG{}7										& {SoftTune}			& \multicolumn{1}{r}{$2.67\pm0.07\,^{+0.42}_{-0.71}$}	& \multicolumn{1}{r}{$4.85\pm0.08\,^{+0.31}_{-0.52}$} 	& \multicolumn{1}{r}{$2.18\pm0.11\,^{+0.36}_{-0.61}$} \\ [\cmsTabSkip]
\MGvATNLO LO \MEmixed, \PYTHIA{}8 				& {CP5}	        		& \multicolumn{1}{r}{$2.50\pm0.08\,^{+0.38}_{-0.69}$} 	& \multicolumn{1}{r}{$5.14\pm0.08\,^{+0.30}_{-0.56}$}	& \multicolumn{1}{r}{$2.64\pm0.11\,^{+0.35}_{-0.62}$}	\\ 
\MGvATNLO LO \MEmixed, \PYTHIA{}8+\VINCIA 		& Standard \PYTHIA{}8.3	& \multicolumn{1}{r}{$2.55\pm0.09\,^{+0.38}_{-0.66}$}	& \multicolumn{1}{r}{$5.23\pm0.08\,^{+0.27}_{-0.53}$}	& \multicolumn{1}{r}{$2.68\pm0.12\,^{+0.33}_{-0.60}$}	\\ 
\MGvATNLO NLO \MEtwo, \PYTHIA{}8 				& {CP5} 				& \multicolumn{1}{r}{$7.13\pm0.08\,^{+0.28}_{-0.42}$} 	& \multicolumn{1}{r}{$11.45\pm0.08\,^{+0.22}_{-0.27}$}	& \multicolumn{1}{r}{$4.32\pm0.11\,^{+0.25}_{-0.36}$}	\\ 
\POWHEG NLO \MEtwo, \PYTHIA{}8 					& {CP5}					& \multicolumn{1}{r}{$4.77\pm0.08\,^{+0.32}_{-0.64}$}	& \multicolumn{1}{r}{$10.89\pm0.08\,^{+0.24}_{-0.48}$}	& \multicolumn{1}{r}{$6.12\pm0.11\,^{+0.28}_{-0.53}$}	\\
\POWHEG NLO \MEthree, \PYTHIA{}8 				& {CP5}					& \multicolumn{1}{r}{$5.40\pm0.07\,^{+0.36}_{-0.67}$}	& \multicolumn{1}{r}{$6.51\pm0.07\,^{+0.29}_{-0.51}$}	& \multicolumn{1}{r}{$1.11\pm0.10\,^{+0.33}_{-0.59}$}	\\   
\end{tabular}
}
\end{table}

\begin{table}[htp!]
\centering
\caption{The values of the DPS cross section $\sigmaDPS$ extracted from data using different SPS models, along with their statistical and systematic uncertainties. The results are shown for the model where the full tune is used, and for the same models where the hard MPI have been removed. The last column shows the net difference between the two first columns, and is interpreted as the amount of DPS inherent to the tune.}
\label{tab:sigmadps_temp}
\resizebox{\textwidth}{!}{
\begin{tabular}{p{40mm}p{28mm}lll} 
MC Model & Tune & \multicolumn{3}{c}{$\sigmaDPS\pm(\text{stat})\pm\text{(syst)}$ (\unit{nb})}  \\
\NA & \NA & Full tune & Hard MPI removed & Inherent DPS \\ \hline
\PYTHIA{}8 										& {CP5}			& \multicolumn{1}{r}{$22.9\pm0.7\,^{+5.7}_{-7.3}$}	& \multicolumn{1}{r}{$38.5\pm0.9\,^{+6.5}_{-9.4}$} 	& \multicolumn{1}{r}{$15.6\pm1.1\,^{+6.1}_{-8.4}$}	\\  	   							
\PYTHIA{}8+\VINCIA								& Standard \PYTHIA{}8.3	& \multicolumn{1}{r}{$14.6\pm0.6\,^{+4.4}_{-5.9}$}	& \multicolumn{1}{r}{$23.3\pm0.7\,^{+4.9}_{-7.1}$}	& \multicolumn{1}{r}{$8.7\pm0.9\,^{+4.7}_{-6.5}$}		\\  
\PYTHIA{}8	& {CDPSTP8S1-4j}	& \multicolumn{1}{r}{$-7.9\pm0.5\,^{+2.3}_{-2.9}$} & \multicolumn{1}{r}{$18.6\pm0.6\,^{+4.6}_{-6.5}$} & \multicolumn{1}{r}{$26.5\pm0.8\,^{+3.5}_{-4.7}$} 	\\
\HERWIG{}7 										& {CH3}			& \multicolumn{1}{r}{$22.6\pm0.7\,^{+5.1}_{-7.3}$} 	& \multicolumn{1}{r}{$38.2\pm0.9\,^{+6.4}_{-9.4}$}	& \multicolumn{1}{r}{$15.6\pm1.1\,^{+5.8}_{-8.4}$}	\\   
\HERWIG{}7	& {SoftTune}	& \multicolumn{1}{r}{$16.2\pm0.6\,^{+4.6}_{-6.3}$}	& \multicolumn{1}{r}{$29.5\pm0.8\,^{+5.6}_{-8.6}$} 	& \multicolumn{1}{r}{$13.3\pm1.0\,^{+5.1}_{-7.5}$} \\ [\cmsTabSkip]
\MGvATNLO LO \MEmixed, \PYTHIA{}8 						& {CP5}	        & \multicolumn{1}{r}{$15.2\pm0.6\,^{+4.2}_{-6.1}$} 	& \multicolumn{1}{r}{$31.2\pm0.8\,^{+5.5}_{-8.1}$}	& \multicolumn{1}{r}{$16.0\pm1.0\,^{+4.9}_{-7.1}$}	\\ 
\MGvATNLO LO \MEmixed, \PYTHIA{}8+\VINCIA 				& Standard \PYTHIA{}8.3	& \multicolumn{1}{r}{$15.5\pm0.6\,^{+4.3}_{-6.0}$}	& \multicolumn{1}{r}{$31.8\pm0.8\,^{+5.3}_{-8.0}$}	& \multicolumn{1}{r}{$16.3\pm1.0\,^{+4.8}_{-7.0}$}	\\ 
\MGvATNLO NLO \MEtwo, \PYTHIA{}8 						& {CP5} 			& \multicolumn{1}{r}{$43.3\pm1.0\,^{+4.5}_{-9.4}$} 	& \multicolumn{1}{r}{$70\pm2\,^{+8}_{-13}$}	& \multicolumn{1}{r}{$26.7\pm2\,^{+6}_{-11}$}	\\ 
\POWHEG NLO \MEtwo, \PYTHIA{}8 	& {CP5}			& \multicolumn{1}{r}{$29.0\pm0.9\,^{+5.3}_{-6.1}$}	& \multicolumn{1}{r}{$66\pm3\,^{+7}_{-12}$}		& \multicolumn{1}{r}{$37.0\pm3.1\,^{+6.2}_{-9.1}$}	\\
\POWHEG NLO \MEthree, \PYTHIA{}8 				& {CP5}			& \multicolumn{1}{r}{$32.8\pm0.9\,^{+5.9}_{-8.3}$}	& \multicolumn{1}{r}{$39.5\pm1.0\,^{+6.3}_{-9.2}$}		& \multicolumn{1}{r}{$6.7\pm1.3\,^{+6.1}_{-8.8}$}	\\   
\end{tabular}
}
\end{table}

\begin{table}[ht!]
\centering
\caption{The values of the effective cross section $\sigmaeff$ extracted from data using different SPS models, along with their statistical and systematic uncertainties. The results are shown for the model where the full tune is used, and for the same models where the hard MPI have been removed}
\label{tab:sigmaeff_temp}
\begin{tabular}{p{40mm}p{28mm}ll} 
MC Model & Tune & \multicolumn{2}{c}{$\sigmaeff\pm(\text{stat})\pm\text{(syst)}$ (\unit{mb})}  \\
\NA & \NA & Full tune & Hard MPI removed \\ \hline
\PYTHIA{}8 										& {CP5}			& \multicolumn{1}{r}{$22.2\pm0.7\,^{+1.2}_{-0.8}$}	& \multicolumn{1}{r}{$13.2\pm0.3\,^{+1.6}_{-1.7}$} 		\\  	
\PYTHIA{}8+\VINCIA								& Standard \PYTHIA{}8.3	& \multicolumn{1}{r}{$34.8\pm1.3\,^{+0.9}_{-3.5}$}	& \multicolumn{1}{r}{$21.8\pm0.7\,^{+1.9}_{-1.2}$}	\\ 
\PYTHIA{}8	& {CDPSTP8S1-4j}  & \multicolumn{1}{r}{-} & \multicolumn{1}{r}{$27.3\pm0.9\,^{+1.5}_{-0.5}$} \\
\HERWIG{}7 										& {CH3}			& \multicolumn{1}{r}{$22.5\pm0.7\,^{+1.7}_{-0.7}$} 	& \multicolumn{1}{r}{$13.3\pm0.3\,^{+1.7}_{-1.7}$}	\\ 
\HERWIG{}7	& {SoftTune}	& \multicolumn{1}{r}{$31.4\pm1.1\,^{+1.1}_{-2.4}$}	& \multicolumn{1}{r}{$17.2\pm0.5\,^{+1.8}_{-1.5}$} \\  [\cmsTabSkip]
\MGvATNLO LO \MEmixed, \PYTHIA{}8 						& {CP5}	        & \multicolumn{1}{r}{$33.4\pm1.3\,^{+1.1}_{-3.1}$} 	& \multicolumn{1}{r}{$16.3\pm0.5\,^{+1.9}_{-1.8}$}		\\ 
\MGvATNLO LO \MEmixed, \PYTHIA{}8+\VINCIA 				& Standard \PYTHIA{}8.3	& \multicolumn{1}{r}{$32.8\pm1.3\,^{+1.2}_{-2.1}$}	& \multicolumn{1}{r}{$16.0\pm0.4\,^{+2.0}_{-1.9}$}		\\ 
\MGvATNLO NLO \MEtwo, \PYTHIA{}8 						& {CP5} 			& \multicolumn{1}{r}{$11.7\pm0.3\,^{+2.3}_{-1.9}$} 	& \multicolumn{1}{r}{$7.3\pm0.2\,^{+1.3}_{-1.4}$}		\\ 
\POWHEG NLO \MEtwo, \PYTHIA{}8 	& {CP5}			& \multicolumn{1}{r}{$17.5\pm0.9\,^{+1.6}_{-1.7}$}	& \multicolumn{1}{r}{$7.7\pm0.3\,^{+1.5}_{-1.6}$}		\\
\POWHEG NLO \MEthree, \PYTHIA{}8 				& {CP5}			& \multicolumn{1}{r}{$15.5\pm0.5\,^{+1.8}_{-1.8}$}	& \multicolumn{1}{r}{$12.9\pm0.3\,^{+1.7}_{-1.8}$}			\\  
\end{tabular}
\end{table}

\section{Summary}
\label{sec:summary}

A study of the inclusive production of four-jet events at low transverse momentum has been presented based on data from proton-proton collisions collected with the CMS detector at a center-of-mass energy of 13\TeV.  Various observables sensitive to double-parton scattering (DPS) are studied and values for its effective cross section have been extracted.

Models based on leading order (LO) \MEtwo matrix elements significantly overestimate the absolute four-jet cross section in the phase space domains studied in this paper.  This excess is related to an abundance of low-\PT and forward jets.  The predictions of the absolute cross section generally improve when next-to-leading order (NLO) and/or higher-multiplicity matrix elements are used.  

The azimuthal angle between the jets with the largest separation in $\eta$, \phiij, has a strong discriminating power for different parton-shower approaches and the  data favor the angular-ordered and dipole-antenna parton-shower models over those with a \pt-ordered parton shower.  The yield of jet pairs with large rapidity separation \DY is, however, overestimated by all models, although models based on NLO and/or higher-multiplicity matrix elements are closer to the data.  

The distribution of the minimal combined azimuthal angular range of three jets, \Dphimin, also exhibits sensitivity to the parton-shower implementation, with data favoring \pt-ordered parton showers with the LO \MEtwo models for this observable.  In the case of models based on NLO and/or higher-multiplicity matrix elements the comparisons are less conclusive.

Other observables, such as the azimuthal angle between the two softest jets, \DphiS, and their transverse momentum balance, \DptS, indicate the need for a DPS contribution in the models to various degrees, as confirmed by the extracted values of \sigmaeff.

The distribution of the azimuthal angle between the hard and soft jet pairs, \DS, is the least sensitive to the details of the parton-shower modeling, and it is used for the extraction of the effective cross section, \sigmaeff.

A dependence is observed in the extracted values of \sigmaeff in the model used to describe the SPS contribution. Models based on NLO \MEtwo matrix elements yield the smallest ($\sim$7\unit{mb}) values of \sigmaeff and need the largest DPS contribution. However, models using a \MEtwo matrix element along with older underlying event descriptions and older PDFs, tend to need the smallest DPS contribution. The sensitivity to the underlying event description, parton showers, and the PDFs is observed to be small when including higher-order matrix elements, since both models using the \MEmixed matrix elements show agreement with each other. 

These results demonstrate the need for further development of models to accurately describe final states with multiple jets in phase space regions with large potential DPS contributions.

\begin{acknowledgments}

	We congratulate our colleagues in the CERN accelerator departments for the excellent performance of the LHC and thank the technical and administrative staffs at CERN and at other CMS institutes for their contributions to the success of the CMS effort. In addition, we gratefully acknowledge the computing centers and personnel of the Worldwide LHC Computing Grid and other centers for delivering so effectively the computing infrastructure essential to our analyses. Finally, we acknowledge the enduring support for the construction and operation of the LHC, the CMS detector, and the supporting computing infrastructure provided by the following funding agencies: BMBWF and FWF (Austria); FNRS and FWO (Belgium); CNPq, CAPES, FAPERJ, FAPERGS, and FAPESP (Brazil); MES and BNSF (Bulgaria); CERN; CAS, MoST, and NSFC (China); MINCIENCIAS (Colombia); MSES and CSF (Croatia); RIF (Cyprus); SENESCYT (Ecuador); MoER, ERC PUT and ERDF (Estonia); Academy of Finland, MEC, and HIP (Finland); CEA and CNRS/IN2P3 (France); BMBF, DFG, and HGF (Germany); GSRI (Greece); NKFIA (Hungary); DAE and DST (India); IPM (Iran); SFI (Ireland); INFN (Italy); MSIP and NRF (Republic of Korea); MES (Latvia); LAS (Lithuania); MOE and UM (Malaysia); BUAP, CINVESTAV, CONACYT, LNS, SEP, and UASLP-FAI (Mexico); MOS (Montenegro); MBIE (New Zealand); PAEC (Pakistan); MSHE and NSC (Poland); FCT (Portugal); JINR (Dubna); MON, RosAtom, RAS, RFBR, and NRC KI (Russia); MESTD (Serbia); SEIDI, CPAN, PCTI, and FEDER (Spain); MOSTR (Sri Lanka); Swiss Funding Agencies (Switzerland); MST (Taipei); ThEPCenter, IPST, STAR, and NSTDA (Thailand); TUBITAK and TAEK (Turkey); NASU (Ukraine); STFC (United Kingdom); DOE and NSF (USA).

	\hyphenation{Rachada-pisek} Individuals have received support from the Marie-Curie program and the European Research Council and Horizon 2020 Grant, contract Nos.\ 675440, 724704, 752730, 758316, 765710, 824093, 884104, and COST Action CA16108 (European Union); the Leventis Foundation; the Alfred P.\ Sloan Foundation; the Alexander von Humboldt Foundation; the Belgian Federal Science Policy Office; the Fonds pour la Formation \`a la Recherche dans l'Industrie et dans l'Agriculture (FRIA-Belgium); the Agentschap voor Innovatie door Wetenschap en Technologie (IWT-Belgium); the F.R.S.-FNRS and FWO (Belgium) under the ``Excellence of Science -- EOS" -- be.h project n.\ 30820817; the Beijing Municipal Science \& Technology Commission, No. Z191100007219010; the Ministry of Education, Youth and Sports (MEYS) of the Czech Republic; the Deutsche Forschungsgemeinschaft (DFG), under Germany's Excellence Strategy -- EXC 2121 ``Quantum Universe" -- 390833306, and under project number 400140256 - GRK2497; the Lend\"ulet (``Momentum") Program and the J\'anos Bolyai Research Scholarship of the Hungarian Academy of Sciences, the New National Excellence Program \'UNKP, the NKFIA research grants 123842, 123959, 124845, 124850, 125105, 128713, 128786, and 129058 (Hungary); the Council of Science and Industrial Research, India; the Latvian Council of Science; the Ministry of Science and Higher Education and the National Science Center, contracts Opus 2014/15/B/ST2/03998 and 2015/19/B/ST2/02861 (Poland); the Funda\c{c}\~ao para a Ci\^encia e a Tecnologia, grant CEECIND/01334/2018 (Portugal); the National Priorities Research Program by Qatar National Research Fund; the Ministry of Science and Higher Education, project no. 14.W03.31.0026 (Russia); the Programa Estatal de Fomento de la Investigaci{\'o}n Cient{\'i}fica y T{\'e}cnica de Excelencia Mar\'{\i}a de Maeztu, grant MDM-2015-0509 and the Programa Severo Ochoa del Principado de Asturias; the Stavros Niarchos Foundation (Greece); the Rachadapisek Sompot Fund for Postdoctoral Fellowship, Chulalongkorn University and the Chulalongkorn Academic into Its 2nd Century Project Advancement Project (Thailand); the Kavli Foundation; the Nvidia Corporation; the SuperMicro Corporation; the Welch Foundation, contract C-1845; and the Weston Havens Foundation (USA).

\end{acknowledgments}

\bibliography{auto_generated}

\cleardoublepage \appendix\section{The CMS Collaboration \label{app:collab}}\begin{sloppypar}\hyphenpenalty=5000\widowpenalty=500\clubpenalty=5000\vskip\cmsinstskip
\textbf{Yerevan Physics Institute, Yerevan, Armenia}\\*[0pt]
A.~Tumasyan
\vskip\cmsinstskip
\textbf{Institut f\"{u}r Hochenergiephysik, Vienna, Austria}\\*[0pt]
W.~Adam, J.W.~Andrejkovic, T.~Bergauer, S.~Chatterjee, M.~Dragicevic, A.~Escalante~Del~Valle, R.~Fr\"{u}hwirth\cmsAuthorMark{1}, M.~Jeitler\cmsAuthorMark{1}, N.~Krammer, L.~Lechner, D.~Liko, I.~Mikulec, P.~Paulitsch, F.M.~Pitters, J.~Schieck\cmsAuthorMark{1}, R.~Sch\"{o}fbeck, M.~Spanring, S.~Templ, W.~Waltenberger, C.-E.~Wulz\cmsAuthorMark{1}
% ...
\vskip\cmsinstskip
% ...
1:  Also at TU Wien, Wien, Austria\\
% ...
\end{sloppypar}
\end{document}